\documentclass[aps,footinbib, notitlepage, longbibliography,eqsecnum]{revtex4-1}
\pdfoutput=1
\usepackage{graphicx}
\usepackage{amssymb}
\usepackage{amsmath}
\usepackage{amsfonts}
\usepackage{hyperref}
\usepackage{color}
\usepackage[normalsize]{subfigure}
\usepackage{xcolor,xspace}
\usepackage{bm}
\usepackage[capitalize]{cleveref}

\crefname{section}{Sec.}{Secs.}

\DeclareFontFamily{OT1}{pzc}{}
\DeclareFontShape{OT1}{pzc}{m}{it}{<-> s * [1.10] pzcmi7t}{}
\DeclareMathAlphabet{\mathpzc}{OT1}{pzc}{m}{it}

\DeclareMathAccent{\ring}{\mathalpha}{operators}{"17}
\providecommand{\st}[1]{_{\text{#1}}}
\providecommand{\sfrac}[2]{#1/#2}
\providecommand{\ut}[1]{^{\text{#1}}}

\def\onehalf{\frac{1}{2}}

\def\bra{\ensuremath{\langle}}
\def\ket{\ensuremath{\rangle}}
\def\eq{\st{eq}}

\def\const{\mathrm{const}}

\def\opt{\st{opt}}

\def\pd{\partial}

\def\im{\mathrm{i}}
\def\sgn{\mathrm{sgn}}

\def\b0{\bv{0}}

\def\Fcal{\mathcal{F}}
\def\Hcal{\mathcal{H}}

\def\Ocal{\mathcal{O}}
\def\Pcal{\mathcal{P}}
\def\Scal{\mathcal{S}}
\def\pbc{\ut{(p)}}
\def\Dbc{\ut{(D)}}

\def\DirNoFl{^{(\mathrm{D}')}}
\def\DirCP{^{(\mathrm{D})}}

\def\hyp13{{_1 F_3}}
\def\bcs{boundary conditions\xspace}
\def\frict{\eta}
\def\kbT{\Theta}
\def\upT{\mathrm{T}}
\def\upx{\mathrm{x}}
\def\upt{\mathrm{t}}
\def\dps{\displaystyle}
\def\d{\mathrm{d}}
\def\mass{\mathcal{A}}

\def\dt{\delta t}
\def\cro{_\times}

\newcommand{\bitem}{\begin{itemize}}
\newcommand{\eitem}{\end{itemize}}
\newcommand{\benum}{\begin{enumerate}}
\newcommand{\eenum}{\end{enumerate}}
\newcommand{\bblock}[1]{\begin{block}{#1}}
\newcommand{\eblock}{\end{block}}
\newcommand{\bmini}[1]{\begin{minipage}{#1}}
\newcommand{\emini}{\end{minipage}}
\newcommand{\btab}[1]{\begin{tabular}{#1}}
\newcommand{\etab}{\end{tabular}}
\newcommand{\btabn}[1]{\begin{tabular}{#1}}
\newcommand{\etabn}{\end{tabular}}
\newcommand{\beq}{\begin{equation}}
\newcommand{\eeq}{\end{equation}}
\newcommand{\bv}[1]{\mathbf{#1}}

\begin{document}
\title{First-passage dynamics of linear stochastic interface models: \\weak-noise theory and influence of boundary conditions}
\author{Markus Gross}
\email{gross@is.mpg.de}
\affiliation{Max-Planck-Institut f\"{u}r Intelligente Systeme, Heisenbergstra{\ss}e 3, 70569 Stuttgart, Germany}
\affiliation{IV.\ Institut f\"{u}r Theoretische Physik, Universit\"{a}t Stuttgart, Pfaffenwaldring 57, 70569 Stuttgart, Germany}
\date{\today}

\begin{abstract}
We consider a one-dimensional fluctuating interfacial profile governed by the Edwards-Wilkinson or the stochastic Mullins-Herring equation for periodic, standard Dirichlet and Dirichlet no-flux \bcs.
The minimum action path of an interfacial fluctuation conditioned to reach a given maximum height $M$ at a finite (first-passage) time $T$ is calculated within the weak-noise approximation. 
Dynamic and static scaling functions for the profile shape are obtained in the transient and the equilibrium regime, i.e., for first-passage times $T$ smaller or lager than the characteristic relaxation time, respectively. 
In both regimes, the profile approaches the maximum height $M$ with a universal algebraic time dependence characterized solely by the dynamic exponent of the model. 
It is shown that, in the equilibrium regime, the spatial shape of the profile depends sensitively on \bcs and conservation laws, but it is essentially independent of them in the transient regime.
\end{abstract}


\maketitle

\section{Introduction}
\label{sec_intro}
Let $h(x,t)$ be a one-dimensional interfacial height profile $h(x,t)$ subject to either the Edwards-Wilkinson (EW) equation \cite{edwards_surface_1982}
\beq \pd_t h = \frict \pd_x^2 h + \zeta,
\label{eq_EW}
\eeq
or the stochastic Mullins-Herring (MH) equation \cite{mullins_theory_1957,herring_effect_1950, krug_origins_1997}
\beq \pd_t h = -\frict \pd_x^4 h + \pd_x \zeta.
\label{eq_MH}
\eeq
The white noise $\zeta$ is a Gaussian random variable with zero mean and correlations 
\beq \bra \zeta(x,t)\zeta(x',t')\ket = 2D\delta(x-x')\delta(t-t')\,.
\label{eq_noise}
\eeq
The friction coefficient $\frict$ and the noise strength $D$ are \textit{a priori} free parameters whose ratio can be fixed by requiring that the Gaussian steady-state distribution resulting from \cref{eq_EW,eq_MH} is characterized by a certain temperature (see, e.g., Refs.\ \cite{majumdar_spatial_2006,majumdar_airy_2005}). 
While $h$ is locally conserved for \cref{eq_MH}, the noise term in \cref{eq_EW} violates this property.

The EW equation describes surface growth caused by random deposition and relaxation.
The Kardar-Parisi-Zhang equation \cite{kardar_dynamic_1986} is a nonlinear extension of the EW equation accounting for the effect of lateral growth.
The noiseless MH equation describes interfacial relaxation under the influence of surface diffusion \cite{mullins_theory_1957}.
If $h$ represents a liquid interface, \cref{eq_MH} can be understood as a linearized stochastic lubrication equation in the absence of disjoining pressure \cite{davidovitch_spreading_2005, gruen_thin-film_2006}.
Furthermore, the stochastic Cahn-Hilliard equation, which is used in the modeling of phase-separation, reduces deep in the super-critical phase to \cref{eq_MH} \cite{elliott_cahn-hilliard_1989} \footnote{We remark that, without a microscopic cutoff, the stochastic EW and MH equations yield a diverging variance of the one-point height distribution for spatial dimensions $d\geq  2$ \cite{smith_local_2017,krug_origins_1997}. In the one-dimensional case considered here, the two models are well defined even without a regularization at small scales.}.

\begin{figure}[t]\centering
    \includegraphics[width=0.43\linewidth]{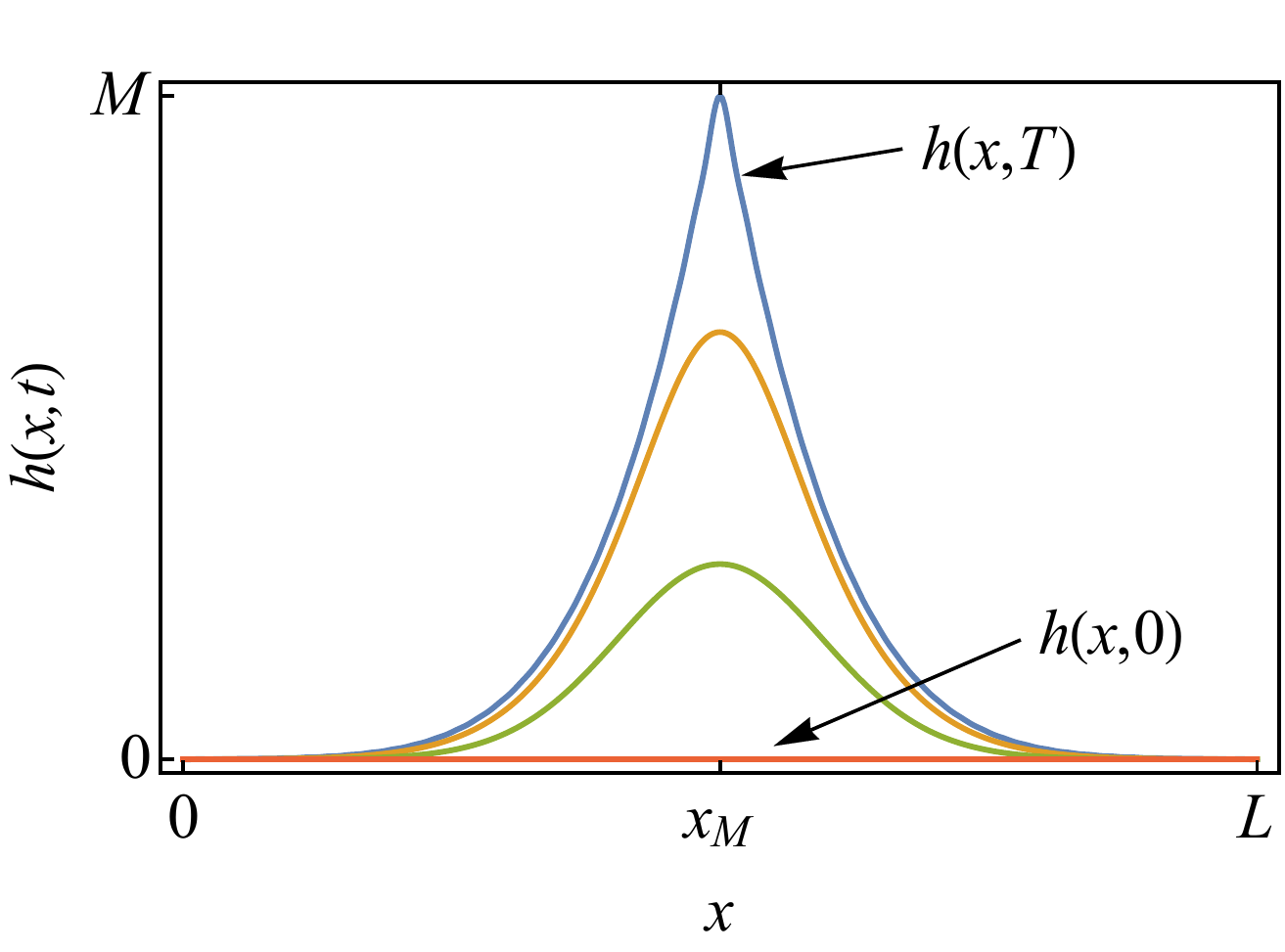}
    \caption{Situation considered in the present study: an initially flat profile $h(x,0)$ on a domain $0\leq x\leq L$ reaches a given maximum height $M$ for the first time at time $T$. $x_M$ denotes the location of this first-passage event. While the actual first-passage dynamics of the interface is stochastic [see \cref{eq_EW,eq_MH}], we focus here on the weak-noise approximation as governed by the (non-stochastic) partial differential equations in \cref{eq_S2_min,eq_S4_min}.}
    \label{fig_sketch}
\end{figure}

Interfacial fluctuations typically exhibit long-ranged correlations and non-Markovian dynamics. 
Roughening of interfaces and the associated dynamic scaling behavior emerging from \cref{eq_EW,eq_MH} has been extensively studied (see, e.g., Refs.\ \cite{abraham_dynamics_1989,racz_scaling_1991,antal_dynamic_1996, barabasi_fractal_1995, majaniemi_kinetic_1996, flekkoy_fluctuating_1995,flekkoy_fluctuating_1996,krug_origins_1997,taloni_generalized_2012, gross_interfacial_2013, halpin-healy_kinetic_1995, pruessner_drift_2004,cheang_edwardswilkinson_2011}).
More recently, extreme events and first-passage properties of interfaces have been investigated \cite{krug_persistence_1997, majumdar_spatial_2001, majumdar_exact_2004, majumdar_airy_2005, majumdar_spatial_2006, schehr_universal_2006, rambeau_extremal_2010, bray_persistence_2013,meerson_macroscopic_2016,meerson_large_2016}.
The present study focuses on the time-evolution of a profile $h(x,t)$ governed by \cref{eq_EW} or \eqref{eq_MH}, under the condition that $h$ reaches a given height $M$ for the first time at time $T$,
\beq h(x_M,T)=M,
\label{eq_firstpsg_cond}\eeq 
given that, initially,
\beq h(x,t=0)=0.
\label{eq_init_cond}\eeq 
The location $x_M$ where the height $M$ is reached first depends on the specific model as well as on the \bcs.
If $T$ is larger than the relaxation time of the interface, the interfacial roughness (i.e., the one-point one-time variance of the height fluctuations) has saturated at the first-passage event \cite{rowlinson_molecular_1982, barabasi_fractal_1995, krug_origins_1997} and the interface is accordingly governed by equilibrium dynamics (the precise meaning of this will be clarified further below). 
We consider profiles on a finite domain $[0,L]$ subject to either periodic boundary conditions (p),
\beq h\pbc(x,t)=h\pbc(x+L,t),
\label{eq_H_pbc}\eeq 
or Dirichlet \bcs (D),
\beq h\Dbc(0,t)= 0=h\Dbc(L,t).
\label{eq_H_Dbc}\eeq 
For the MH equation with Dirichlet \bcs, two further conditions are needed to completely determine the solution. 
We impose in this case a no-flux boundary condition (see also \cref{sec_eigenv_mh}):
\beq \pd_x^3 h\DirNoFl(0,t) = 0 = \pd_x^3 h\DirNoFl(L,t),
\label{eq_H_noflux}\eeq 
and henceforth indicate \cref{eq_H_Dbc,eq_H_noflux} by a superscript $\mathrm{(D')}$ \footnote{Results for the MH equation with standard Dirichlet \bcs are briefly summarized in \cref{app_std_Dirichlet_MH}.}.
We denote by the ``mass'' $\mass$ the total area under the profile:
\beq \mass([h],t) \equiv \int_0^L \d x\, h(x,t). \label{eq_mass}
\eeq 
For the EW equation with periodic \bcs, $\mass([h\pbc],t)$ is not constant in time, but instead behaves diffusively at large times \cite{krug_origins_1997}.
In this case, we consider instead of $h\pbc$ the relative height fluctuation
\beq \tilde h\pbc(x,t)\equiv h\pbc(x,t) - \mass([h\pbc],t)/L\,,
\label{eq_height_redef}\eeq
which fulfills $\mass([\tilde h\pbc],t)=0$.
We henceforth drop the tilde on $\tilde h\pbc$ in order to simplify notation.
Global conservation of the mass with
\beq \mass([h],t) =0
\label{eq_zero_vol}
\eeq
holds also for the MH equation with either periodic or Dirichlet no-flux \bcs [given \cref{eq_init_cond}]. 
For the EW equation with standard Dirichlet \bcs\ \footnote{For standard Dirichlet \bcs, the chemical potential $\mu=- \pd_x^2 h$, instead of the flux $-\pd_x \mu$, vanishes at the boundaries [see \cref{app_DirCP_eigenf}]}, the mass vanishes only after averaging over time.
\Cref{eq_height_redef}, which is rather artificial from a physical point of view, is imposed here mainly in order to compare the different models under the common mass constraint, \cref{eq_zero_vol}. 
The basic situation and the relevant quantities considered in the present study are illustrated in \cref{fig_sketch}.
In passing, we introduce the \emph{dynamic index} $z$, which describes the dependence of the relaxation time $\tau$ of a typical fluctuation governed by Eq.\ \eqref{eq_EW} or \eqref{eq_MH} on the system size $L$ via $\tau\propto L^z$, with
\beq z=2\qquad \text{(EW equation)},\qquad  \qquad  z=4\qquad \text{(MH equation)}.
\label{eq_dynindex}\eeq 

Large deviations of stochastic processes are formally described by Freidlin-Wentzel theory \cite{freidlin_random_1998, e_minimum_2004, luchinsky_analogue_1998}, which is equivalent to a Martin-Siggia-Rose/Janssen/de Dominicis path-integral formulation \cite{martin_statistical_1973, janssen_lagrangean_1976, de_dominicis_techniques_1976,tauber_critical_2014} in the limit of weak noise \cite{fogedby_minimum_2009, ge_analytical_2012, grafke_instanton_2015}.
This approach provides an action functional, the minimization of which yields the \emph{most probable} (``optimal'') path connecting two states [e.g., Eqs.\ \eqref{eq_init_cond} and \eqref{eq_firstpsg_cond}].
For an explicit derivation of the corresponding weak-noise theory (WNT) for the EW and MH equation see, e.g., Refs.\ \cite{meerson_macroscopic_2016, smith_local_2017}.
A related large deviation formalism in the context of lattice gases is reviewed in Ref.\ \cite{bertini_macroscopic_2015}.

An important predecessor to the present work is Ref.\ \cite{meerson_macroscopic_2016}, where the WNT of \cref{eq_MH} with periodic \bcs has been solved. 
Here, we extend that study by discussing further aspects of the first-passage dynamics, focusing, in particular, on the effect of \bcs. 
Within the WNT of \cref{eq_EW,eq_MH}, we obtain minimum-action paths describing extremal fluctuations of the profile fulfilling \cref{eq_init_cond,eq_firstpsg_cond}, without conditioning on the first-passage. 
We remark that the solution of WNT for Dirichlet no-flux \bcs [\cref{eq_H_Dbc,eq_H_noflux}] is technically involved since it requires the consideration of an adjoint eigenproblem [see \cref{app_DirNoFl_eigenf}].
Predictions of WNT will be compared to Langevin simulations in an accompanying paper \cite{gross_first-passage_2017-1}.

The first-passage problem for the MH equation discussed here and in Ref.\ \cite{gross_first-passage_2017-1} is relevant, \emph{inter alia}, for the rupture of liquid wetting films.
In contrast to previous studies \cite{bausch_lifetime_1994, bausch_critical_1994,blossey_nucleation_1995,foltin_critical_1997, seemann_dewetting_2001,thiele_dewetting:_2001,thiele_importance_2002,tsui_views_2003,becker_complex_2003,gruen_thin-film_2006, fetzer_thermal_2007, croll_hole_2010, blossey_thin_2012, nguyen_coexistence_2014, duran-olivencia_instability_2017}, we focus here on the case where disjoining pressure is negligible and film rupture is solely driven by noise.
A related WNT describing the noise-induced breakup of a liquid thread has been analyzed in Ref.\ \cite{eggers_dynamics_2002}.
Rare-event trajectories of the kind considered here are furthermore relevant for the understanding of chemical reaction pathways \cite{e_transition-path_2010,kim_mean_2015,delarue_ab_2017}, phase transitions \cite{e_minimum_2004,li_numerical_2012} as well as for certain aspects in interfacial wetting (see Ref.\ \cite{belardinelli_thermal_2016} and references therein). 

The main results of the present study are contained in \cref{sec_EW,sec_MH}, in which the necessary formalism of WNT for the EW and MH equation, respectively, is introduced and the exact analytical solution for the first-passage profile is discussed. The determination of the analytical solution as well as further mathematical details are deferred to \crefrange{app_min_eqaction}{app_WNT_sol}.  
In the main part (\cref{sec_fullsol_nc,sec_fullsol_c}), we focus on the time-evolution of the first-passage profile in the case of periodic and Dirichlet (no-flux) \bcs. For first-passage times $T\ll \tau$ (transient regime) we find that the profile shape essentially depends only on the type of bulk dynamics, while the influence of \bcs and mass conservation is negligible. In contrast, at late times $T\gg \tau$ (equilibrium regime), the profile evolves over the whole domain and strongly depends on the specific \bcs.
In both temporal regimes, simple analytical expressions for the asymptotic dynamic and static scaling profiles are derived. 
These scaling forms indicate that, within WNT, the peak height $h(x_M,t)$ of the profile approaches the first-passage height $M$ in time with a \emph{universal} exponent $1/z$. 
Moreover, it is shown that, in the presence of a microscopic cutoff, the dynamic scaling exponent eventually crosses over to a value of 1 close to the first-passage event.

\section{Edwards-Wilkinson equation}
\label{sec_EW}

\subsection{Macroscopic fluctuation theory}
\label{sec_ncons_dyn}
The Martin-Siggia-Rose field-theoretical action pertaining to \cref{eq_EW} is given by \cite{tauber_critical_2014,smith_local_2017}
\beq \Scal[h,p] = \int_0^T \d t \int_{0}^{L} \d x\,\left[ p(\pd_t h - \frict \pd_x^2 h) - Dp^2\right],
\label{eq_S2}
\eeq
where $p$ is an auxiliary (``conjugate'') field.
The most-probable (optimal) path emerging from the stochastic dynamics is the one that minimizes $\Scal$:
\begin{subequations}
\begin{align}
0 &= \frac{\delta \Scal}{\delta p} = \pd_t h - \frict \pd_x^2 h -2Dp, \label{eq_S2_min_h}\\
0 &= \frac{\delta \Scal}{\delta h} = -\pd_t p - \frict \pd_x^2 p\,.\label{eq_S2_min_p}
\end{align}\label{eq_S2_min}
\end{subequations}
The field $p$, which can be interpreted as the typical noise magnitude, is governed by an anti-diffusion equation [\cref{eq_S2_min_p}]. This indicates that the creation of a rare event requires the local accumulation of noise intensity. 
We consider either periodic \bcs [\cref{eq_H_pbc}],
\beq h\pbc(x,t)=h\pbc(x+L,t),\qquad p\pbc(x,t)=p\pbc(x+L,t),
\label{eq_h2_pbc}\eeq 
or Dirichlet \bcs [\cref{eq_H_Dbc}],
\beq h\Dbc(0,t)=0=h\Dbc(L,t),\qquad  p\Dbc(0,t)=0=p\Dbc(L,t).
\label{eq_h2_Dbc}\eeq 
Note that, since $\pd_x^2$ is self-adjoint on $[0,L]$ for the considered \bcs, $p$ fulfills the same \bcs as $h$ (see also \cref{sec_eigenv_mh,app_WNT_sol}).
Inserting the mean-field equations \eqref{eq_S2_min} into the action in \cref{eq_S2} yields the optimal action
\beq \Scal\opt = D \int_0^{T} \d t \int_{0}^{L} \d x\, p^2 .
\label{eq_Sopt2}\eeq

\Cref{eq_S2_min} admits a special solution which can be identified with thermal \emph{equilibrium}.
In equilibrium, the most-likely noise-activated trajectory $h(x,t)$ is the time-reversed of the corresponding relaxation trajectory $h_r(x,t)$ --- a property known as Onsager-Machlup symmetry \cite{onsager_fluctuations_1953}.
In order to exhibit this symmetry for the dynamics described by \cref{eq_S2_min}, consider the solution $h_r(x,t)$ of the noise-free analog of \cref{eq_S2_min_h}, i.e., the diffusion equation
\beq \pd_t h_r = \frict \pd_x^2 h_r,
\label{eq_diffus_nc}\eeq 
with initial condition $h_r(x,t=0)= h_0(x)$, where $h_0(x)$ is a given profile [e.g. the equilibrium first-passage profile $h(x,T\to\infty)$, which can be determined independently, see \cref{eq_opt2_finalprof_eq} below].
Then, the solution $h(x,t)$, $p(x,t)$ of \cref{eq_S2_min}, fulfilling $h(x,T)=h_0(x)$ at some final time $T$, is given by 
\beq h(x,t) = h_r(T-t),\quad p(x,t) = -\frac{\frict}{D}\pd_x^2 h(x,t).
\label{eq_actrel_sol_nc}\eeq 
Indeed, it is readily checked that \cref{eq_actrel_sol_nc} solves \cref{eq_S2_min}, as
\beq \pd_t h = -\pd_t h_r = -\frict \pd_x^2 h= \frict \pd_x^2 h + 2Dp,
\label{eq_heq_act_antidiff}\eeq
which is precisely \cref{eq_S2_min_h}; furthermore $\pd_t p= -(\frict/D) \pd_x^2 \pd_t h= (\frict^2/D) \pd_x^4 h = -\frict \pd_x^2 p$, which is \cref{eq_S2_min_p}.
According to \cref{eq_heq_act_antidiff}, $h$ effectively obeys an anti-diffusion equation in the equilibrium regime.
Note that the ansatz in \cref{eq_actrel_sol_nc} implies that the time evolution starts at time $t=0$ from the initial configuration $h(x,0)= h_r(T)$, which is flat only for $T\to \infty$.  
Accordingly, under requirement of \cref{eq_init_cond}, the equilibrium regime corresponds to \emph{large} first-passage times $T$---as anticipated in \cref{sec_intro}.
The general solution of \cref{eq_S2_min} fulfilling \cref{eq_init_cond,eq_firstpsg_cond} for arbitrary $T$ is presented below.

In the equilibrium regime, upon using \cref{eq_actrel_sol_nc,eq_heq_act_antidiff}, the optimal action in \cref{eq_Sopt2} reduces to
\beq \begin{split}
\Scal\st{opt,eq} &= \frac{\frict^2}{D} \int_{0}^{T} \d t \int_{0}^{L} \d x\, (\pd_x^2 h)^2 = -\frac{\frict^2}{D} \int_{0}^{T} \d t \int_{0}^{L} \d x\, (\pd_x h)(\pd_x^3 h) \\
&= \frac{\frict}{D} \int_{0}^{T} \d t \int_{0}^{L} \d x\, (\pd_x h)(\pd^2_{xt} h) = \frac{\frict}{D} \int_{0}^{L} \d x\, (\pd_x h)^2\Big|_{0}^{T} - \frac{\frict}{D} \int_{0}^{T} \d t \int_{0}^{L} \d x\, (\pd^2_{xt} h)(\pd_x h) \\
&= \frac{\frict}{2D} \int_{0}^{L} \d x\, (\pd_x h)^2\Big|_{0}^{T}.
\end{split} \label{eq_Sopt2_eq}\eeq 
In the partial integrations above we made use of the fact that the spatial boundary terms generally vanish for periodic and Dirichlet \bcs\ \footnote{Note that standard Dirichlet \bcs imply $\pd_x^2 h(x)=0$ for $x\in\{0,L\}$, as can be inferred from the series representation in \cref{eq4_DirCP0_hsol}.}.
\Cref{eq_Sopt2_eq} provides a fluctuation-dissipation relation, from which the temperature $\kbT$ (in units of $k_B$) can be identified via $\frict/(2D) = 1/(4 \kbT)$.

We henceforth consider time to be rescaled by the friction coefficient $\frict$, i.e., $\tilde t=\frict t$,  and define new fields $\tilde h$, $\tilde p$ via 
\beq h(x,t)\equiv \tilde h(x,\frict t),\qquad p(x,t)= (\frict/D)\tilde p(x,\frict t).
\label{eq_h_rescaled}\eeq 
The Euler-Lagrange equations in \cref{eq_S2_min} can then be cast into the form
\begin{subequations}
\begin{align}
\pd_{\tilde t} \tilde h &=  \pd_x^2 \tilde h +2 \tilde p, \label{eq_EW_MFTr_h} \\
\pd_{\tilde t} \tilde p &= - \pd_x^2 \tilde p\,. \label{eq_EW_MFTr_p}
\end{align}\label{eq_EW_MFTr}
\end{subequations}
Analogously, $\Scal\opt$ in \cref{eq_Sopt2} can be expressed in terms of the  rescaled action 
\beq \tilde\Scal\opt\equiv  \int_0^{\tilde T}\d \tilde t\int_0^L \d x\, \tilde p(x,\tilde t)^2
\label{eq_Sopt2_resc}\eeq 
as
\beq \Scal\opt = \frac{\eta}{D} \tilde\Scal\opt,
\label{eq_Sopt2_resc_relation}\eeq 
with $\tilde T\equiv \frict T$. It is useful to remark that the dimension of $\eta/D$ is the same as of $L/M^2$.
\Cref{eq_Sopt2_resc_relation} makes it obvious that the saddle-point solution of the action dominates the dynamics in the weak-noise limit $D\to 0$.
We proceed with the analysis of \cref{eq_EW_MFTr,eq_Sopt2_resc} and henceforth drop the tilde in order to simplify the notation.

\subsection{Exact solution}
\label{sec_fullsol_nc}
\begin{figure}[t]\centering
    \subfigure[]{\includegraphics[width=0.49\linewidth]{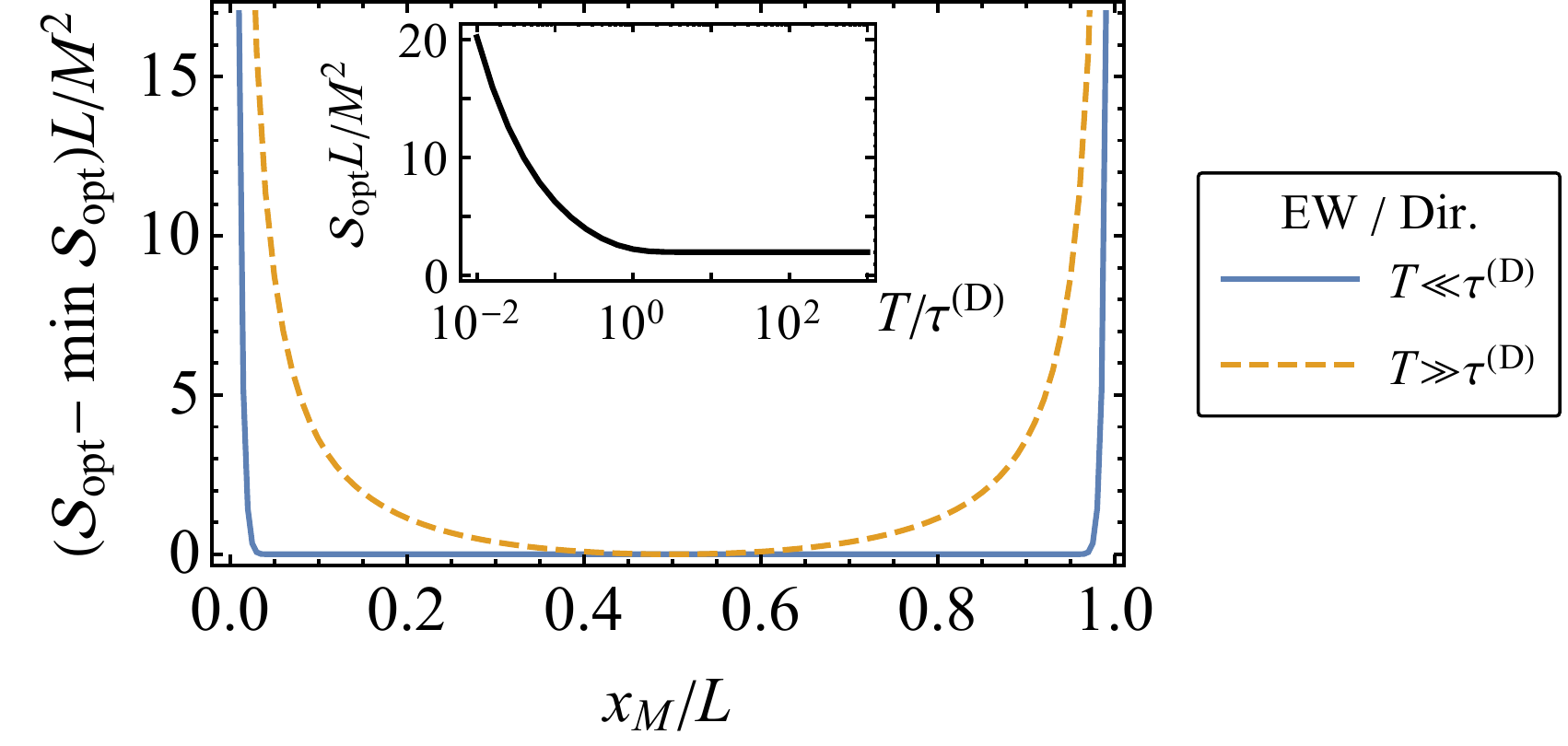} \label{fig_Sopt_EW}}\quad
    \subfigure[]{\includegraphics[width=0.47\linewidth]{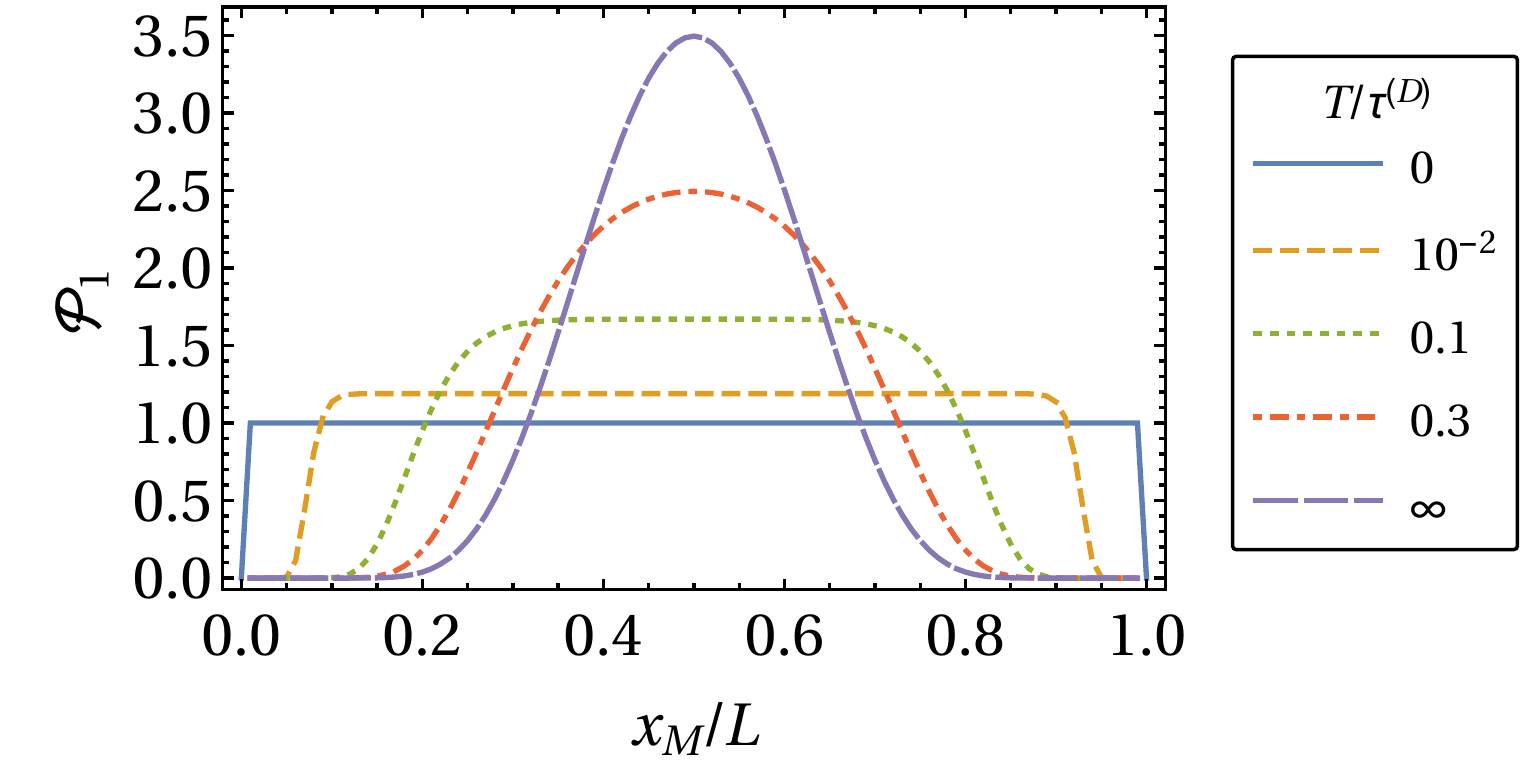} \label{fig_fpsg_EW}}
    \caption{(a) Optimal action $\Scal\opt\Dbc$ [\cref{eq_Sopt_expr}] for the EW equation with Dirichlet \bcs. The curves for $\Scal\opt\Dbc$ are shifted such that their respective minima are zero. For sufficiently large or small $T$, $\Scal\opt\Dbc-\mathrm{min}\Scal\opt\Dbc$ becomes independent of $T$. Asymptotically for $T\to 0$ in the transient regime, $\Scal\opt\Dbc$ is spatially constant (and nonzero) for $0<x_M<L$. Due to Dirichlet \bcs, $\Scal\opt\Dbc$ diverges for $x_M=0,L$. The inset shows $\Scal\opt\Dbc$ evaluated for $x_M=L/2$, which approaches a nonzero constant for $T\gg\tau\Dbc$ and diverges $\propto T^{-1/z}$ as $T\to 0$ [see \cref{eq6_Sopt_shortT}]. (b) Probability distribution $\Pcal_1\Dbc$ [\cref{eq_P1_EW}] of the first-passage location $x_M$ for Dirichlet \bcs, $M^2/L=2$ (in units of $\eta/D$) and various values of $T/\tau\Dbc$. The curves labeled by $T/\tau\Dbc=0$ and $\infty$ pertain to the asymptotic transient and the equilibrium regime, respectively, where $\Pcal_1\Dbc$ is independent of $T$. Upon increasing $M^2/L$, the width of the curves (except the one corresponding to $T/\tau\Dbc\to 0$) decrease and their peak height increases.}
    \label{fig_Sopt_EW_extra}
\end{figure}

The solution of \cref{eq_EW_MFTr} subject to the initial and final conditions in \cref{eq_firstpsg_cond,eq_init_cond} as well as to the \bcs in \cref{eq_h2_pbc} or \cref{eq_h2_Dbc} can be determined exactly [see \cref{app_WNT_sol}] and is summarized below.
It turns out that initial and final conditions for $p$ do not have to be specified additionally, but instead implicitly follow from the ones imposed on $h$.
Two characteristic regimes can be distinguished: a transient regime, corresponding to first-passage times $T \ll \tau$, and an equilibrium regime, corresponding to $T\gg\tau$. The relaxation time $\tau$ is given by ($z=2$)
\begin{subequations}\begin{align}
\tau\pbc &= \left(\frac{L}{2\pi}\right)^z  \label{eq_nc_timescale_pbc} 
\intertext{for periodic and by}
\tau\Dbc &= \left(\frac{L}{\pi}\right)^z \label{eq_nc_timescale_Dir}
\end{align}\label{eq_nc_timescales}\end{subequations}
for Dirichlet boundary conditions.
Within WNT, $\tau$ is in fact the characteristic time scale for the creation of a first-passage event.
Asymptotically for $T\to\infty$, the profile in the equilibrium regime fulfills \cref{eq_actrel_sol_nc}.

\begin{figure}[t]\centering
    \subfigure[]{\includegraphics[width=0.43\linewidth]{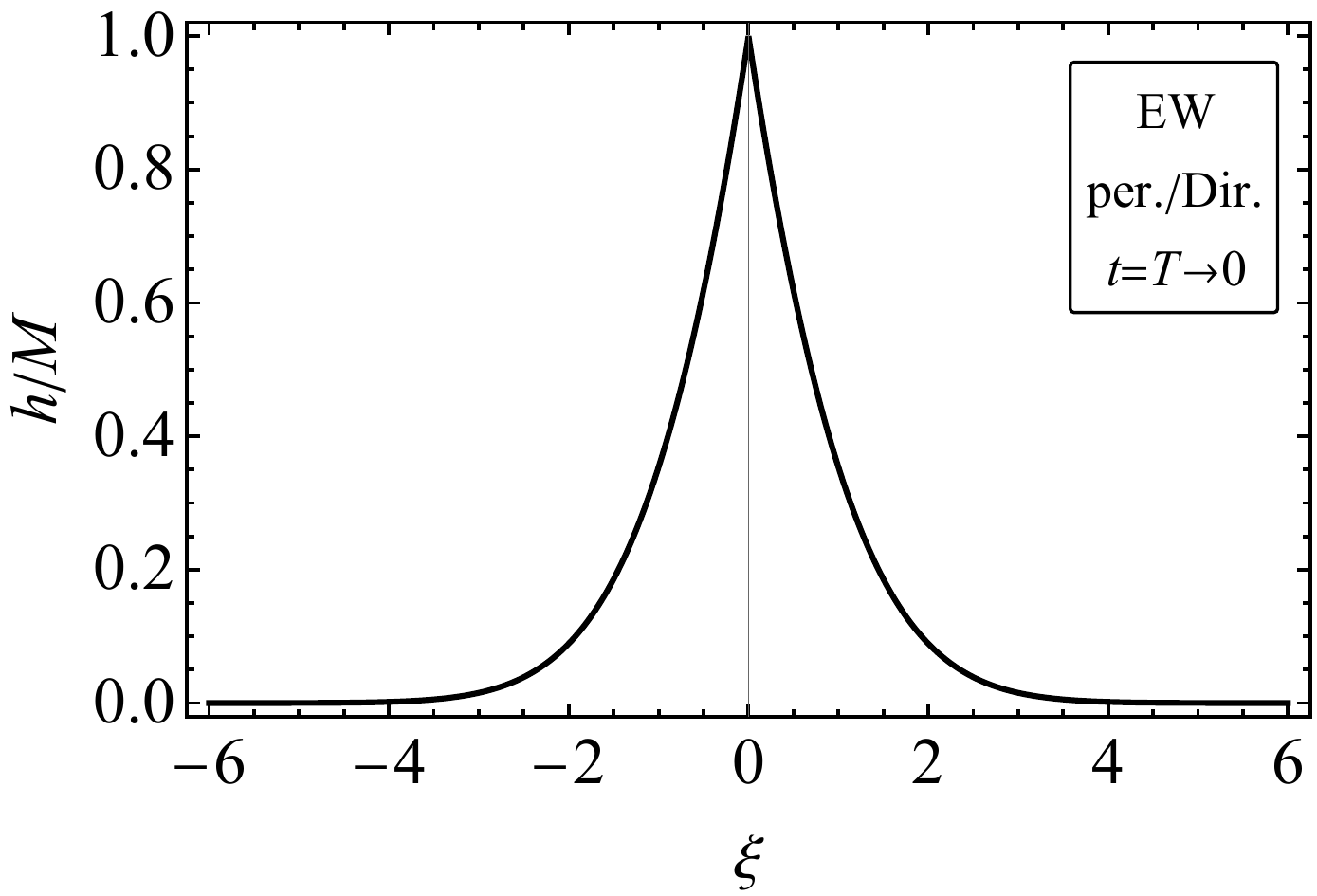}}\qquad
    \subfigure[]{\includegraphics[width=0.44\linewidth]{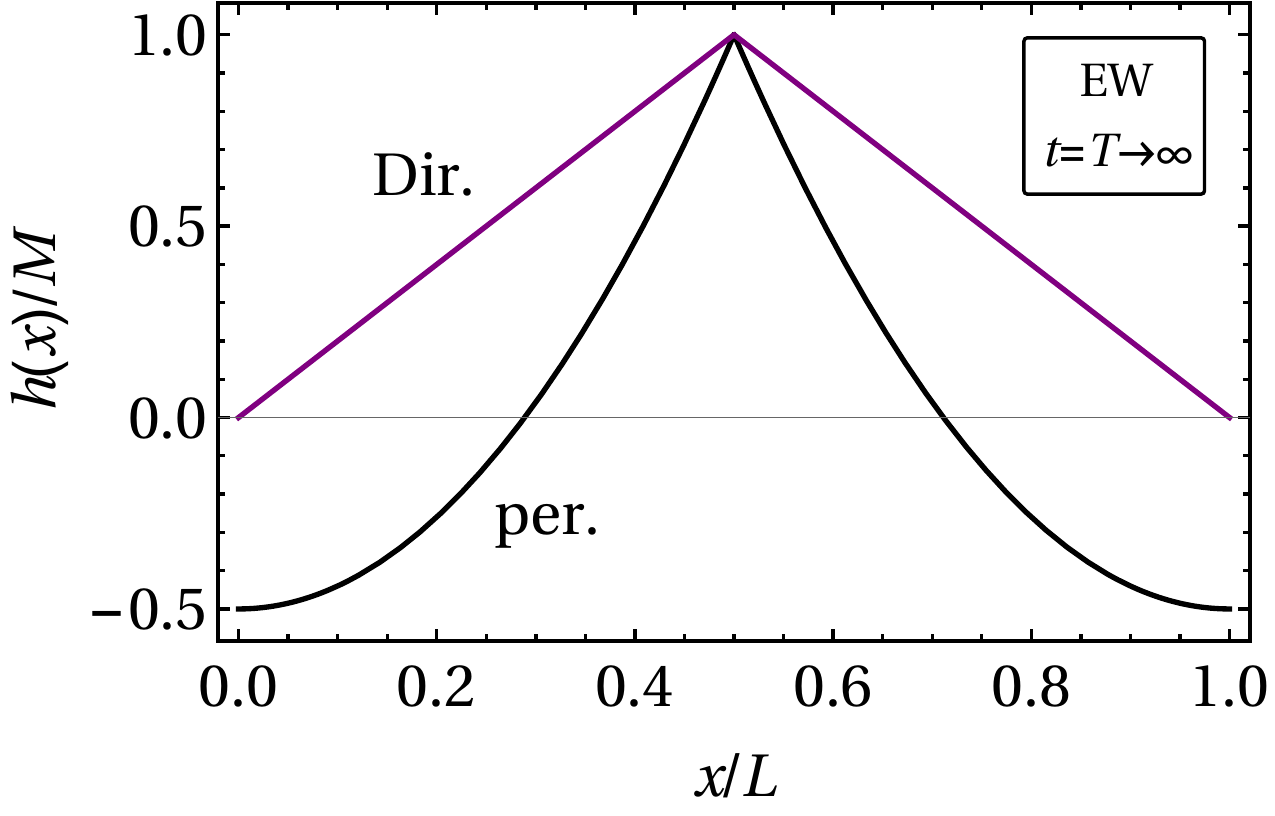}\label{fig_opt_prof_nc_limit_eq}}
    \caption{Asymptotic first-passage profiles $h(x,t=T)$ (normalized by $M$) obtained within WNT of the EW equation [\cref{eq_EW}] in (a) the transient regime, $T\to 0$ [\cref{eq_h_shortT_EW}], and (b) the equilibrium regime, $T\to\infty$ [\cref{eq_opt2_finalprof_eq}]. In the transient regime, the profiles depend on the scaling variable $\xi\equiv (x-L/2)/(2T)^{1/2}$ and are identical for periodic and Dirichlet \bcs. In the equilibrium regime, the (normalized) profile is a function of $x/L$ and is specific to each boundary condition. }
    \label{fig_opt_prof_nc_limits}
\end{figure}

The optimal action [\cref{eq_Sopt2_resc}] has the following formal scaling property [see \cref{app_WNT_sol}]: 
\beq \Scal\opt(x_M,M, T,L) = \frac{M^2}{L}\Scal\opt\left( \frac{x_M}{L}, 1, \frac{T}{L^z}, 1 \right).
\label{eq_Sopt_expr}\eeq 
Recalling \cref{eq_Sopt2_resc_relation}, \cref{eq_Sopt_expr} accordingly demonstrates that, \emph{within} WNT, the weak-noise limit $D\to 0$ is equivalent to the limit of large heights $M^2/L\to\infty$. 
Furthermore, $\Scal\opt$ determines the probability distribution of the first-passage coordinate $x_M$, 
\beq \mathcal{P}_1(x_M) \sim \exp[-\Scal\opt(x_M,M,T,L)],
\label{eq_P1_EW}\eeq 
which is assumed to be normalized such that $\int_0^L \d x_M\, \Pcal_1(x_M)=1$.
For the purpose of numerical evaluation it is convenient to use the relation $\Scal\opt(x_M,M, T,L)= \sfrac{M^2}{[2Q(x_M,T,L)]}$, where the function $Q$ is reported in \cref{eq6_Q}. 
\Cref{fig_Sopt_EW} displays $\Scal\opt\Dbc$ as a function of $x_M$ for Dirichlet \bcs in the asymptotic transient ($T\ll \tau\Dbc$) and equilibrium regimes ($T\gg \tau\Dbc$). 
In equilibrium, $\Scal\st{opt}$ generally simplifies to $\Scal\st{opt,eq}$ in \cref{eq_Sopt2_eq}. Minimization of $\Scal\st{opt,eq}\Dbc$ yields [see \cref{app_min_eqaction}]
\beq x_M\DirCP=L/2.
\label{eq_xM_EW}\eeq 
Asymptotically for $T\to 0$ one has $\Scal\opt \propto T^{-1/z}$ [see \cref{eq6_Sopt_shortT}].
Specifically, for $T\to 0$ and Dirichlet \bcs, $\Scal\opt\Dbc$ becomes independent of $x_M$ for $0<x_M<L$ and diverges for $x_M\in\{0,L\}$.
For definiteness, we shall henceforth take for $x_M$ in the transient regime the same value as in \cref{eq_xM_EW}. 
In fact, since the short-time profile is strongly localized for $T\to 0$ [see, e.g., \cref{fig_opt_prof_nc_limits}(a)], its shape is independent of the precise value of $x_M$.
In \cref{fig_fpsg_EW}, the first-passage distribution in \cref{eq_P1_EW} is illustrated for Dirichlet \bcs and an (arbitrarily chosen) reduced height $M^2/L=2$ [in units of $\eta/D$, see \cref{eq_Sopt2_resc_relation}]. One observes a smooth transition between the shapes pertaining to the asymptotic transient and equilibrium regimes, in both of which $\Pcal_1\Dbc$ is independent of $T$.
Upon increasing the value of $M^2/L$ for nonzero $T/\tau\Dbc$, the maximum height of the distribution increases and, correspondingly, its width decreases. In the limit $M^2/L\to\infty$, $\Pcal_1$ approaches a Dirac delta-function.

The profile $h(x,t)$ solving \cref{eq_EW_MFTr} can be brought into the following scaling form: 
\beq h(x,t,T,M,L) = M \mathpzc{h}\left(\frac{x}{L},\frac{t}{\tau}, \frac{T}{\tau}\right),
\label{eq_opt2_hscalform}\eeq 
where, for periodic \bcs, the scaling function $\mathpzc{h}$ is given by [see \cref{eq6_h_pbc,eq6_Q_pbc}] 
\beq \mathpzc{h}\pbc(\upx,\upt,\upT) = \frac{1}{\mathpzc{Q}\pbc(\upT)}\sum_{k=1}^\infty \frac{1-\exp\left(-2 k^2 \upT\right)}{k^2} \frac{\sinh\left(k^2 \upt\right)}{\sinh\left(k^2 \upT\right)} \cos\left(2\pi k(\upx-1/2)\right)
\label{eq_opt2_solh_pbc}\eeq 
with 
\beq \mathpzc{Q}\pbc(\upT) \equiv \sum_{k=1}^\infty \frac{1-\exp\left(-2 k^{2} \upT\right)}{k^2}.
\label{eq_opt2_solh_Q_pbc}\eeq
Although $\Scal\opt\pbc$ [\cref{eq_Sopt2_resc}] is manifestly independent of $x_M$ owing to translational invariance, for definiteness we choose $x_M\pbc=L/2$, which also simplifies the expressions for $h$ somewhat. 
As a consequence of explicitly enforcing the mass constraint [\cref{eq_zero_vol}] in this case, the zero-mode ($k=0$) is absent from \cref{eq_opt2_solh_pbc,eq_opt2_solh_Q_pbc} [see \cref{eq6_zeromode_pbc}].
Indeed, since $\int_0^L \d x\, \cos(2\pi k(x/L-1/2))=0$ for $k\geq 1$, the mass vanishes identically for $h\pbc$.
For Dirichlet \bcs, using \cref{eq_xM_EW}, one has [see \cref{eq6_h_Dir,eq6_Q_Dir,eq6_eigen_Dir_halfpt,eq6_eigen_Dir_prodrep}]
\beq \mathpzc{h}\Dbc(\upx,\upt,\upT) = \frac{1}{\mathpzc{Q}\Dbc(\upT)}\sum_{k=1,3,5,\ldots}^\infty \frac{1-\exp\left(-2 k^2 \upT\right)}{k^2} \frac{\sinh\left(k^2 \upt\right)}{\sinh\left(k^2 \upT\right)} \cos\left(\pi k (\upx-1/2)\right)
\label{eq_opt2_solh_Dir}\eeq 
with
\beq \mathpzc{Q}\Dbc(\upT) \equiv \sum_{k=1,3,5,\ldots}^\infty \frac{1-\exp\left(-2 k^{2} \upT\right)}{\lambda_k^2}.
\label{eq_opt2_solh_Q_Dir}\eeq
Since $x_M\Dbc=L/2$, the above sums run only over the odd eigenmodes $k=1,3,5,\ldots$, which have nonzero mass, $\int_0^L \d x\, \sin(k\pi x/L)=L/(k\pi)$ (eigenfunctions for even $k$ have vanishing mass).
The general expression for the conjugate field $p(x,t)$ is provided in \cref{eq6_p_expr}.

\begin{figure}[t]\centering
    \subfigure[]{\includegraphics[width=0.42\linewidth]{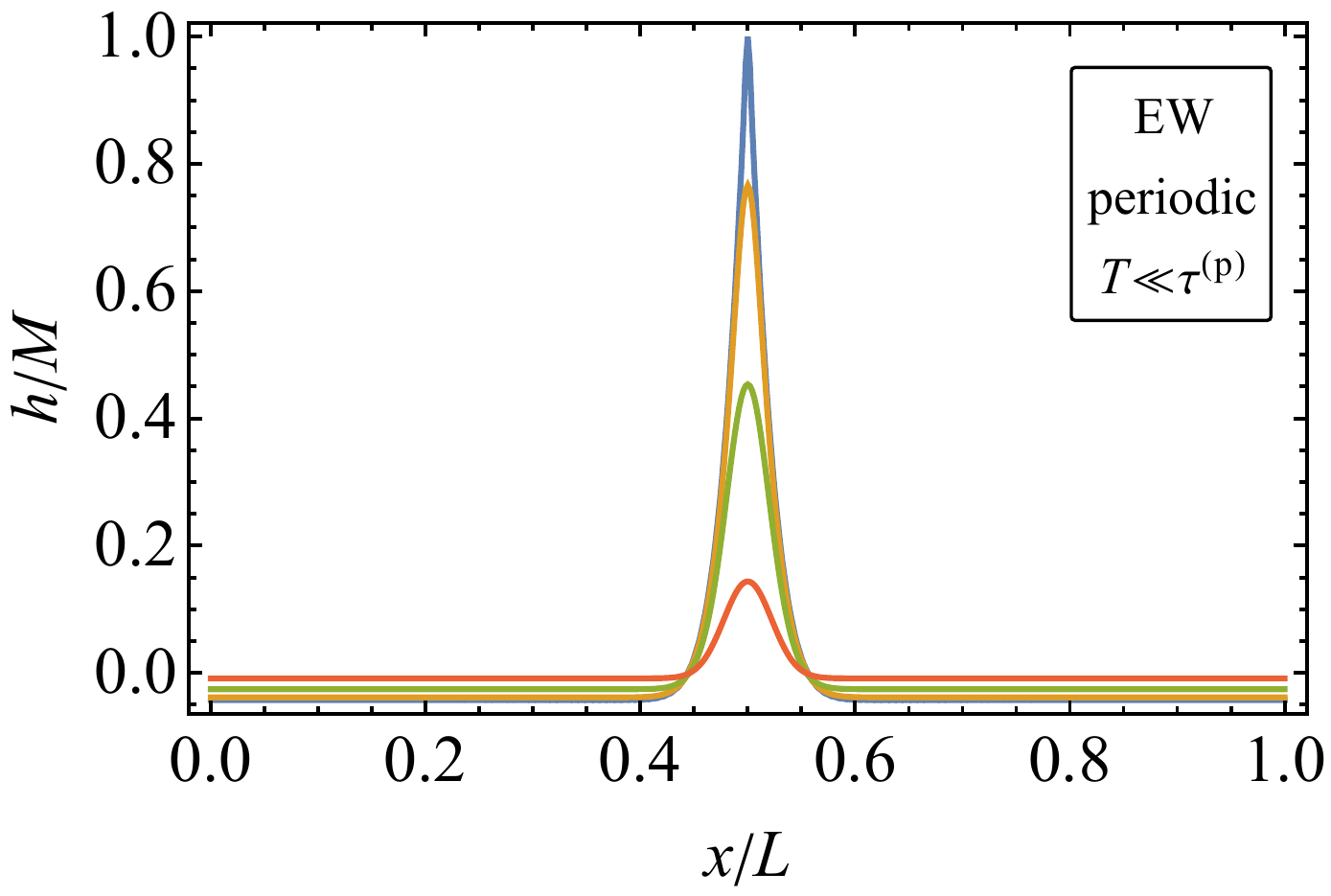}}\qquad
    \subfigure[]{\includegraphics[width=0.43\linewidth]{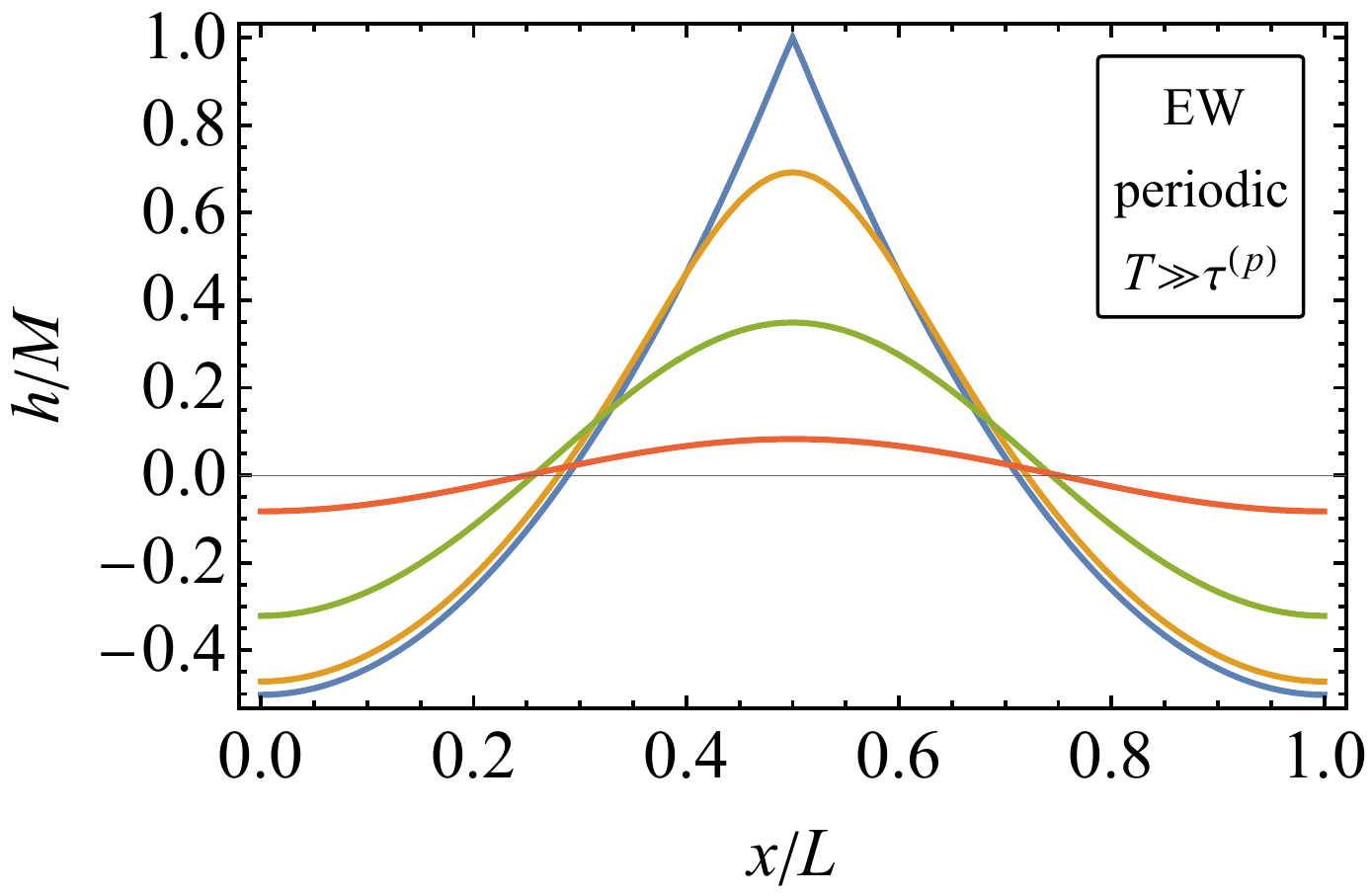}}
    \caption{Time evolution of the optimal profile [\cref{eq_opt2_hscalform,eq_opt2_solh_pbc}] for the EW equation with periodic \bcs in (a) the transient and (b) the equilibrium regime. The curves correspond, from center top to bottom, to (a) $1-t/T=0, 0.1,0.4,0.8$ with $T=10^{-2}\tau\pbc$, and (b) $1-t/T = 0, 0.001, 0.006, 0.02$ with $T=100\tau\pbc$. Decreasing $T$ in (a) leads essentially to a reduction of the width of the curves [see also \cref{eq_h2_shortTdyn_asympt}]. The fundamental time scale $\tau\pbc$ is reported in \cref{eq_nc_timescale_pbc}.}
    \label{fig_mft_prof_nc_pbc}
\end{figure}

\begin{figure}[t]\centering
    \subfigure[]{\includegraphics[width=0.42\linewidth]{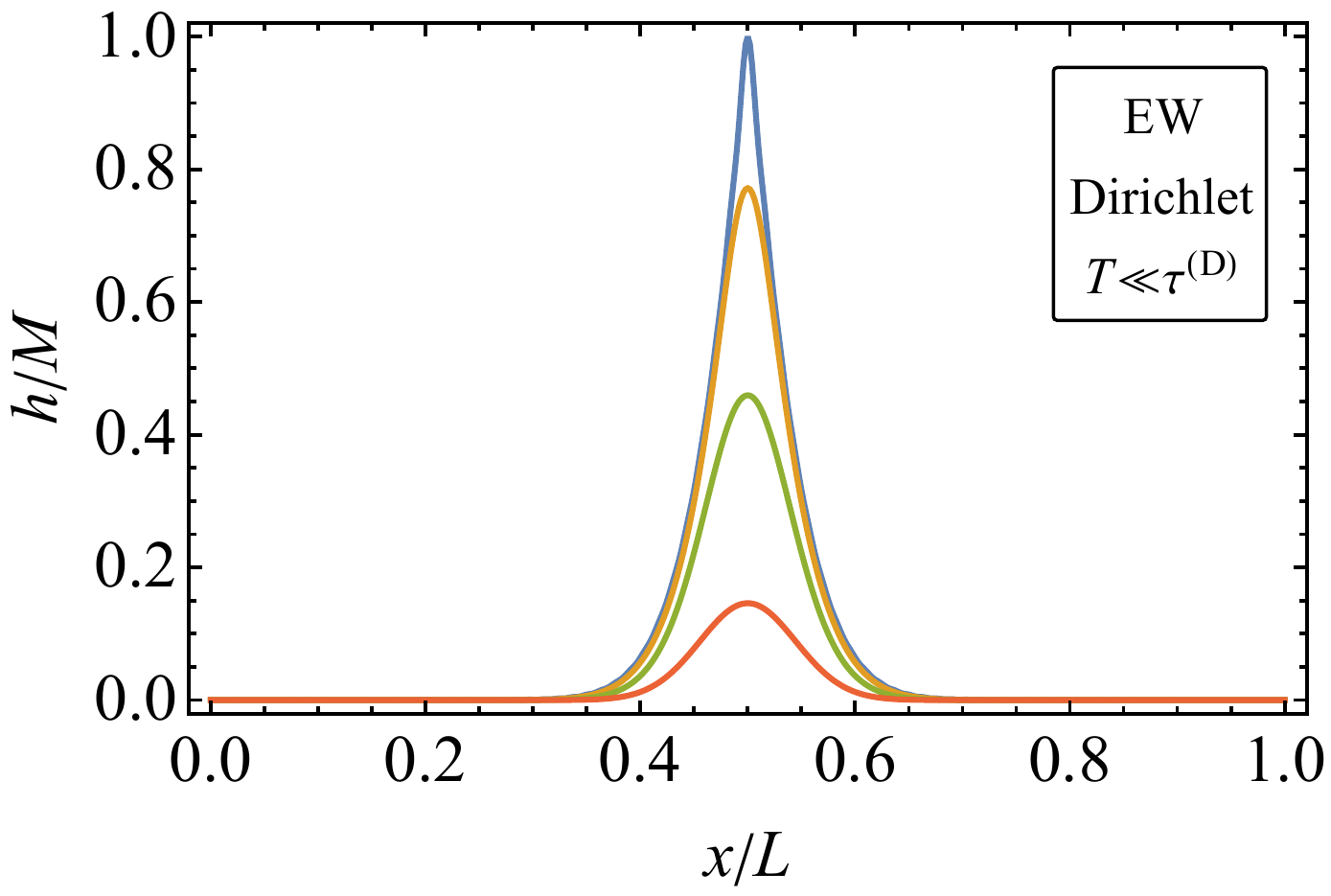}}\qquad
    \subfigure[]{\includegraphics[width=0.42\linewidth]{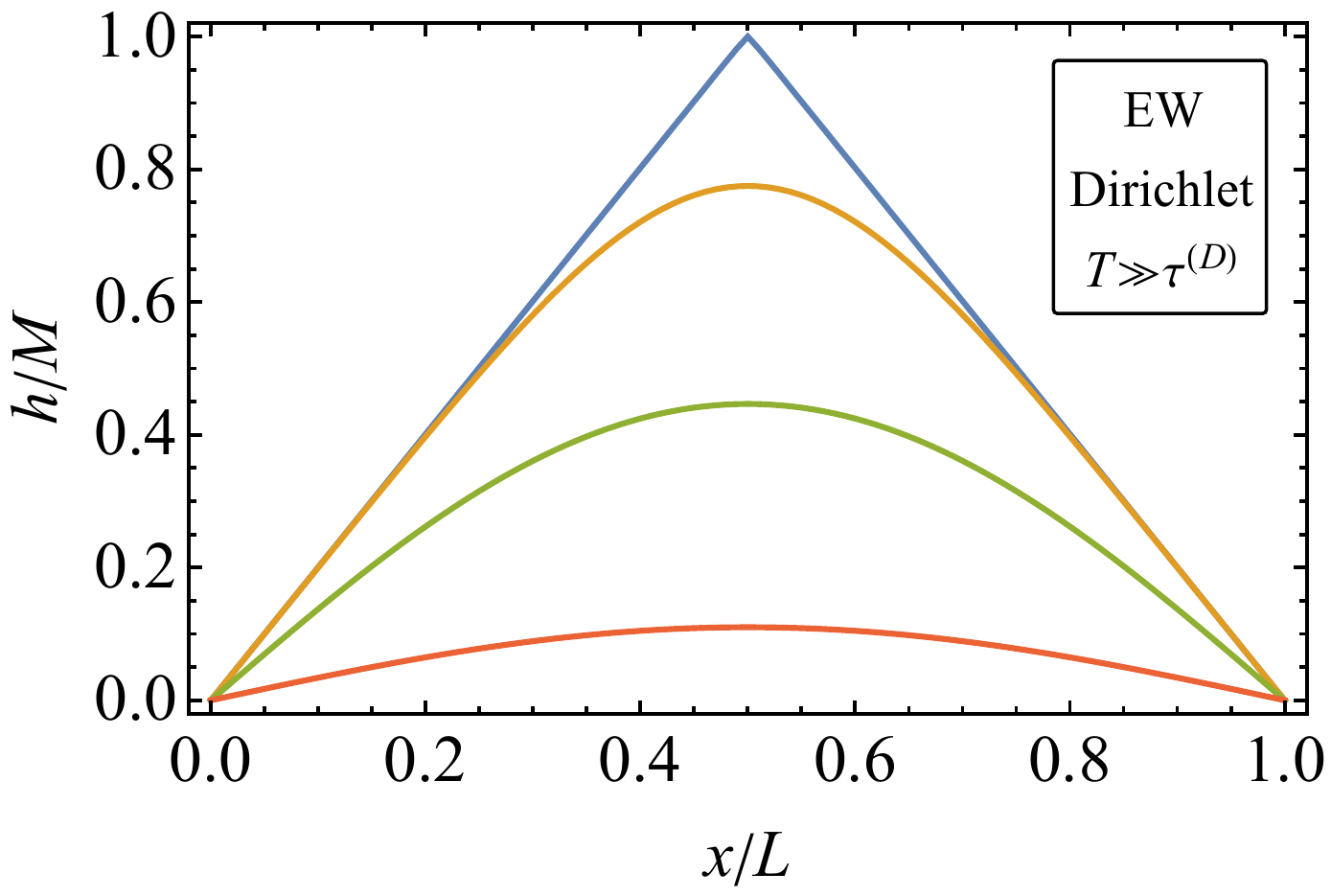}}
    \caption{Time evolution of the optimal profile [\cref{eq_opt2_hscalform,eq_opt2_solh_Dir}] for the EW equation with Dirichlet \bcs in (a) the transient and (b) the equilibrium regime. The curves correspond, from center top to bottom, to (a) $1-t/T=0, 0.1, 0.4, 0.8$ with $T=10^{-2}\tau\Dbc$, and (b) $1-t/T = 0, 0.001, 0.006, 0.02$ with $T=100\tau\Dbc$. Decreasing $T$ in (a) leads essentially to a reduction of the width of the curves [see also \cref{eq_h2_shortTdyn_asympt}]. The fundamental time scale $\tau\Dbc$ is reported in \cref{eq_nc_timescale_Dir}.}
    \label{fig_mft_prof_nc_Dir}
\end{figure}

The typical spatio-temporal evolution of $h(x,t)$ is illustrated in \cref{fig_mft_prof_nc_pbc,fig_mft_prof_nc_Dir} for periodic and Dirichlet \bcs, respectively.
In the equilibrium regime ($T\gg \tau$), the profile at time $t=T\to\infty$ can be readily calculated from \cref{eq_opt2_solh_pbc,eq_opt2_solh_Dir} [see \cref{eq6_h_eq_limprof} in \cref{sec_optprof_lateT}]:
\begin{subequations}\begin{align}
h\pbc(x,T)\big|_{T\to\infty}/M &=   1-6\Bigg|\frac{x}{L}-\onehalf\Bigg| + 6\left(\frac{x}{L}-\onehalf\right)^2, \label{eq_opt2_finalprof_eq_pbc} \\
h\Dbc(x,T)\big|_{T\to\infty}/M  &= 1-\left|1-\frac{2x}{L}\right|. \label{eq_opt2_finalprof_eq_Dir}
\end{align}\label{eq_opt2_finalprof_eq}\end{subequations}
The same results are obtained via minimization of the equilibrium action in \cref{eq_Sopt2_eq}, using the fact that $h(x,0)=0$ [see \cref{app_min_eqaction}].
For times $t=T-\delta t<T$ with $\delta t\ll T$ and $T\gg\tau$, \cref{eq_opt2_hscalform} adopts a reduced dynamic scaling form [see \cref{eq6_h_lateT_full}]:
\beq h(x,T-\delta t)\big|_{T\gg\tau} \simeq M -  M (\delta t)^{1/z} \Gamma(1-1/z) \tilde\Hcal\left(\frac{x-L/2}{\delta t^{1/z}}\right),\qquad z=2,
\label{eq_h_lateT_dynscal_EW}\eeq 
with the scaling function
\beq \tilde \Hcal(\xi) = 
\exp\left(-\frac{\xi^2}{4}\right) + \onehalf \sqrt{\pi} \, \xi\,  \mathrm{erf}\left(\frac{\xi}{2}\right), \label{eq_hscalf_asympt_EW}\eeq 
both for periodic and Dirichlet \bcs.
It is convenient to carry along the dynamic index $z$ [\cref{eq_dynindex}] in these and the following expressions.
Note that $\eta$ has the same dimension as $L^z/T$, such that, upon re-instating the unscaled quantities [see \cref{eq_h_rescaled}], the argument of $\tilde\Hcal$ in \cref{eq_h_lateT_dynscal_EW} is seen to be dimensionless. 

In the transient regime ($T\ll \tau$), the scaling profile at time $t=T$ is given by [see \cref{eq6_h_shortT}]:
\beq h(x,T)\big|_{T\ll \tau} = M \Hcal\left(\frac{x-L/2}{(2T)^{1/z}}\right),\qquad z=2,
\label{eq_h_shortT_EW}\eeq 
with the scaling function
\beq \Hcal(\xi)=
\exp\left(-\frac{\xi^2}{4}\right) + \onehalf \sqrt{\pi} |\xi| \left[\mathrm{erf}\left(\frac{|\xi|}{2}\right)-1 \right].
\label{eq_h_shortT_scalF_EW}\eeq 
Since there is no risk of confusion, we use the same symbol $\xi$ for the scaling variables in \cref{eq_hscalf_asympt_EW,eq_h_shortT_scalF_EW}.
For times $t=T-\delta t<T$ in the limit $\delta t/T\to 0$ (with $T\ll \tau$), a dynamic scaling profile follows as [see \cref{eq6_h_shortTdyn_asympt}]
\beq h(x,T-\delta t)\Big|_{\substack{T\ll \tau\\\delta t\ll T}} = M - M \left(\frac{\delta t}{2T}\right)^{1/z} \tilde \Hcal\left(\frac{x-L/2}{\delta t^{1/z}}\right),\qquad z=2,
\label{eq_h2_shortTdyn_asympt}\eeq 
with the same scaling function as in \cref{eq_hscalf_asympt_EW}. 
The above scaling profiles are independent of the specific boundary condition and apply for values of the scaling variable $|\xi|\lesssim \Ocal(1)$, i.e. in an ``inner'' region near the first-passage location $x_M$.
The accuracy of the approximations involved in \cref{eq_h2_shortTdyn_asympt} is further illustrated in \cref{fig_mft_shortT_scaling_test} in \cref{app_WNT_sol}. 
[A short-time scaling profile for finite nonzero $\delta t\ll T$, which entails a scaling function different from $\tilde\Hcal$, is provided in \cref{eq6_h_shortTdyn}.]
Note that the final profile in the transient regime [\cref{eq_h_shortT_EW}] still depends on $T$ via the scaling variable $\xi$, whereas the final profile in the equilibrium regime [\cref{eq_opt2_finalprof_eq}] is independent of $T$ for $T\gg \tau$.
We remark that, in contrast to the exact expression in \cref{eq_opt2_solh_pbc}, $h\pbc$ as given in \cref{eq_h_shortT_EW} has nonzero mass. This, however, constitutes a negligible error in the asymptotic limit $T\to 0$, as the profile becomes sharply peaked.
The final profiles in the transient and the equilibrium regime are illustrated in \cref{fig_opt_prof_nc_limits}.

\begin{figure}[t]\centering
    {\includegraphics[width=0.45\linewidth]{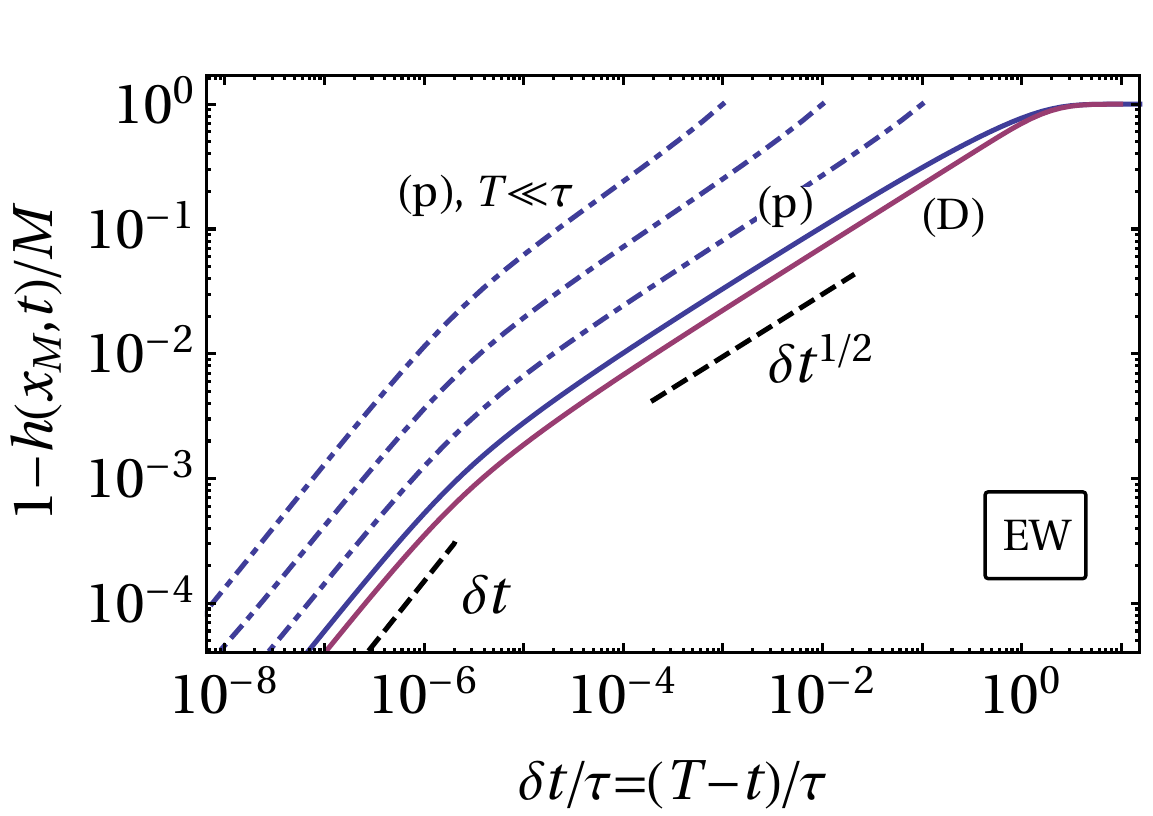}}
    \caption{Time evolution of the peak of the profile, $h(x_M,t)$, which reaches the height $M$ at the first-passage time $T$, for the EW equation as a function of $T-t$. The solid curves correspond to $h\ut{(p,D)}(x_M,t)$ in the equilibrium regime, while the dash-dotted curves illustrate the time evolution of $h\pbc(x_M,t)$ in the transient regime for $T/\tau=10^{-1},10^{-2},10^{-3}$ (from bottom to top). The corresponding behavior of $h\DirCP(x_M,t)$ for $t\ll T$ is similar and not shown. Both in the transient and the equilibrium regime, a power law $M-h(x_M, T -\dt)\propto \delta t^{1/2}$ is predicted [see \cref{eq_peakevol_EW}]. In the presence of an upper bound to the number of (eigen-)modes in the system, a linear behavior in $\dt$ emerges for times $\dt\lesssim \tau\cro=\tau/k\cro^z$ [see \cref{eq_peakevol_EW_linear}], where $k\cro$ is the largest mode index ($k\cro=\infty$ in the continuum limit). For illustrative purposes, we have chosen here $k\cro=1000$, corresponding to $\tau\cro/\tau\simeq 10^{-6}$. The fundamental time scale $\tau$ is defined in \cref{eq_nc_timescales} for the respective \bcs.}
    \label{fig_opt_peakevol_nc}
\end{figure}

According to \cref{eq_h_lateT_dynscal_EW,eq_h2_shortTdyn_asympt} the maximum $h(x_M,t)$ of the profile approaches the height $M$ at the first-passage time $T$ via a power law,
\beq 1- h(x_M,T-\delta t)/M \propto \delta t^{1/z},\qquad z=2.
\label{eq_peakevol_EW}\eeq 
This behavior applies both in the transient and the equilibrium regime and is independent of the \bcs.
If the system considered can accommodate only a finite number of modes---which, for instance, is the case when \cref{eq_EW,eq_MH} are discretized on a lattice---the sums in \crefrange{eq_opt2_solh_pbc}{eq_opt2_solh_Q_Dir} are bounded by a largest mode $k\cro$.
In this case, \cref{eq_peakevol_EW} is eventually superseded by a linear behavior,
\beq 1- h(x_M,T-\delta t)/M \propto \delta t\qquad \text{for}\qquad \dt\lesssim \tau\cro\equiv  \tau / k\cro^z,
\label{eq_peakevol_EW_linear}\eeq 
where $\tau\cro$ denotes the corresponding cross-over time [see \cref{eq6_h_cutoff_small_dt}].
The time evolution of the peak $h(x_M,t)$ is illustrated in \cref{fig_opt_peakevol_nc}, where the time is rescaled by the characteristic relaxation time $\tau$ in \cref{eq_nc_timescales}.
Note that, in the equilibrium regime, the evolution of the profile towards the first-passage event happens on a time scale of $\tau$, independently from the value of $T$. For times $t\ll T-\tau$ the equilibrium profile thus remains near its initial configuration [\cref{eq_init_cond}; see also panels (b) in \cref{fig_mft_prof_nc_pbc,fig_mft_prof_nc_Dir}].
In the transient regime [dash-dotted lines in  \cref{fig_opt_peakevol_nc} and panels (a) in \cref{fig_mft_prof_nc_pbc,fig_mft_prof_nc_Dir}], the evolution proceeds over the whole time interval between 0 and $T$ (where, however, $T\ll\tau$).

According to \cref{eq_peakevol_EW}, the distance $M$ is traversed within a time $\delta t^{1/z}$. Consequently, the requirement $\delta t\ll \tau$ for the transient regime implies $\sfrac{M}{L} \ll 1/(c\pi)$, with $c\pbc=2$ and $c\DirCP=1$ [see \cref{eq_nc_timescales}].
Hence, in the transient regime, the weak-noise limit of \cref{eq_Sopt_expr} is obtained if
$ L\gg \frac{D}{\eta} \left(\frac{L}{M}\right)^2 \gg \frac{D}{\eta} (c\pi)^{2}$,
where we re-instated all dimensional factors.
Conversely, the equilibrium regime is realized if $M/L\gg 1/(c\pi)$, such that in this case the weak-noise limit requires  
$ L\gg \frac{D}{\eta} \left(\frac{L}{M}\right)^2$ and $\left(\frac{L}{M}\right)^2 \ll (c\pi)^{2}$.

\section{Mullins-Herring dynamics}
\label{sec_MH}

We now turn to the optimal first-passage dynamics emerging from the MH equation. The analysis in this section proceeds in essentially the same fashion as for the EW equation in \cref{sec_EW}.
However, at the expense of some redundancy, the subsequent discussion is kept largely self-contained.

\subsection{Macroscopic fluctuation theory}
The Martin-Siggia-Rose action pertaining to the stochastic MH equation [\cref{eq_MH}] is given by \cite{tauber_critical_2014}
\beq \Scal[h,p] = \int_0^T \d t \int_{0}^{L} \d x\, p\left[\pd_t h+\frict \pd_x^4 h + D \pd_x^2 p \right].
\label{eq_S4}\eeq
The Euler-Lagrange equations describing the most-likely path of the profile $h$ and of the conjugate field $p$ follow as (see also Ref.\ \cite{smith_local_2017})
\begin{subequations}\begin{align}
0 &= \frac{\delta \Scal}{\delta p} = \pd_t h + \frict \pd_x^4 h + 2D\pd_x^2 p, \label{eq_S4_min_h}\\
0 &= \frac{\delta \Scal}{\delta h} = -\pd_t p + \frict \pd_x^4 p. \label{eq_S4_min_p}
\end{align}\label{eq_S4_min}\end{subequations}
We consider either periodic \bcs [\cref{eq_H_pbc}],
\beq h\pbc(x,t)=h\pbc(x+L,t),\qquad p\pbc(x,t)=p\pbc(x+L,t),
\label{eq_h4_pbc}\eeq 
or Dirichlet \bcs with a no-flux condition [\cref{eq_H_Dbc,eq_H_noflux}],
\beq h\DirNoFl(0,t)=0=h\DirNoFl(L,t),\qquad  \pd_x^3 h\DirNoFl(0,t)=0=\pd_x^3 h\DirNoFl(L,t).
\label{eq_h4_Dbc}\eeq 
In the latter case, the bi-harmonic operator $\pd_x^4$ is not self-adjoint on $[0,L]$, which renders the solution of \cref{eq_S4_min} technically more involved than in the self-adjoint case (see \cref{app_WNT_sol}). 
If Dirichlet no-flux \bcs are imposed on $h$, the conjugate field $p$ must fulfill the associated \emph{adjoint} \bcs (see \cref{sec_eigenv_mh})
\beq \pd_x p\DirNoFl(0,t) = 0 = \pd_x p\DirNoFl(L,t),\qquad \pd_x^2 p\DirNoFl(0,t) = 0 = \pd_x^2 p\DirNoFl(L,t).
\eeq
The mass-conserving property of the noise in \cref{eq_MH} is reflected by the presence of a derivative of $p$ in \cref{eq_S4_min_h}.
Indeed, it is readily proven that the considered \bcs ensure conservation of the mass [\cref{eq_zero_vol}].
Initial and final conditions on the profile $h$ are given in \cref{eq_init_cond,eq_firstpsg_cond} and suffice to determine also the conjugate field $p$.
Inserting \cref{eq_S4_min} into \cref{eq_S4} renders the optimal action 
\beq \Scal\opt = -D \int_0^T \d t \int_{0}^{L} \d x\, p\pd_x^2 p ,
\label{eq_Sopt4}\eeq 
which describes the most-likely activation dynamics \cite{meerson_macroscopic_2016, smith_local_2017}.

As was the case for the EW equation (see \cref{sec_ncons_dyn}), \cref{eq_S4_min} admits,  as a manifestation of the Onsager-Machlup time-reversal symmetry \cite{onsager_fluctuations_1953}, a specific solution corresponding to thermal \emph{equilibrium}.
In fact, consider a profile $h_r(x,t)$ obeying the (deterministic) fourth-order diffusion equation
\beq \pd_t h_r =- \frict \pd_x^4 h_r,
\label{eq_diffus_c}\eeq 
with the initial condition $h_r(x,t=0)= h_0(x)$, where $h_0(x)$ is a given profile [e.g., $h_0(x)=h(x,T\to\infty)$, where $h(x,T\to\infty)$ is a known first-passage profile].
Then the fields $h$, $p$ defined by 
\begin{subequations}\begin{align}
 h(x,t) &= h_r(x,T-t)\qquad \text{and}\label{eq_actrel_sol_c_h}\\
 p(x,t) &= -\frac{\frict}{D}\pd_x^2 h \label{eq_actrel_sol_c_p}
\end{align}\label{eq_actrel_sol_c}\end{subequations} 
fulfill the relations
\beq 
\pd_t h = -\pd_t h_r = \eta \pd_x^4 h = -\eta \pd_x^4 h - 2D \pd_x^2 h
\label{eq_antidiffus_c}\eeq 
as well as $\pd_t p = -(\eta/D) \pd_t\pd_x^2 h = -(\eta^2/D) \pd_x^6 h_r = \eta \pd_x^4 p$, which coincide with \cref{eq_S4_min_h,eq_S4_min_p}, respectively.
Accordingly, the fields defined in \cref{eq_actrel_sol_c} solve \cref{eq_S4_min} subject to the final condition $h(x,T)=h_0(x)$. 
\Cref{eq_actrel_sol_c_h} implies that $h(x,t=0)=h_r(x,T)$, which is generally nonzero for non-vanishing $h_0(x)$ and finite $T$. Hence, only for $T\to\infty$, equilibrium dynamics is strictly compatible with the initial condition in \cref{eq_init_cond}.
Using \cref{eq_actrel_sol_c,eq_antidiffus_c} in \cref{eq_Sopt4} renders the equilibrium action:
\beq\begin{split} 
\Scal\st{opt,eq} &= -\frac{\frict^2}{D} \int_0^T \d t\int_{0}^{L} \d x (\pd_x^2 h)(\pd_x^4 h)
=  \int_0^T \d t \left[-\frac{\frict}{D}(\pd_x h)(\pd_t h)\Big|_{0}^{L} + \frac{\frict^2}{D}\int_{0}^{L} \d x (\pd_x h)(\pd_x^5 h)\right] \\
&= -\frac{\frict}{D} \left[ h(\pd_x h) \Big|_{0}^{L} \right]_{t=0}^{T} + \frac{\frict}{D} \int_0^T \d t\, h(\pd_{tx}^2 h)\Big|_{0}^{L} + \frac{\frict}{D}\int_0^T \d t \int_{0}^{L}\d x (\pd_x h)(\pd_{tx}^2 h) \\
&= \frac{\frict}{D} \left[ \int_{0}^{L} \d x (\pd_x h)^2 \Bigg|_{t=0}^{T} - \int_0^T \d t \int_{0}^{L} \d x (\pd_{tx}^2 h)(\pd_x h) \right] \\
&= \frac{\frict}{2D} \int_{0}^{L} \d x (\pd_x h)^2 \Bigg|_{t=0}^{T}, 
\end{split}\label{eq_Sopt4_eq}\eeq 
where we made use of the fact that the boundary terms vanish for the \bcs in \cref{eq_h4_pbc,eq_h4_Dbc}.
In \cref{eq_Sopt4_eq} the temperature $\kbT$ can be identified via $\frict/(2D) = 1/(4 \kbT)$.
As expected, the final expression in \cref{eq_Sopt4_eq} coincides with the one in \cref{eq_Sopt2_eq} and shows that, in thermal equilibrium, the action essentially reduces to a free energy difference. 

Upon rescaling time by $\frict$ and redefining the fields $h$ and $p$ as in \cref{eq_h_rescaled}, \cref{eq_S4_min} becomes
\begin{subequations}\begin{align}
\pd_t h &= -\pd_x^4 h - 2\pd_x^2 p, \label{eq_MH_MFTr_h}\\
\pd_t p &= \pd_x^4 p. \label{eq_MH_MFTr_p}
\end{align}\label{eq_MH_MFTr}\end{subequations}
We henceforth consider also $\Scal\opt$ to be rescaled as in \cref{eq_Sopt2_resc_relation} and proceed by analyzing \cref{eq_MH_MFTr}.

\subsection{Exact solution}
\label{sec_fullsol_c}

\begin{figure}[t]\centering
    \subfigure[]{\includegraphics[width=0.5\linewidth]{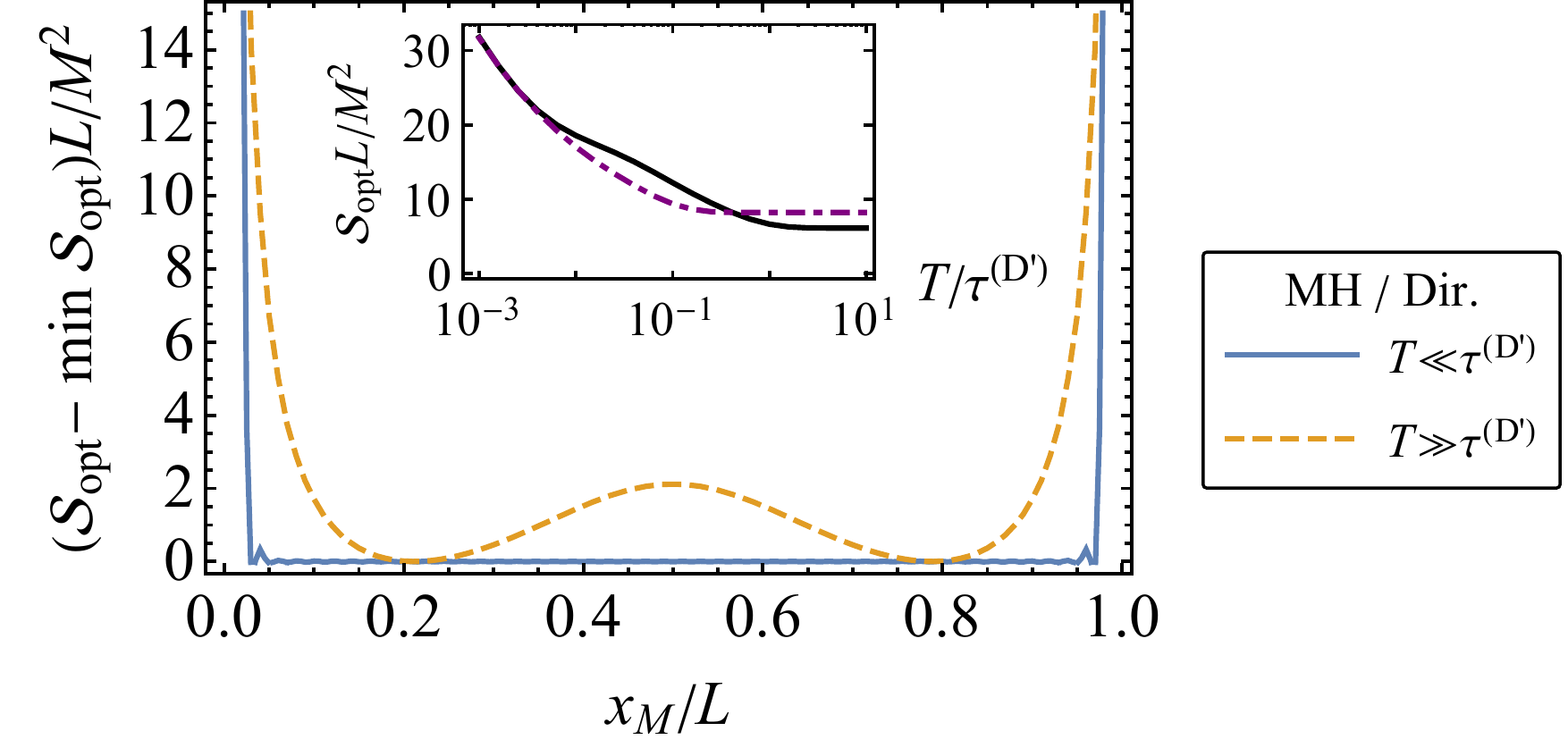}}\hfill
    \subfigure[]{\includegraphics[width=0.47\linewidth]{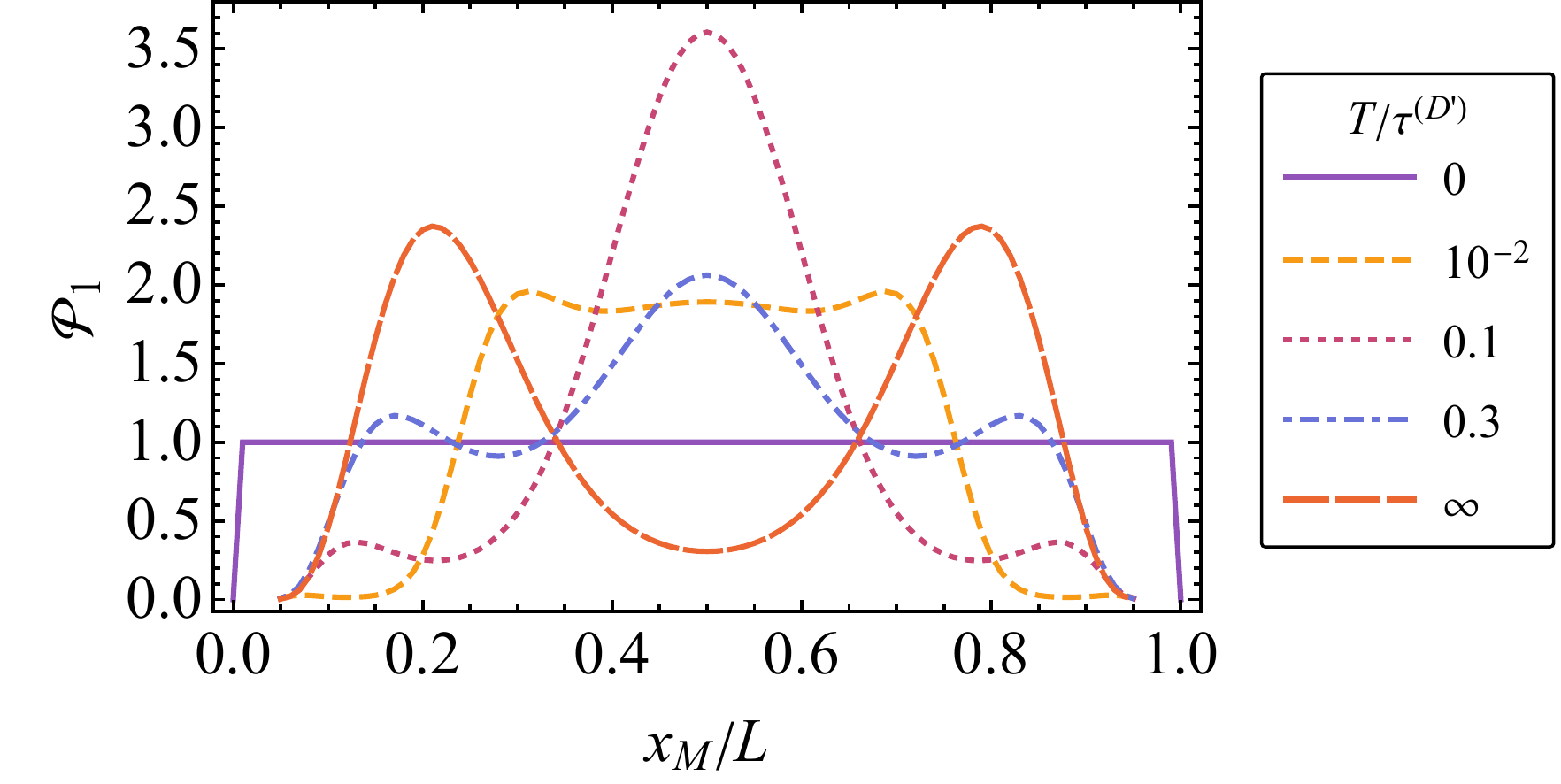}}
    \caption{(a) Optimal action $\Scal\opt\DirNoFl$ [\cref{eq_Sopt4_expr}, in units of $\eta/D$] for the MH equation with Dirichlet no-flux \bcs. The curves representing $\Scal\opt\DirNoFl$ are shifted such that their respective minima are zero.  For sufficiently large or small $T$, $\Scal\opt\DirNoFl-\mathrm{min}\Scal\opt\DirNoFl$ becomes independent of $T$. Asymptotically for $T\to 0$, $\Scal\opt\DirNoFl$ is spatially constant for $0<x_M<L$ and diverges at the boundaries. The inset shows $\Scal\opt\DirNoFl(x_M=x_M\DirNoFl)$ (solid curve) and $\Scal\opt\DirNoFl(x_M=L/2)$ (dash-dotted curve) as functions of $T$. These quantities diverge $\propto T^{-1/z}$ as $T\to 0$ [see \cref{eq6_Sopt_shortT}] and attain a nonzero constant for $T\gg\tau\DirNoFl$. (b) Probability distribution $\Pcal_1\DirNoFl$ [\cref{eq_P1_MH}] of the first-passage location $x_M$ for Dirichlet no-flux \bcs with $M^2/L=1$ (in units of $\eta/D$) and various values of $T$. The curves labeled by $T/\tau\DirNoFl=0$ and $\infty$ represent the asymptotic shapes in the transient and the equilibrium regime, respectively, where $\Pcal_1$ is independent of $T$.}
    \label{fig_Sopt_MH}
\end{figure}

The exact analytic solution of \cref{eq_MH_MFTr} subject to the the initial and final conditions in \cref{eq_init_cond,eq_firstpsg_cond} as well as to the \bcs in \cref{eq_h4_pbc} or \cref{eq_h4_Dbc} is determined in detail in \cref{app_WNT_sol} and summarized below.
The characteristic time scale for the creation of a rare event is given by ($z=4$)
\begin{subequations}\begin{align}
\tau\pbc &= \left(\frac{L}{2\pi}\right)^z  \label{eq_c_timescale_pbc}
\intertext{for periodic and by}
\tau\DirNoFl &= \left(\frac{L}{\omega_1}\right)^z  \label{eq_c_timescale_DirNoFl}
\end{align}\label{eq_c_timescales}\end{subequations}
for Dirichlet no-flux \bcs, respectively, where $\omega_1\simeq 4.73$ is the smallest positive solution of the eigenvalue equation $\cos(\omega)\cosh(\omega)=1$ [see \cref{eq_DirFl0_det}]. 
As was the case for the EW equation, the dynamics emerging from \cref{eq_MH_MFTr} is distinct in the transient ($T\ll\tau$) and the equilibrium ($T\gg\tau$) regime. In the latter case, \cref{eq_actrel_sol_c} applies.

Analogously to \cref{eq_Sopt_expr}, the optimal action [see \cref{eq_Sopt4,eq6_Sopt_expr}; expressed in units of $\eta/D$] fulfills the formal scaling property 
\beq \Scal\opt(x_M, M, T, L)=  \frac{M^2}{L} \Scal\opt\left(\frac{x_M}{L}, 1, \frac{T}{L^z}, 1\right) .
\label{eq_Sopt4_expr}\eeq 
The value of the first-passage location $x_M$ [see \cref{eq_firstpsg_cond}] follows from minimizing $\Scal\opt$ evaluated on the general solution in \cref{eq_MH_MFTr}.
For periodic \bcs, one may simply set $x_M\pbc=L/2$ owing to translational invariance.
For Dirichlet no-flux \bcs, the optimal action $\Scal\opt\DirNoFl$ is shown as a function of $x_M$ in \cref{fig_Sopt_MH}(a).
\Cref{fig_Sopt_MH}(b) displays the corresponding (normalized) probability distribution of the first-passage location $x_M$,
\beq \Pcal_1(x_M) \sim \exp\left[-\Scal\opt(x_M,M,T,L)\right].
\label{eq_P1_MH}\eeq
For illustrative purposes, we have chosen $M^2/L=1$ (in units of $\eta/D$) in the plot, and remark that, upon increasing $M^2/L$, the peak height of the distribution grows and, correspondingly, its characteristic width decreases---except in the limit $T\to 0$, where the form of $\Pcal_1$ is invariant.
In the equilibrium regime ($T\gg\tau$), $\Scal\opt$ and hence also $\Pcal_1(x_M)$ are generally independent of $T$ [see inset to \cref{fig_Sopt_MH}(a)].
For $T\to\infty$, $\Scal\opt$ reduces to the expression in \cref{eq_Sopt4_eq}, which can be evaluated analytically [see \cref{app_min_eqaction}].
In the case of Dirichlet no-flux \bcs, $\Scal\st{opt,eq}$ is minimal for the two values [see \cref{eq5_constr_minLoc}]
\beq 
x_M\DirNoFl\big|_{T\gg \tau\DirNoFl} = \frac{L}{2}\left(1 \pm \frac{1}{\sqrt{3}}\right).
\label{eq_xM_DirNoFl}
\eeq 
Accordingly, $\Pcal_1\DirNoFl$ shows two peaks, the sharpness of which increases with growing $M$ according to \cref{eq_Sopt4_expr}.
Asymptotically for $T\to 0$, $\Scal\opt$ scales $\propto T^{-1/z}$, independently of the \bcs [see \cref{eq6_Sopt_shortT}]. Furthermore, $\Scal\opt\DirNoFl$ becomes independent of $x_M$ for $0<x_M<L$. The corresponding distribution $\Pcal_1\DirNoFl$ is thus flat and independent of $M$ and $T$ in this limit.
One may therefore set $x_M\DirNoFl\big|_{T\ll \tau\DirNoFl}=L/2$ in order to evaluate the first-passage profile in this case. 
As illustrated in \cref{fig_Sopt_MH}(b), $\Pcal_1\DirNoFl$ assumes rather intricate shapes between its asymptotic transient and equilibrium limits. In particular, as $T/\tau\DirNoFl$ grows from small values, $\Pcal_1\DirNoFl$ develops a pronounced peak in the central region. For $T/\tau\DirNoFl\gtrsim 0.1$, this peak diminishes while two maxima grow near the locations given in \cref{eq_xM_DirNoFl}. 

The profile solving \cref{eq_MH_MFTr} can be written in scaling form,
\beq h(x,t,T,M,L) = M \mathpzc{h}\left(\frac{x}{L},\frac{t}{\tau}, \frac{T}{\tau}\right),
\label{eq_opt4_hscalform}\eeq 
where, for periodic \bcs (setting $x_M=L/2$) the dimensionless scaling function $\mathpzc{h}$ is given by [see \cref{eq6_h_pbc,eq6_Q_pbc}] 
\beq \mathpzc{h}\pbc(\upx,\upt,\upT) = \frac{1}{\mathpzc{Q}\pbc(\upT)}\sum_{k=1}^\infty \frac{1-\exp\left(-2 k^4 \upT\right)}{k^2} \frac{\sinh\left(k^4 \upt\right)}{\sinh\left(k^4 \upT\right)} \cos\left(2\pi k(\upx-1/2)\right)
\label{eq_opt4_solh_pbc}\eeq 
with
\beq \mathpzc{Q}\pbc(\upT) \equiv \sum_{k=1}^\infty \frac{1-\exp\left(-2k^{4} \upT\right)}{k^2}.
\label{eq_opt4_solh_Q_pbc}\eeq
These expressions have been previously obtained in Ref.\ \cite{meerson_macroscopic_2016}.
For Dirichlet no-flux \bcs, keeping $\upx_M\equiv x_M/L$ general here, one has [see \cref{eq6_h_Dir,eq6_Q_Dir}]
\beq \mathpzc{h}\DirNoFl(\upx,\upt,\upT) = \frac{1}{\mathpzc{Q}\DirNoFl(\upT)}\sum_{k=1}^\infty \frac{1-\exp\left(-2(\omega_k/\omega_1)^4 \upT\right)}{\omega_k^{2}\kappa_k} \frac{\sinh\left((\omega_k/\omega_1)^4 \upt\right)}{\sinh\left((\omega_k/\omega_1)^4 \upT\right)} \hat\sigma_k\DirNoFl(\upx_M) \hat\sigma_k\DirNoFl(\upx)
\label{eq_opt4_solh_DirNoFl}\eeq 
with
\beq \mathpzc{Q}\DirNoFl(\upT) \equiv \sum_{k=1}^\infty [\hat\sigma_k\DirNoFl(\upx_M)]^2 \frac{1-\exp\left(-2(\omega_k/\omega_1)^4 \upT\right)}{\omega_k^{2} \kappa_k}.
\label{eq_opt4_solh_Q_DirNoFl}\eeq
Here, $\hat\sigma_k\DirNoFl(\upx)\equiv \sigma_k\DirNoFl(\upx L)$ and the eigenfunctions $\sigma_k\DirNoFl$ are reported in \cref{eq4_DirFl0_eigenf} [see also \cref{eq6_h_Dir,tab_eigenfunc}];
furthermore $\kappa_k=[1-(-1)^k/\cosh(\omega_k)]/3$ and $\omega_k$ denotes the $k$th positive solution of the equation $\cos(\omega)\cosh(\omega)=1$ [see \cref{eq4_DirFl0_eigenval}].
Since $\int_0^L dx\, \cos(2\pi k(x/L-1/2))=0$ for $k\geq 1$, the profile for periodic \bcs in \cref{eq_opt4_solh_pbc} exactly fulfills mass conservation [\cref{eq_zero_vol}]. 
Note that, in contrast to the EW case, this property is not enforced explicitly [cf.\ \cref{eq_height_redef}] but follows readily from the fact that \cref{eq_MH} conserves $h$ locally.
Global mass conservation applies, by construction, also to the profile for Dirichlet no-flux \bcs in \cref{eq_opt4_solh_DirNoFl} [see \cref{eq4_DirFl0_mass}].
The general expression for the conjugate field $p$ is reported in \cref{eq6_p_expr}.

\begin{figure}[t]\centering
    \subfigure[]{\includegraphics[width=0.43\linewidth]{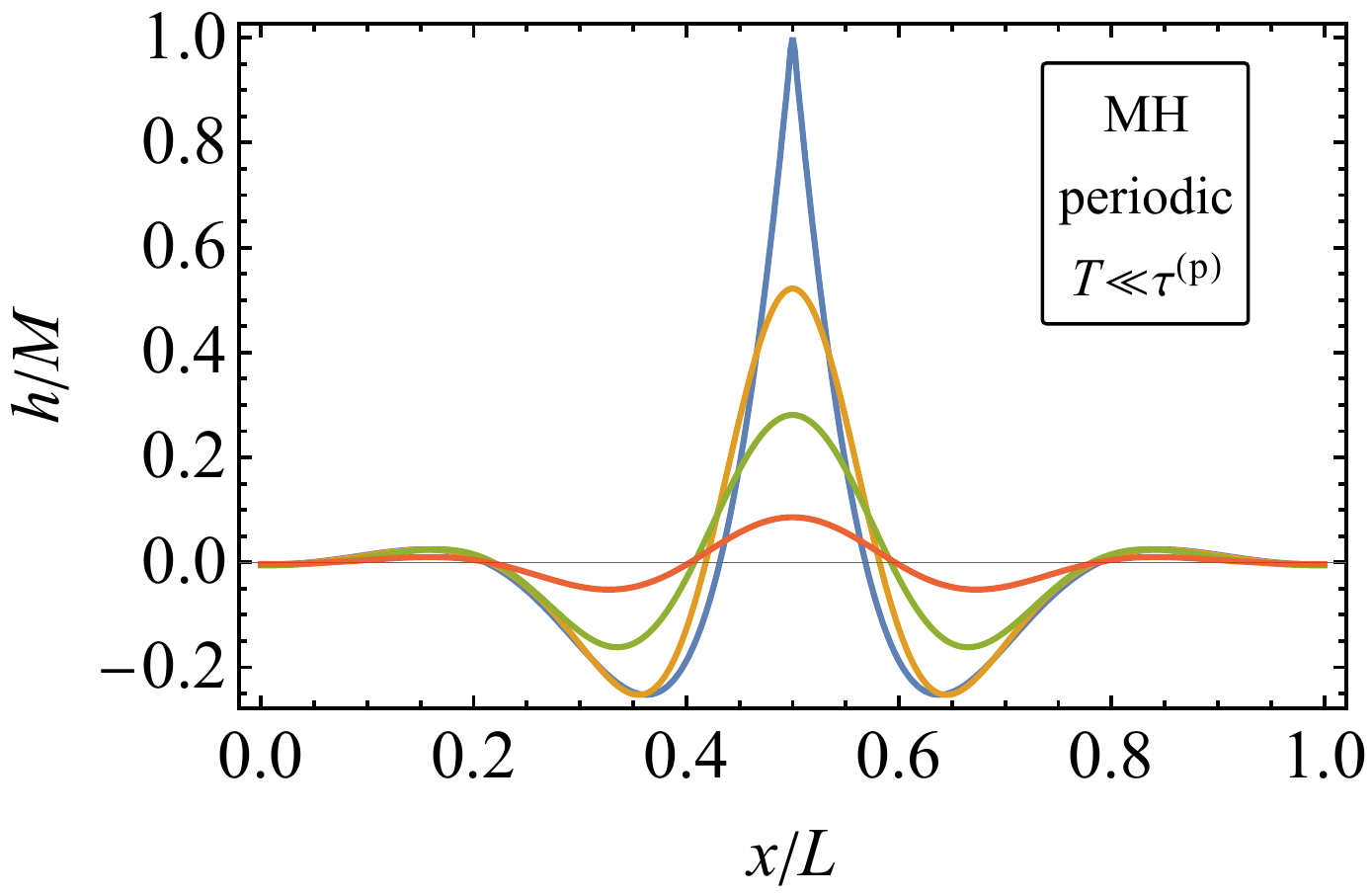}}\qquad
    \subfigure[]{\includegraphics[width=0.43\linewidth]{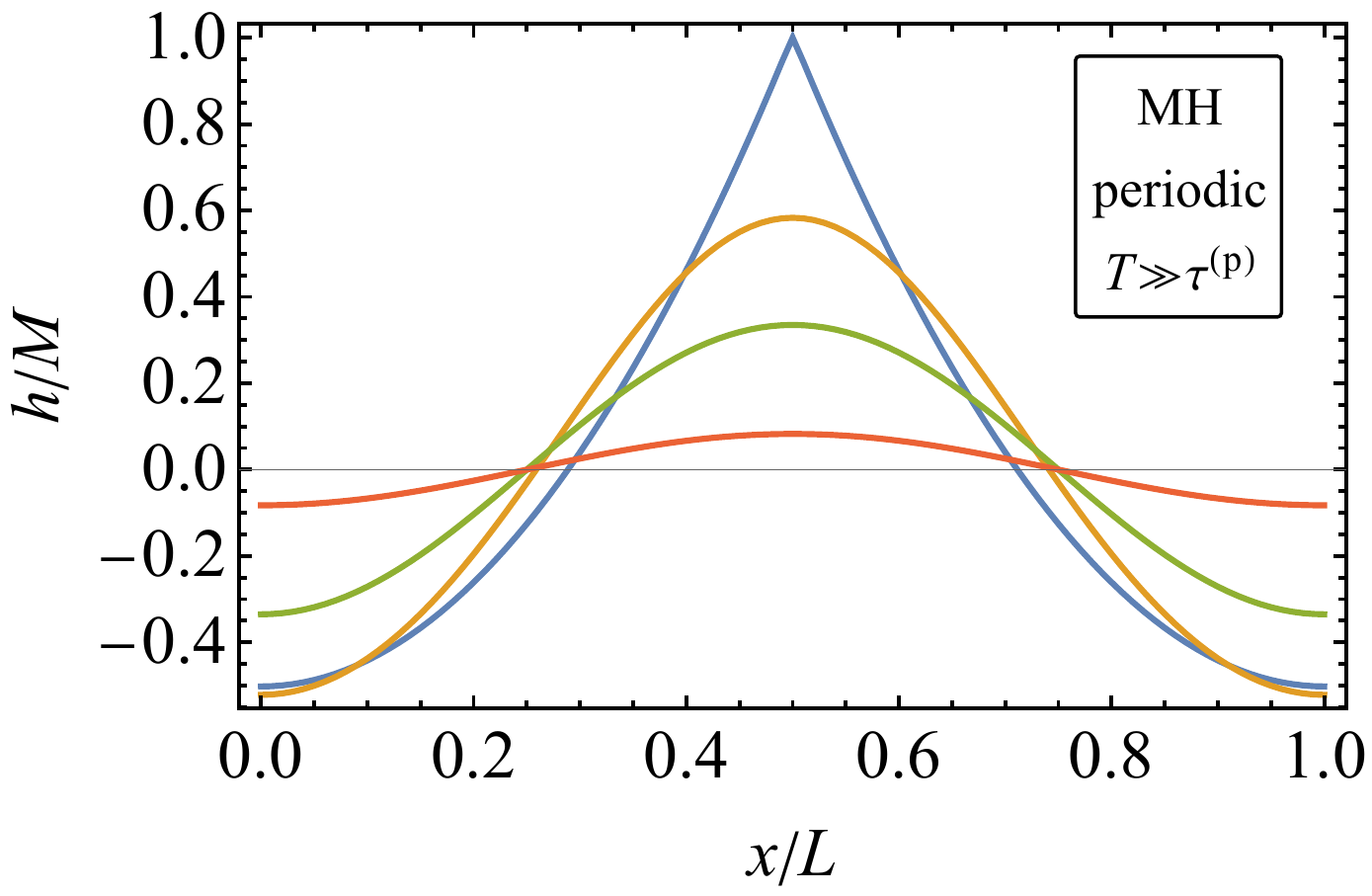}}
    \caption{Time evolution of the optimal profile [\cref{eq_opt4_hscalform,eq_opt4_solh_pbc}] for the MH equation with periodic \bcs in (a) the transient and (b) the equilibrium regime. The curves correspond, from center top to bottom, to (a) $1-t/T=0, 0.1,0.4,0.8$ with $T=10^{-2}\tau\pbc$, and (b) $1-t/T = 0, 0.001, 0.006, 0.02$ with $T=100\tau\pbc$. Decreasing $T$ in (a) leads essentially to a reduction of the width of the curves [see also \cref{eq_h4_shortTdyn_asympt}]. The fundamental time scale $\tau\pbc$ is reported in \cref{eq_c_timescale_pbc}.}
    \label{fig_mft_prof_c_pbc}
\end{figure}

\begin{figure}[t]\centering
    \subfigure[]{\includegraphics[width=0.44\linewidth]{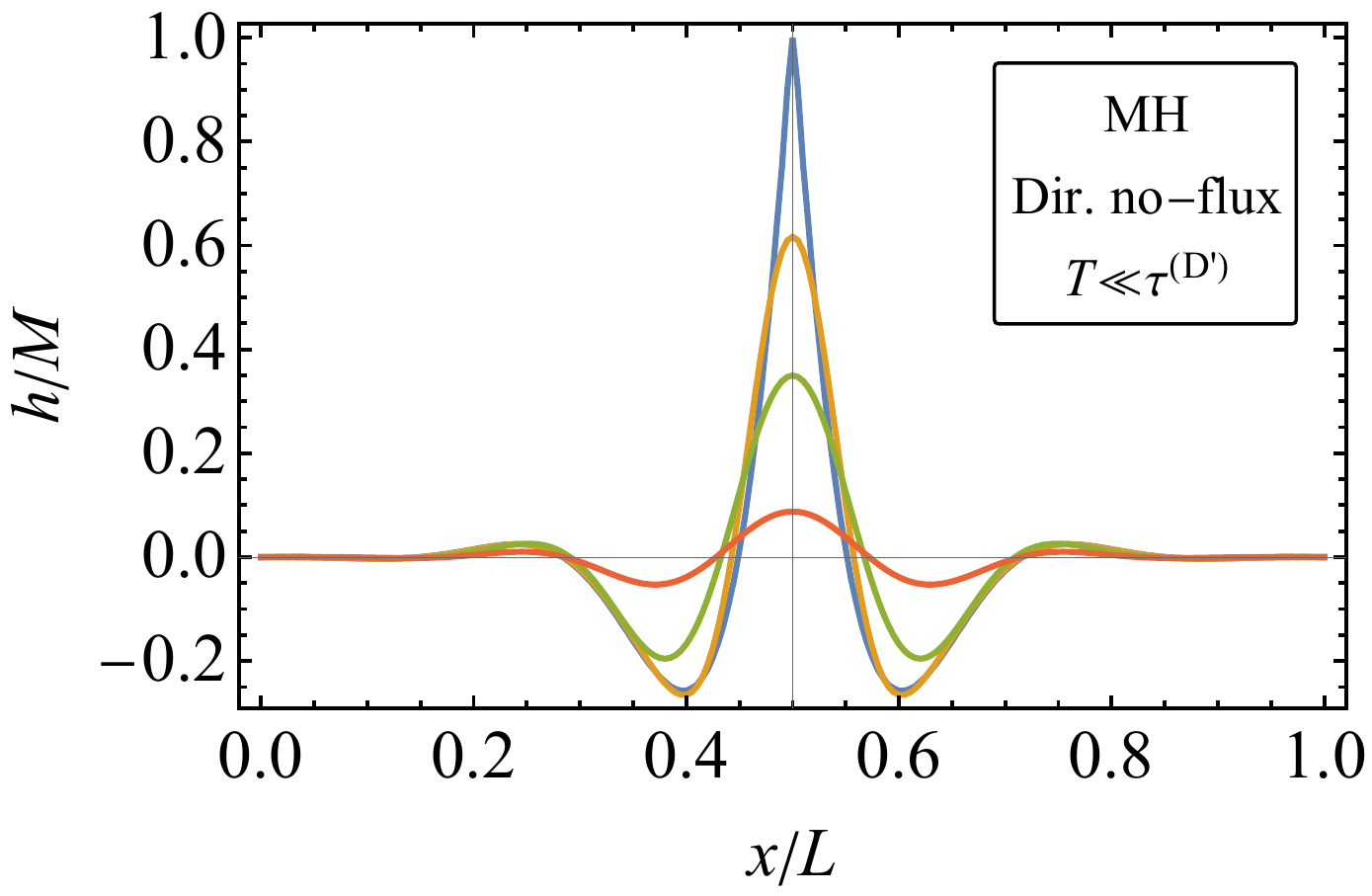}}\qquad 
    \subfigure[]{\includegraphics[width=0.43\linewidth]{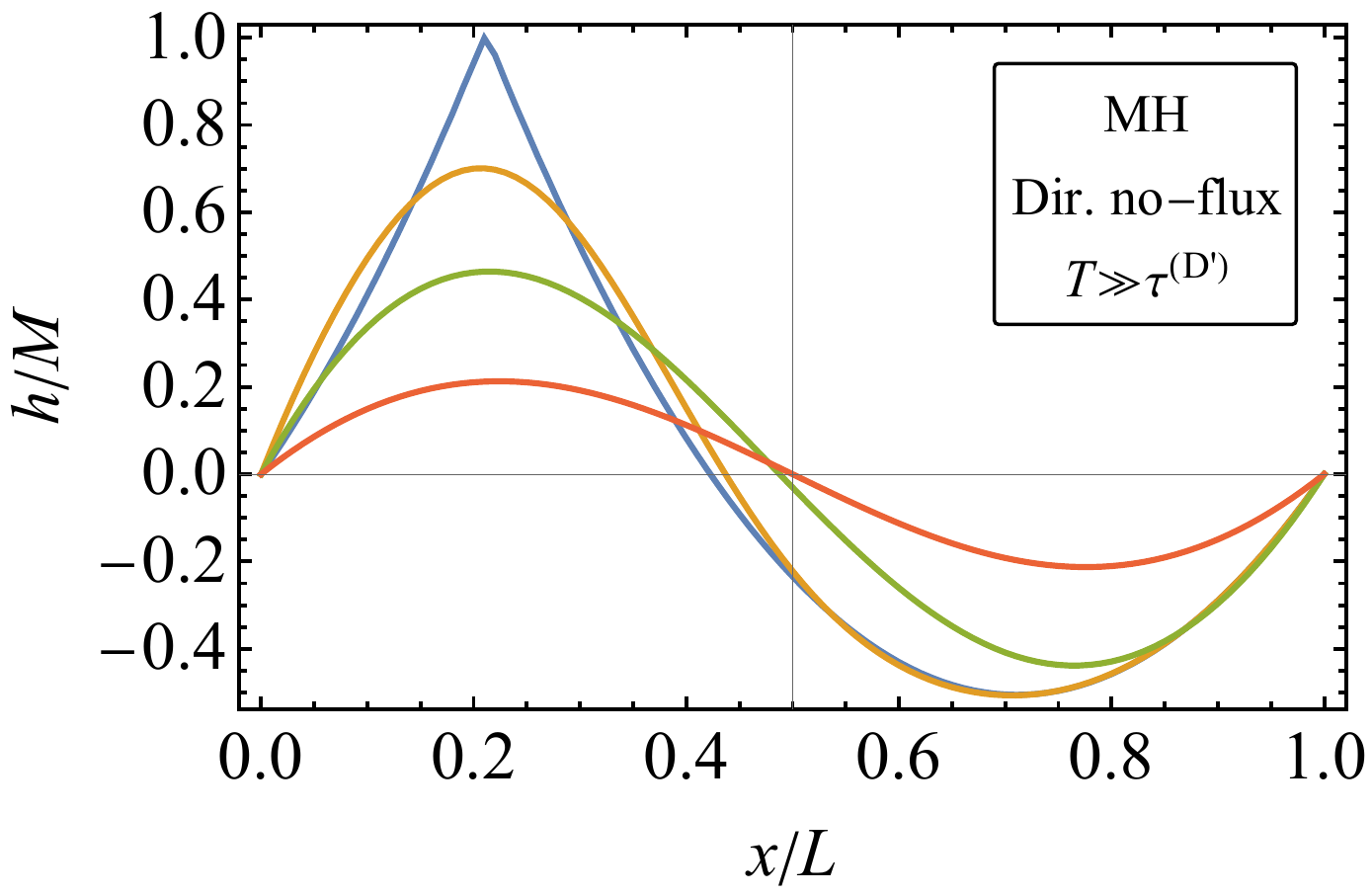}}
    \caption{Time evolution of the optimal profile [\cref{eq6_h_Dir}] for the MH equation with Dirichlet no-flux \bcs in (a) the transient and (b) the equilibrium regime. The curves correspond, from center top to bottom, to (a) $1-t/T = 0, 0.05, 0.3, 0.8$ with $T=10^{-3}\tau\DirNoFl$ and (b) $1-t/T=0, 10^{-4}, 0.0025, 0.01$ with $T=100\tau\DirNoFl$. Decreasing $T$ in (a) leads essentially to a reduction of the width of the curves [see also \cref{eq_h4_shortTdyn_asympt}]. The fundamental time scale $\tau\DirNoFl$ is reported in \cref{eq_c_timescale_DirNoFl}.}
    \label{fig_mft_prof_c_DirNofl}
\end{figure}

The spatio-temporal evolution of the optimal profile for periodic and Dirichlet no-flux \bcs is illustrated in \cref{fig_mft_prof_c_pbc,fig_mft_prof_c_DirNofl}, respectively. 
(For completeness, in \cref{fig_mft_prof_c_DirCP} in \cref{app_WNT_sol} also the profile obtained for the MH equation with standard Dirichlet \bcs is discussed.)
In contrast to the EW equation, the transient first-passage profiles emerging from the MH equation show an oscillatory decay in space [see panels (a) of \cref{fig_mft_prof_c_pbc,fig_mft_prof_c_DirNofl}].
In the equilibrium regime, the first-passage profile generally develops on a time scale of $\Ocal(\tau)$.
In the case of periodic \bcs, the time-dependent equilibrium profiles are qualitatively similar for EW and MH dynamics [compare panels (b) of \cref{fig_mft_prof_nc_pbc,fig_mft_prof_c_pbc}]. 

\begin{figure}[t]\centering
    \subfigure[]{\includegraphics[width=0.43\linewidth]{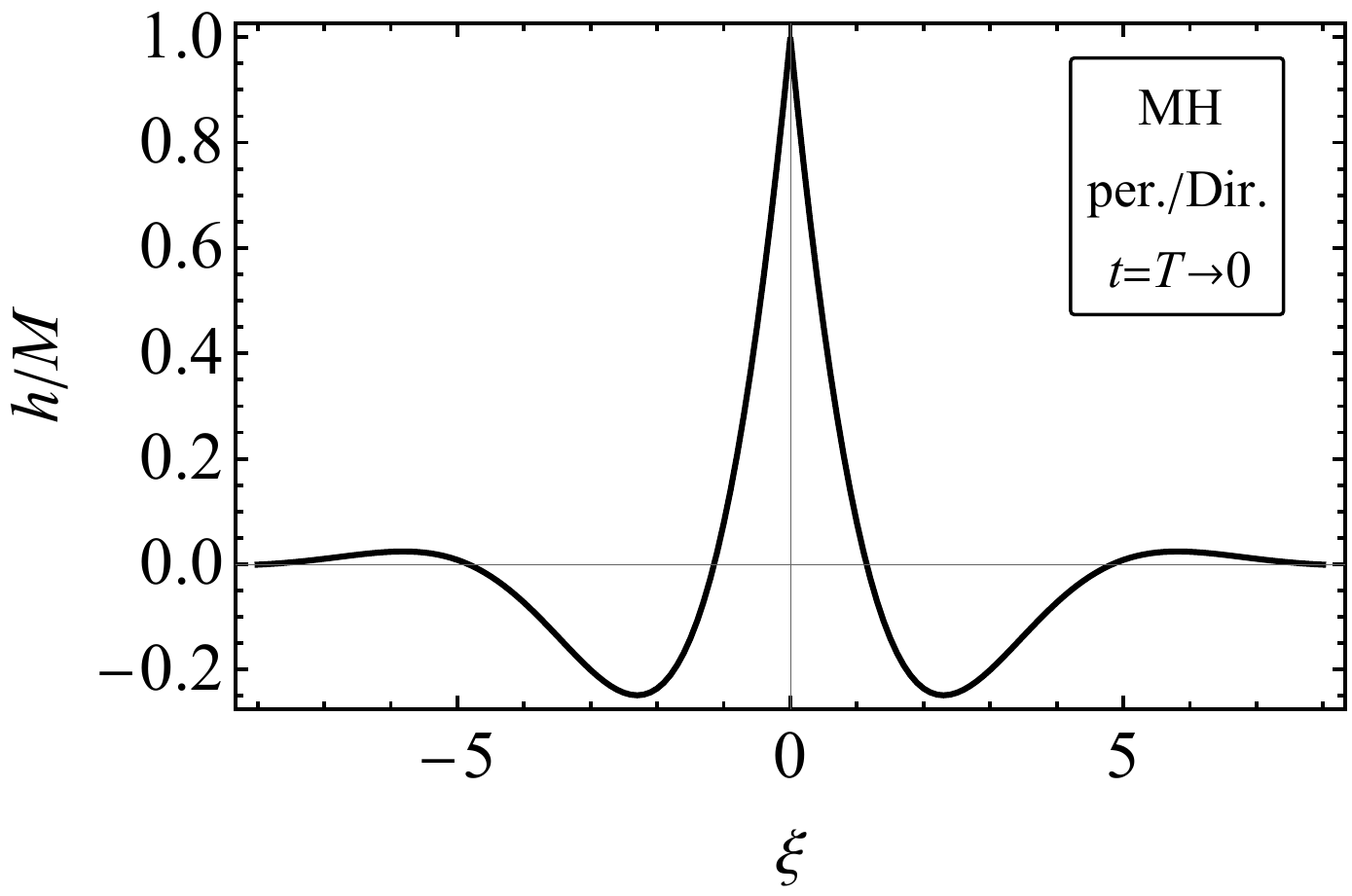}}\qquad
    \subfigure[]{\includegraphics[width=0.44\linewidth]{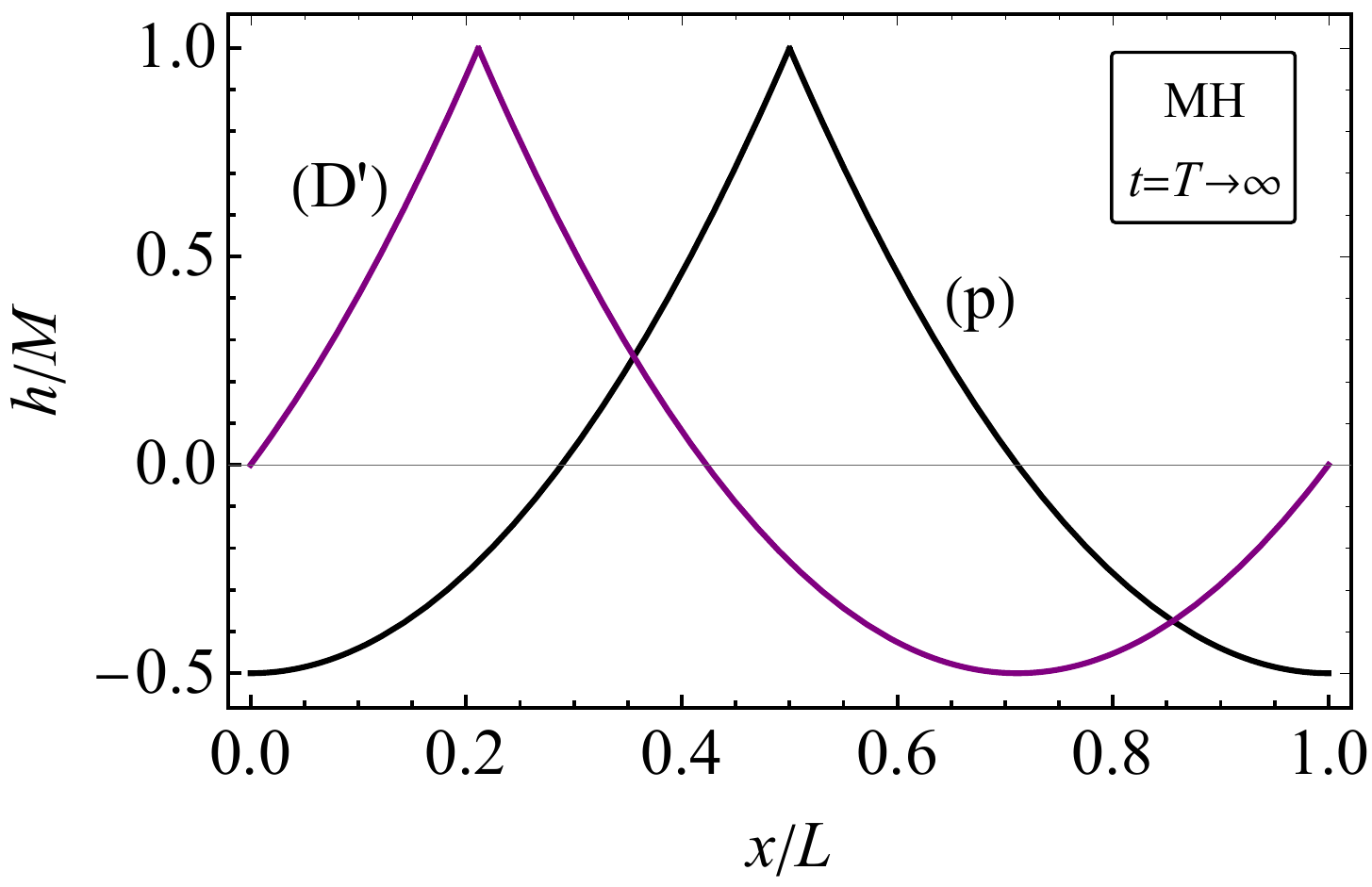} \label{fig_opt_prof_c_limit_eq}}
    \caption{Asymptotic first-passage profiles $h(x,t=T)$ obtained within WNT of the MH equation for (a) the transient regime, $T\to 0$ [\cref{eq_h_shortT_MH}], and (b) the equilibrium regime, $T\to\infty$ [\cref{eq_opt2_finalprof_eq_pbc,eq_opt4_finalprof_eq_DirNoFl}]. In the transient regime, the profiles depend on the scaling variable $\xi\equiv (x-L/2)/(2T)^{1/4}$ and coincide for periodic and Dirichlet no-flux \bcs. The profiles for $T\to\infty$ follow from constrained minimization of the equilibrium action in \cref{eq_Sopt4_eq}. The  equilibrium profiles at time $t=T$ for periodic and Dirichlet no-flux \bcs are related via a shift along $x$ [see \cref{eq_opt4_finalprof_eq_DirNoFl}].}
    \label{fig_opt_prof_c_limits}
\end{figure}

For $T\gg\tau$, the profile at time $t=T$ minimizes the equilibrium action $\Scal\st{opt,eq}$ [\cref{eq_Sopt4_eq}].
Since the latter quantity is independent of the specific dynamics, the expression for the profile $h\pbc(x,T)|_{T\to\infty}$ subject to periodic \bcs coincides with the one in \cref{eq_opt2_finalprof_eq_pbc}. Alternatively, it can be directly derived from the expression  in \cref{eq_opt4_solh_pbc} [see \cref{eq6_eq_prof_pbc}].
In contrast to standard Dirichlet \bcs [see \cref{eq_opt2_finalprof_eq_Dir} as well as \cref{fig_mft_prof_c_DirCP} in \cref{app_WNT_sol}], for Dirichlet no-flux \bcs one has to additionally take into account the constraint of zero mass [\cref{eq_zero_vol}] in the minimization of $\Scal\st{opt,eq}$.
Accordingly, using the fact that $h(x,0)=0$, one obtains [see \cref{eq5_prof_DirNoFl,eq5_prof_Dir_pbc_map}]
\beq h\DirNoFl(x,T\to\infty)/M = h\pbc(x + L/2-x_M,T\to\infty)/M = \begin{cases}\displaystyle 6\frac{x}{L} \left(\frac{x}{L} + \frac{1}{\sqrt{3}}\right) ,\qquad &x\leq x_M\DirNoFl,\\
\displaystyle 6\left(\frac{x}{L}-1\right)\left(\frac{x}{L}-1+\frac{1}{\sqrt{3}}\right) ,&x > x_M\DirNoFl,
\end{cases}\label{eq_opt4_finalprof_eq_DirNoFl}
\eeq
with $x_M\DirNoFl$ given in \cref{eq_xM_DirNoFl} and the last expression in \cref{eq_opt4_finalprof_eq_DirNoFl} applying to the smaller of the two possible values of $x_M\DirNoFl$.
Note that, while, at the time $t=T$, $h\DirNoFl$ can be expressed in terms of $h\pbc$, this is not possible at arbitrary times $t<T$, as, e.g., a close inspection of \cref{fig_mft_prof_c_pbc}(b) and \cref{fig_mft_prof_c_DirNofl}(b) near $h\approx 0$ reveals.
In the equilibrium regime for nonzero but small time differences $\delta t\equiv T-t \ll T$, \cref{eq_opt4_hscalform} can be cast into a dynamic scaling form [see \cref{eq6_h_lateT_full}]:
\beq h(x,T-\delta t)\big|_{T\gg\tau} \simeq M -  M (\delta t)^{1/z} \Gamma(1-1/z) \tilde\Hcal\left(\frac{x-x_M}{\delta t^{1/z}}\right),\qquad z=4,
\label{eq_h_lateT_dynscal_MH}\eeq 
with the scaling function
\beq \tilde \Hcal(\xi) = 
 {}_1 F_3\left(-\frac{1}{4}; \frac{1}{4},\frac{1}{2}, \frac{3}{4}; \frac{\xi^4}{256}\right) + \xi^2 \frac{\Gamma\left(\frac{1}{4}\right)}{8\Gamma\left(\frac{3}{4}\right)} {}_1 F_3\left(\frac{1}{4}; \frac{3}{4}, \frac{5}{4}, \frac{3}{2}; \frac{\xi^4}{256}\right) .
 \label{eq_hscalf_asympt_MH}\eeq 
We recall that, in terms of the unscaled time variable, the argument of $\tilde\Hcal$ in \cref{eq_h_lateT_dynscal_MH} is given by $(x-x_M)/(\eta\, \delta t)^{1/z}$, which is dimensionless since $\eta$ and $L^z/T$ have the same dimensions.
Asymptotically for $T\to 0$ in the transient regime, the profile at time $t=T$ is given by [see \cref{eq6_h_shortT}]:
\beq h(x,T)\big|_{T\ll \tau} = M \Hcal\left(\frac{x-L/2}{(2T)^{1/z}}\right),\qquad z=4,
\label{eq_h_shortT_MH}\eeq 
with the scaling function
\beq \Hcal(\xi)={}_1 F_3\left(-\frac{1}{4}; \frac{1}{4},\frac{1}{2}, \frac{3}{4}; \frac{\xi^4}{256}\right) + \xi^2 \frac{\Gamma\left(\frac{1}{4}\right)}{8\Gamma\left(\frac{3}{4}\right)} {}_1 F_3\left(\frac{1}{4}; \frac{3}{4}, \frac{5}{4}, \frac{3}{2}; \frac{\xi^4}{256}\right) - \frac{\pi}{2\Gamma\left(\frac{3}{4}\right)}|\xi| .
\label{eq_h_shortT_scalF_MH}\eeq 
For nonzero time differences $\delta t=T-t$ in the transient regime, a dynamic scaling profile follows at leading order in $\delta t/T\ll 1$ as [see \cref{eq6_h_shortTdyn_asympt}]
\beq h(x,T-\delta t)\big|_{\substack{T\ll \tau\\\delta t\ll T}} = M - M \left(\frac{\delta t}{2T}\right)^{1/z} \tilde \Hcal\left(\frac{x-L/2}{\delta t^{1/z}}\right),\qquad z=4,
\label{eq_h4_shortTdyn_asympt}\eeq 
where the scaling function takes the same form as in \cref{eq_hscalf_asympt_MH}.
The scaling forms in \cref{eq_h_lateT_dynscal_MH,eq_h_shortT_MH,eq_h4_shortTdyn_asympt} apply to both periodic and Dirichlet \bcs and are valid for values of the scaling variable $|\xi|\lesssim \Ocal(1)$. [A comparison of the approximative profile in \cref{eq_h4_shortTdyn_asympt} with the exact one is provided in \cref{fig_mft_shortT_scaling_test} in \cref{app_WNT_sol}, while a scaling form improving \cref{eq_h4_shortTdyn_asympt} beyond leading order in $\delta t/T$ is reported in \cref{eq6_h_shortTdyn}.]
In the case of periodic \bcs, the expressions in \cref{eq_h_shortT_scalF_MH,eq_opt2_finalprof_eq_pbc} have been previously obtained in Ref.\ \cite{meerson_macroscopic_2016}.
Note that the static profile in the transient regime [\cref{eq_h_shortT_MH}] still depends on $T$ via the scaling variable $\xi$, whereas the static profile in the equilibrium regime [\cref{eq_opt2_finalprof_eq_pbc,eq_opt4_finalprof_eq_DirNoFl}] is independent of $T$ for sufficiently large $T$.
The scaling profiles in \cref{eq_h_shortT_MH,eq_h4_shortTdyn_asympt} have [in contrast to the exact solution in \cref{eq_opt4_solh_pbc}] nonzero mass [\cref{eq_mass}], which, however, constitutes a negligible error in the asymptotic limit $T\to 0$, where the profiles become sharply peaked.
The profiles at time $t=T$ in the transient and the equilibrium regime are illustrated in \cref{fig_opt_prof_c_limits}.

\begin{figure}[t]\centering
    {\includegraphics[width=0.45\linewidth]{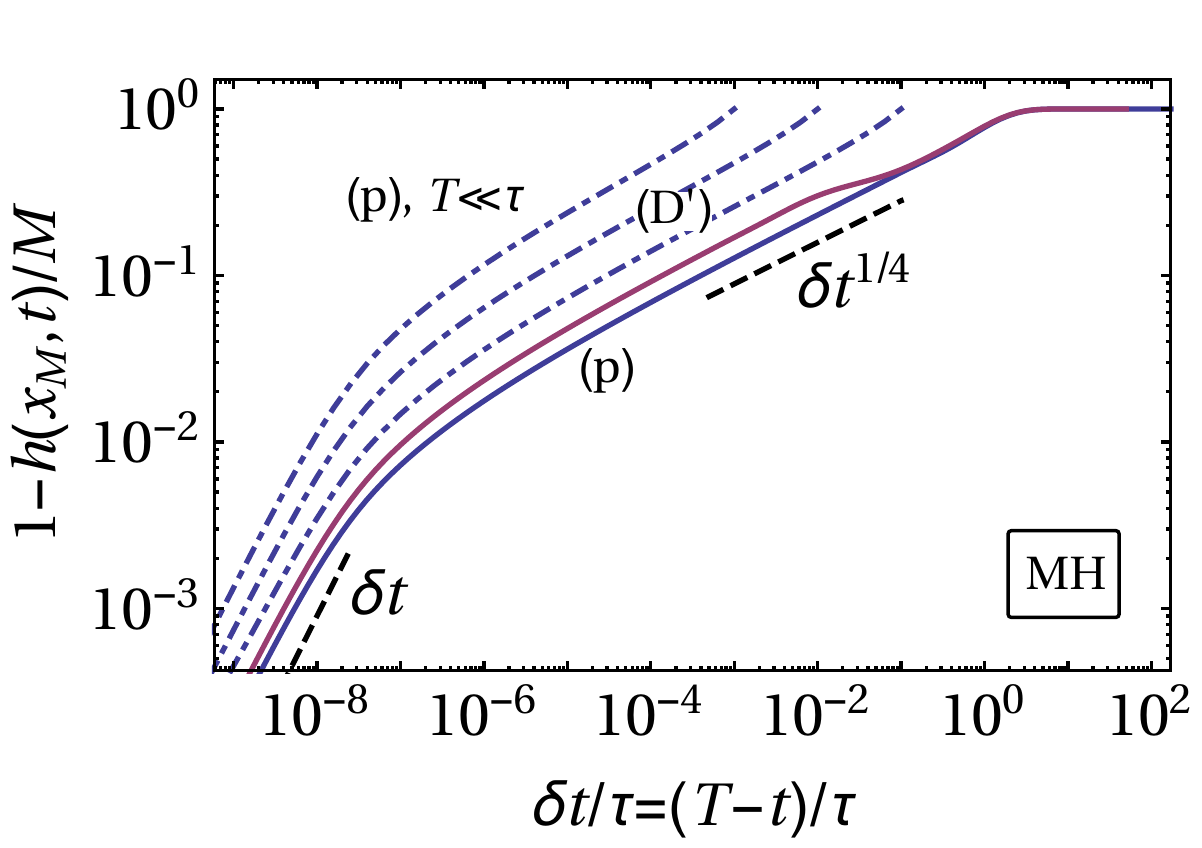}}
    \caption{Time evolution of the peak of the profile $h(x_M,t)$, which reaches the height $M$ at the first-passage time $T$, for the MH equation as a function of $T-t$. The solid curves correspond to $h\ut{(p,D')}(x_M,t)$ in the equilibrium regime ($T\gg \tau$), while the dash-dotted curves illustrate the time evolution of $h\pbc(x_M,t)$ in the transient regime for $T/\tau=10^{-1},10^{-2},10^{-3}$ (from bottom to top). Both in the transient and the equilibrium regime, a power law $M-h(x_M,t)\propto \delta t^{1/4}$ is obtained as an intermediate asymptotic [see \cref{eq_peakevol_MH}; dashed line]. If the number of modes in the system is finite, a linear behavior in $\dt$ emerges for times $\dt\lesssim \tau\cro$ [see \cref{eq_peakevol_MH_linear}], where $\tau\cro$ is the crossover time ($\tau\cro=0$ in the continuum limit). For illustrative purposes we have chosen here $\tau\cro/\tau\simeq 10^{-8}$. The fundamental time scale $\tau$ is defined in \cref{eq_c_timescales} for the respective \bcs.}
    \label{fig_opt_peakevol_c}
\end{figure}

According to \cref{eq_h_lateT_dynscal_MH,eq_h4_shortTdyn_asympt}, noting that $\tilde\Hcal(0)=1$, the peak $h(x_M,t)$ of the profile approaches the maximum height $M$ via a power law
\beq 1- h(x_M,T-\delta t)/M \propto \delta t^{1/z},\qquad z=4.
\label{eq_peakevol_MH}\eeq 
This behavior applies to a continuum system both in the transient and the equilibrium regime and is independent of the specific \bcs.
If, due to a microscopic cutoff, the mode spectrum of the system is bounded from above, \cref{eq_peakevol_MH} crosses over to a linear law,
\beq 1- h(x_M,T-\delta t)/M \propto \delta t\qquad \text{for}\qquad \dt\lesssim \tau\cro,
\label{eq_peakevol_MH_linear}\eeq 
where $\tau\cro$ is the crossover time. For periodic \bcs, $\tau\cro=\tau\pbc/k\cro^z$, while for Dirichlet no-flux \bcs, $\tau\cro\DirNoFl=\tau\DirNoFl (\omega_1/\omega_{k\cro})^z$, where $k\cro$ is the maximum mode index and $\omega_k$ denotes the eigenvalues in \cref{eq4_DirFl0_eigenval}.
The time evolution of $h(x_M,t)$ is illustrated in \cref{fig_opt_peakevol_c}, where the time is rescaled by the characteristic relaxation time $\tau$ defined in \cref{eq_c_timescales}.
As noted previously, in the equilibrium regime, the actual evolution of the profile towards the maximum occurs within a time interval $\tau$ before $T$. 
In the case of Dirichlet no-flux \bcs, the intermediate asymptotic regime described by \cref{eq_peakevol_MH} is seen to be of somewhat smaller size than for periodic \bcs.
In the transient regime, a condition determining the weak-noise limit of \cref{eq_Sopt4_expr} follows from \cref{eq_peakevol_MH} as 
$ L\gg \frac{D}{\eta} \left(\frac{L}{M}\right)^2 \gg \frac{D}{\eta} \omega_1^{2}$,
with $\omega_1\pbc=2\pi$ and $\omega_1\DirNoFl=4.73$ [see \cref{eq_c_timescale_DirNoFl}].
In contrast, in the equilibrium regime, the weak-noise limit is realized for
$ L\gg \frac{D}{\eta} \left(\frac{L}{M}\right)^2$ and $\left(\frac{L}{M}\right)^2 \ll \omega_1^2$.

\section{Summary}
In the present study, first-passage events of a one-dimensional interfacial profile $h(x,t)$, subject to the Edwards-Wilkinson (EW) or the (stochastic) Mullins-Herring (MH) equation, have been investigated analytically. 
The approach here is based on the weak-noise approximation of a Martin-Siggia-Rose/Janssen/de Dominicis path integral formulation of the corresponding Langevin equations [\cref{eq_EW,eq_MH}] \cite{bertini_macroscopic_2015, fogedby_minimum_2009, ge_analytical_2012, grafke_instanton_2015, meerson_macroscopic_2016,tauber_critical_2014}.
A comparison to numerical solutions of the EW and MH equation beyond the weak-noise approximation will be provided in a separate paper.
Minimization of the associated action yields the most-probable (``optimal'') profile which, starting from a flat initial configuration [\cref{eq_init_cond}], realizes the first-passage event $h(x_M,T)=M$ at a specified time $T$ and a location $x_M$.
Note that here the rare event dynamics is purely fluctuation-induced, i.e., there is no deterministic driving force involved --- in contrast to, e.g., the classical problem \cite{zwanzig_non-equilibrium_2001} of determining noise-activated transitions between energy minima. 

The first-passage problem of the MH equation for periodic \bcs has been studied previously in Ref.\ \cite{meerson_macroscopic_2016}. Extending that work, here we have investigated the influence of various \bcs on the spatio-temporal evolution of the optimal profile and discussed in detail its dynamic scaling behavior.
Since the optimal profile is provided here in terms of a generic eigenfunction expansion [see \cref{app_WNT_sol}], the corresponding expressions can be readily specialized to other \bcs.
We point out that, in order to ensure mass conservation [\cref{eq_mass}] for the MH equation with Dirichlet \bcs, a no-flux condition must be imposed [see \cref{eq_H_Dbc,eq_H_noflux}].
This renders the solution of the corresponding WNT technically involved, as the bi-harmonic operator is not self-adjoint anymore.
Standard Dirichlet \bcs, instead, do not conserve mass and are studied here mainly in conjunction with the EW equation.

The ensuing rare event dynamics is phenomenologically distinct for first-passage times $T\ll\tau$ and $T\gg\tau$, corresponding to the transient (non-equilibrium) and the equilibrium regime, respectively. $\tau$ denotes the fundamental relaxation time of the model, which coincides with the characteristic time scale for the evolution of the first-passage event.
In the equilibrium regime, the optimal profile at time $t=T$ minimizes the equilibrium action and depends sensitively on the \bcs as well as on possible conservation laws. 
In contrast, in the transient regime, \bcs and mass conservation have a negligible influence and the optimal profile is strongly localized. In fact, in the transient regime, the profile shape close to the first-passage event (i.e., for $t\to T$) depends only on the type of bulk dynamics.
The peak of the profile is predicted to approach the first-passage height $M$ algebraically in time, $M-h(x_M,t)\propto (T-t)^{\alpha}$, with an exponent $\alpha=1/z$, where $z=2$ for the EW and $z=4$ for the MH equation.
Notably, this behavior applies both in the transient and the equilibrium regimes and is independent of the specific \bcs or conservation laws.

\appendix

\section{Equilibrium profiles}
\label{app_min_eqaction}
Here, we determine static profiles $h(x)$ ($0\leq x \leq L$) which minimize the equilibrium action [see \cref{eq_Sopt2_eq,eq_Sopt4_eq}] 
\beq \Scal\eq[h] = \frac{\frict}{2 D} \int_0^L \d x\, [\pd_x h(x)]^2,
\label{eq5_Seq}\eeq 
under the constraint of attaining a maximum height $M$ at a certain location $x_M$,
\beq M=h(x_M).
\label{eq5_constr_height}\eeq 
In certain cases, we additionally impose a mass constraint:
\beq \mass=\int_0^L \d x\, h(x).
\label{eq5_constr_area}\eeq 
The profile $h$ is furthermore required to fulfill either periodic \bcs,
\begin{subequations}\begin{align}
h(x)=h(x+L),
\intertext{or Dirichlet \bcs,}
h(0)= 0=h(L). \label{eq5_bcs_Dir}
\end{align}\label{eq5_bcs}\end{subequations}
Introducing Lagrange multipliers $\lambda$ and $\beta$, we obtain the augmented action
\beq \tilde \Scal\eq([h],\lambda,\beta) \equiv \Scal\eq[h] - \lambda \left[\int_0^L \d x\, h(x)-\mass \right] - \beta \left[ \int_0^L \d x\, h(x)\,\delta\left(x-x_M\right) - M\right],
\label{eq5_S_constr}\eeq 
the minimization of which results in the Euler-Lagrange equation
\beq 0=\frac{\delta \tilde \Scal\eq}{\delta h} =\frac{\frict}{D} \pd_x^2 h + \lambda + \beta\delta\left(x-x_M\right) .
\label{eq5_ELE}\eeq 
We remark that integration of \cref{eq5_ELE} over an infinitesimal interval centered at $x_M$ yields the relation $h'(x_M^+) - h'(x_M^-) = \beta$, which, however, is not needed to determine the constrained profile.
Instead, \cref{eq5_ELE} is solved separately in the domains $x\lessgtr x_M$, subject to the \bcs in \cref{eq5_bcs} and the requirement of continuity at $x_M$ [see \cref{eq5_constr_height}], i.e., 
\beq h(x_M^+)=h(x_M^-)=M.
\label{eq5_midpoint}\eeq 
Subsequently, the mass constraint in \cref{eq5_constr_area} is imposed.
The expressions for the constrained profiles turn out to be independent of the factor $\eta/2D$ present in \cref{eq5_Seq}.

For $\mass=0$ and periodic \bcs, setting $x_M=L/2$, one obtains the constrained profile \cite{meerson_macroscopic_2016}
\beq
h\ut{(p)}(x)/M = 1-6\Bigg|\frac{x}{L}-\onehalf\Bigg| + 6\left(\frac{x}{L}-\onehalf\right)^2.
\label{eq5_prof_pbc}\eeq
For Dirichlet zero-$\mu$ \bcs [cf.\ \cref{app_DirCP_eigenf}], we do not enforce the mass constraint [\cref{eq5_constr_area}]. Accordingly, the Lagrange multiplier $\lambda$ is absent and one simply solves $0=\pd_x^2 h$, subject to \cref{eq5_bcs_Dir,eq5_constr_height}, in each domain.
The resulting solution still depends on $x_M$; the associated action, which is displayed in \cref{fig_Sopt_EW} in the main text, follows as
\beq \frac{2D}{\eta} \frac{L}{M^2} \Scal\eq\DirCP(x_M) = \frac{1}{\zeta_M} + \frac{1}{1-\zeta_M},\qquad \text{with}\quad  \zeta_M\equiv x_M/L.
\label{eq5_Seq_DirCP}\eeq 
$\Scal\eq\DirCP$ is minimal for a value of 
\beq x_M\DirCP = \frac{L}{2},
\label{eq5_xM_DirCP}\eeq 
which finally leads to the constrained profile 
\beq h\DirCP(x)/M = 1- \left|1-\frac{2x}{L}\right|.
\label{eq5_prof_DirCP}\eeq 
For Dirichlet no-flux \bcs, instead, the mass constraint is respected and, for $\mass=0$, one obtains 
\beq h\DirNoFl(x;x_M)/M = \begin{cases}\displaystyle \frac{\zeta \left[1+3\, \zeta_M(\zeta-1)\right]}{\zeta_M\left[1+3\zeta_M(\zeta_M-1)\right]} ,\qquad &x\leq x_M,\\
 h\DirNoFl(L-x,L-x_M),&x>x_M,
\end{cases}\label{eq5_prof_DirNoFl0}\eeq 
with $\zeta\equiv x/L$ and $\zeta_M \equiv x_M/L$.
Inserting \cref{eq5_prof_DirNoFl0} into \cref{eq5_Seq} results in 
\beq \frac{2D}{\frict} \frac{L}{M^2} S\eq\DirNoFl(x_M) = \frac{1}{\zeta_M} +  \frac{1}{1-\zeta_M} + \frac{3}{1+3\zeta_M(\zeta_M-1)},
\label{eq5_Seq_DirNoFl}\eeq 
which is illustrated in \cref{fig_Sopt_MH}.
This free energy has two symmetric minima, located at
\beq x_M\DirNoFl = \frac{L}{2}\left(1 \pm \frac{1}{\sqrt{3}}\right).
\label{eq5_constr_minLoc}\eeq 
The resulting optimal profile for Dirichlet no-flux \bcs and $\mass=0$ is related to be $h\pbc$ [\cref{eq5_prof_pbc}] via
\beq h\DirNoFl(x) = h\pbc(x + L/2-x_M\DirNoFl).
\label{eq5_prof_Dir_pbc_map}\eeq 
Specifically, upon choosing the smaller value for $x_M\DirNoFl$, one obtains
\beq h\DirNoFl(x)/M = \begin{cases}\displaystyle 6\zeta \left(\zeta + \frac{1}{\sqrt{3}}\right) ,\qquad &x\leq x_M\DirNoFl,\\
6(\zeta-1)\left(\zeta-1+\frac{1}{\sqrt{3}}\right) ,&x>x_M\DirNoFl.
\end{cases}\label{eq5_prof_DirNoFl}\eeq
Note that, since the above constrained profiles are polynomials of at most second order, one has $\pd_x^{(n)} h(x) = 0$ for $n\geq 3$ in each domain $x\lessgtr x_M$, such that no-flux \bcs [see \cref{eq_H_noflux}] are indeed fulfilled by $h\DirNoFl$.
In passing, we mention that, in the context of dewetting of thin films, related free-energy minimizing profiles have been considered in Refs.\ \cite{bausch_lifetime_1994, bausch_critical_1994,blossey_nucleation_1995,foltin_critical_1997}.

\section{Eigenvalue problem for the Mullins-Herring equation}
\label{sec_eigenv_mh}
Consider the noiseless MH equation,
\beq \pd_t h(x,t) =-  \pd_x^4 h(x,t),
\label{eq4_diffus_c}\eeq 
on the interval $[0,L]$ with 
\begin{subequations}\begin{align}
\text{periodic:}\qquad & h(x,t)=h(x+L,t), \label{eq4_bcs_per}\\
\text{Dirichlet:}\qquad & h(0,t)=0=h(L,t),\label{eq4_bcs_Dir} \\
\text{or Neumann:}\qquad & \pd_x h(0,t)=0=\pd_x h(L,t), \label{eq4_bcs_Neu}
\end{align}\label{eq4_bcs}\end{subequations}
\bcs.
The separation ansatz 
\beq h(x,t)=\sigma(x)\psi(t)
\eeq 
leads to 
\begin{subequations}\begin{align}
\pd_t \psi(t) = -\gamma \psi(t),\label{eq4_sep_t}\\
\pd_x^4 \sigma(x) = \gamma \sigma(x), \label{eq4_eigenvaleq}
\end{align}\end{subequations}
with a constant $\gamma\geq 0$.
While \cref{eq4_sep_t} is solved by 
\beq \psi(t)\sim e^{-\gamma t},
\eeq 
the general solution of the eigenvalue equation \eqref{eq4_eigenvaleq} is given by
\beq \sigma(x) = c_1 e^{x\gamma^{1/4}} + c_2 e^{-x\gamma^{1/4}} + c_3 \sin(x \gamma^{1/4}) + c_4 \cos(x \gamma^{1/4})
\label{eq4_ansatz}\eeq
with constants $c_i$, which are determined below for the specific \bcs. 

To proceed, it is useful to introduce the free energy functional $\Fcal[h] \equiv \int_0^L dx\, (\pd_x h)^2$ and the associated chemical potential $\mu\equiv \delta\Fcal/\delta h = -\pd_x^2 h$, which allows one to rewrite \cref{eq4_diffus_c} as a ``gradient-flow'' equation \cite{safran_statistical_1994}:
\beq \pd_t h =  \pd_x^2 \frac{\delta \Fcal}{\delta h} =  \pd_x^2 \mu = -\pd_x \left[ -\pd_x \mu\right].
\label{eq4_diffus_chempot}\eeq
In the last step we have identified $- \pd_x\mu$ as the flux, such that \cref{eq4_diffus_chempot} takes the form of a continuity equation.
Being a fourth order differential equation, \cref{eq4_diffus_c} requires two additional conditions on $h$ beside those specified in \cref{eq4_bcs}. 
Here, one typically chooses either a vanishing chemical potential at the boundaries:
\beq \mu(0,t) = 0 = \mu(L,t) \qquad \Leftrightarrow \qquad \sigma''(0) = 0 = \sigma''(L),
\label{eq4_zerochempot_bcs}\eeq 
or a vanishing flux:
\beq
\pd_x\mu(0,t) = 0 = \pd_x\mu(L,t) \qquad \Leftrightarrow \qquad \sigma'''(0)= 0 = \sigma'''(L).
\label{eq4_noflux_bcs}\eeq 
In contrast to the zero-chemical potential \bcs in \cref{eq4_zerochempot_bcs}, no-flux \bcs ensure mass conservation for the MH equation in a finite domain.

The type of boundary condition determines whether the operator $\pd_x^4$ is \emph{self-adjoint} on the interval $[0,L]$ (see, e.g., Refs.\ \cite{esposito_fourth-order_1999, birkhoff_boundary_1908, smith_spectral_2011}).
Since, for two arbitrary functions $\sigma(x)$ and $\varphi(x)$, one has
\beq \int_0^L \d x\, \sigma^{(4)}(x) \varphi(x) = [\sigma \varphi''']_0^L - [\sigma' \varphi'']_0^L + [\sigma'' \varphi']_0^L - [\sigma''' \varphi]_0^L + \int_0^L \d x\, \sigma(x) \varphi^{(4)}(x),
\label{eq4_selfadjoint}\eeq
the operator $\pd_x^4$ is self-adjoint only if both $\sigma$ and $\varphi$ fulfill either (i) periodic \bcs [\cref{eq4_bcs_per}], (ii) Dirichlet zero-chemical potential \bcs [\cref{eq4_bcs_Dir,eq4_zerochempot_bcs}], or (iii) Neumann no-flux \bcs [\cref{eq4_bcs_Neu,eq4_noflux_bcs}].
In these cases, the eigenfunctions $\sigma_m$ defined by \cref{eq4_eigenvaleq}, with $m\in\mathbb{Z}$ enumerating the spectrum, are orthogonal:
\beq \int_0^L \d x\, \sigma_m^*(x) \sigma_n(x) =0,\qquad m\neq n.
\label{eq4_ortho_eigen}\eeq 
In contrast, for Dirichlet no-flux \bcs [\cref{eq4_bcs_Dir,eq4_noflux_bcs}], the boundary terms in \cref{eq4_selfadjoint} do not vanish. 
Consequently, $\pd_x^4$ is not self-adjoint on $[0,L]$ and the ensuing eigenfunctions $\sigma_m$ are not guaranteed to be orthogonal.
This issue can be dealt with by introducing a set of eigenfunctions $\varphi_m(x)$ which solve the associated \emph{adjoint} eigenproblem \cite{birkhoff_boundary_1908}.
In the case of Dirichlet no-flux \bcs, this is defined by the eigenvalue equation
\beq \pd_x^4 \varphi(x) = \tilde\gamma \varphi(x)
\label{eq4_adj_eigenvaleq}\eeq 
and the \bcs 
\begin{subequations}\begin{align}
\varphi'(0) = &\, 0 = \varphi'(L), \\
\varphi''(0) = &\, 0 = \varphi''(L).
\end{align}\label{eq4_adj_bcs}\end{subequations}
Note that these \bcs are indeed such that, upon using \cref{eq4_noflux_bcs}, all boundary terms in \cref{eq4_selfadjoint} vanish.
In general, the (suitably ordered) proper and adjoint eigenvalues, $\gamma_m$ and $\tilde\gamma_m$, coincide \cite{birkhoff_boundary_1908},
\beq \gamma_m = \tilde \gamma_m.
\label{eq4_adj_eigenv_id}\eeq 
This result is proven explicitly in \cref{app_DirNoFl_eigenf}.
Upon using this fact, \cref{eq4_selfadjoint} readily yields the mutual orthogonality of the proper and adjoint eigenfunctions $\sigma_m$, $\varphi_n$:
\beq \int_0^L \d x\, \sigma_m^*(x) \varphi_n(x) = 0,\qquad m\neq n.
\label{eq4_adj_ortho_eigen}\eeq 
This equation replaces \cref{eq4_ortho_eigen} in the non-self-adjoint case and is crucial in constructing the eigenfunction solution of \cref{eq4_diffus_c} or \eqref{eq_MH_MFTr} for Dirichlet no-flux \bcs.
We now proceed by discussing the eigenproblem of the MH equation for various \bcs.

\subsection{Dirichlet \bcs}
\subsubsection{Vanishing flux}
\label{app_DirNoFl_eigenf}
We consider here the proper eigenproblem defined by \cref{eq4_eigenvaleq} and turn to the adjoint problem in the next subsection.
Defining 
\beq \omega\equiv L\gamma^{1/4},
\eeq 
the four conditions in \cref{eq4_bcs_Dir,eq4_noflux_bcs} result in the requirement
\beq \begin{pmatrix}
      1 		& 1 			& 0 			& 1 \\
      e^{\omega} & e^{-\omega} & \sin(\omega) & \cos(\omega) \\
      1			& -1			& -1		& 0 \\
      e^{\omega} & -e^{-\omega} & -\cos(\omega) & \sin(\omega)
     \end{pmatrix} 
     \begin{pmatrix}
      c_1 \\ c_2 \\ c_3 \\ c_4
     \end{pmatrix} = 
     \begin{pmatrix}
      0 \\ 0 \\ 0 \\ 0
     \end{pmatrix}
\label{eq4_DirFl0_matrix}\eeq
for the coefficients $c_i$ defined in \cref{eq4_ansatz}.
For a nontrivial solution of \cref{eq4_DirFl0_matrix} to exist, the determinant of the coefficient matrix must vanish, which implies
\beq \cos(\omega) \cosh(\omega) = 1.
\label{eq_DirFl0_det}\eeq 
In general, the solutions of \cref{eq_DirFl0_det} cannot be represented in a simple form. Numerically, one obtains
\beq \omega_k = 0,\, \pm 4.7300,\, \pm 7.8532,\, \pm 10.9956,\ldots \qquad (k=0,\pm 1,\pm 2,\ldots).
\label{eq4_DirFl0_eigenval}\eeq 
For $k\gtrsim 4$ the eigenvalues are well approximated by
\beq |\omega_k| \simeq \pi \left( k+ \onehalf \right),
\label{eq4_DirFl0_eigenval_approx}\eeq 
which becomes exact in the limit $\omega\to\pm \infty$.
Using \cref{eq_DirFl0_det}, it can be shown that the eigenvalues $\omega_k$ fulfill the relation
\beq \sin(\omega_k) = \mathrm{sgn}\left(\omega_k\right) (-1)^k \sqrt{1-\frac{1}{\cosh^2(\omega_k)}}.
\label{eq4_DirFl0_sinhlp}\eeq
Accordingly, \cref{eq4_DirFl0_matrix} reduces to
\beq \begin{pmatrix}
      1 		& 1 			& 0 			& 1 \\
      e^{\omega_k} & e^{-\omega_k} & (-1)^k\tanh(\omega_k) & 1/\cosh(\omega_k) \\
      1			& -1			& -1		& 0 \\
      e^{\omega_k} & -e^{-\omega_k} & -1/\cosh(\omega_k) & (-1)^k\tanh(\omega_k)
     \end{pmatrix} 
     \begin{pmatrix}
      c_1 \\ c_2 \\ c_3 \\ c_4
     \end{pmatrix} = 
     \begin{pmatrix}
      0 \\ 0 \\ 0 \\ 0
     \end{pmatrix},
\label{eq4_DirFl0_matrix2}\eeq
which yields for the $c_i$ the nontrivial solutions
\beq \begin{pmatrix}
       c_1,  c_2, c_3, c_4
     \end{pmatrix}_k
     =
     (\sgn \omega_k)^k \begin{pmatrix}\displaystyle
       - \frac{(-1)^k}{\sqrt{3+3 e^{2\omega_k}}}, & \displaystyle
       -\frac{\sqrt{1+\tanh(\omega_k)}}{\sqrt{6}}, & \displaystyle
       \frac{-(-1)^k+e^{\omega_k}}{\sqrt{3+3e^{2\omega_k}}}, & \displaystyle
       \frac{(-1)^k+e^{\omega_k}}{\sqrt{3+3e^{2\omega_k}}}
     \end{pmatrix}.
\label{eq4_DirFl0_coeffsol}\eeq 
The eigenfunctions $\sigma_k(x)$ [\cref{eq4_ansatz}] of the operator $\pd_x^4$ for Dirichlet no-flux \bcs thus result as
\beq \sigma_k\DirNoFl(x) = c_{1,k} e^{x\gamma_k^{1/4}} + c_{2,k} e^{-x\gamma_k^{1/4}} + c_{3,k} \sin(x \gamma_k^{1/4}) + c_{4,k} \cos(x \gamma^{1/4}_k),
\label{eq4_DirFl0_eigenf}\eeq 
with the $c_{i,k}$ given in \cref{eq4_DirFl0_coeffsol}.
It is straightforward to show that $\sigma_{k=0}\DirNoFl(x)=0$ as well as $\sigma_k\DirNoFl(x) = \sigma_{-k}\DirNoFl(x)$ [cf.\ \cref{eq4_DirFl0_eigenval}]. Hence, we can restrict $k$ to strictly positive values, such that the general solution of \cref{eq4_diffus_c} reads
\beq h\DirNoFl(x,t) = \sum_{k=1}^\infty a_k e^{-\gamma_k t}  \sigma_k\DirNoFl(x) ,
\label{eq4_DirFl0_sol}\eeq 
with constants $a_k$.
It is furthermore useful to note that $\sigma_k\DirNoFl(L/2)=0$ for odd $k$. 
The eigenfunctions $\sigma_k\DirNoFl$ are not normalized here, but instead one has
\beq \int_0^L \d x \left[\sigma_k\DirNoFl(x)\right]^2 = \frac{L}{3}\left(1+\frac{(-1)^k}{\cosh\omega_k} -\frac{2}{\omega}\tanh(\omega_k) \right).
\label{eq4_DirFl0_norm}\eeq
Upon using \cref{eq_DirFl0_det,eq4_DirFl0_sinhlp} it can be shown that the mass identically vanishes:
\beq \int_0^L \d x\, \sigma_k\DirNoFl(x) = 0.
\label{eq4_DirFl0_mass}\eeq 
Consequently, the solution in \cref{eq4_DirFl0_sol} is only compatible with initial conditions having zero mass. [A nonzero mass can be trivially introduced by adding a constant to the r.h.s.\ of \cref{eq4_DirFl0_sol}.]
Moreover, it can be readily checked that, as a consequence of the non-self-adjoint character of $\pd_x^4$ for Dirichlet no-flux \bcs, the eigenfunctions $\sigma_k\DirNoFl(x)$ are in general not orthogonal.
This is the reason for considering an additional adjoint set of eigenfunctions (see below).
In \cref{fig_eigenfunc_mh_DirNoFlux}(a), the first few eigenfunctions defined by \cref{eq4_DirFl0_eigenf} are illustrated.

\begin{figure}[t]\centering
    \subfigure[]{\includegraphics[width=0.45\linewidth]{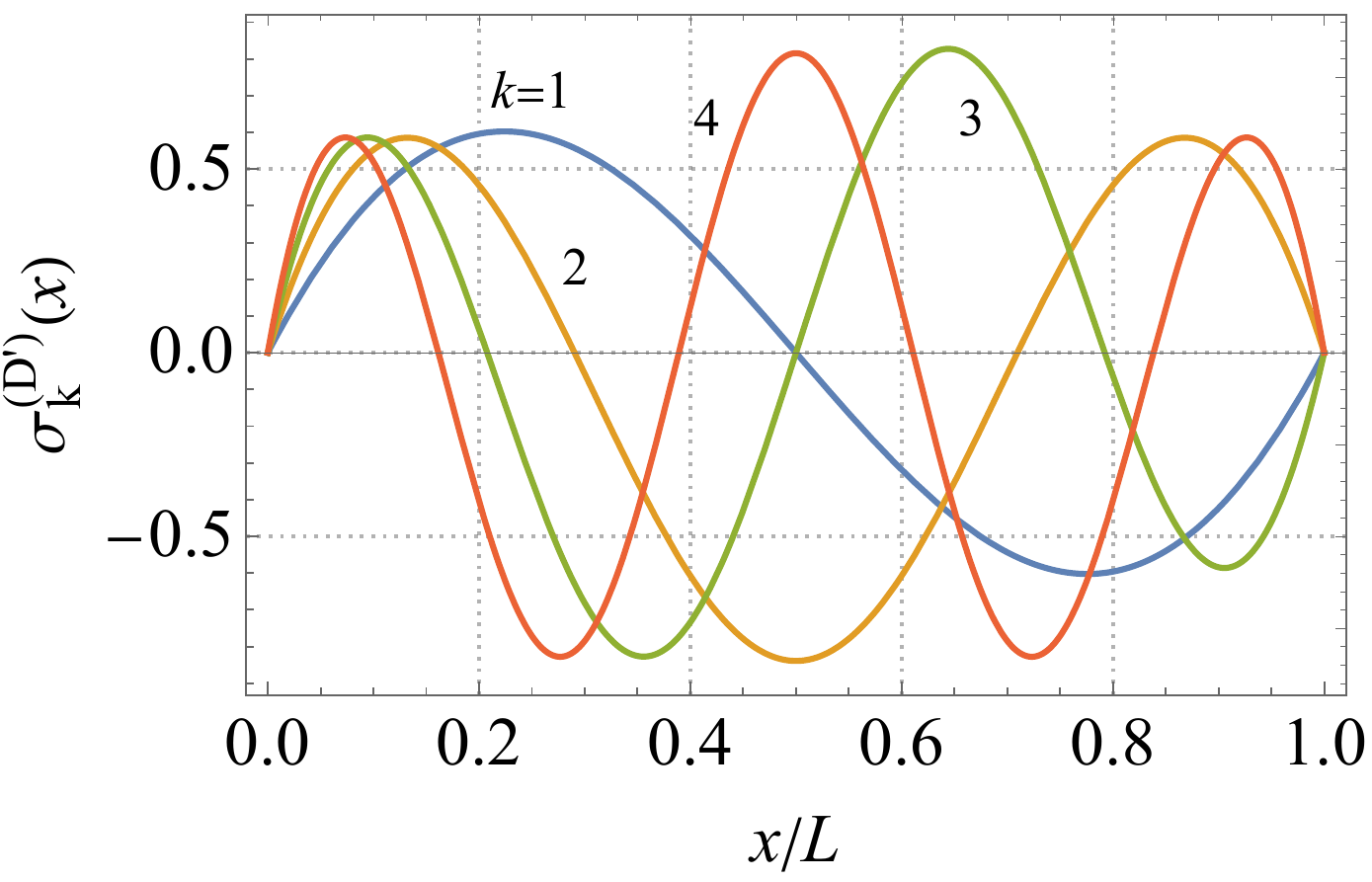}}\qquad
    \subfigure[]{\includegraphics[width=0.45\linewidth]{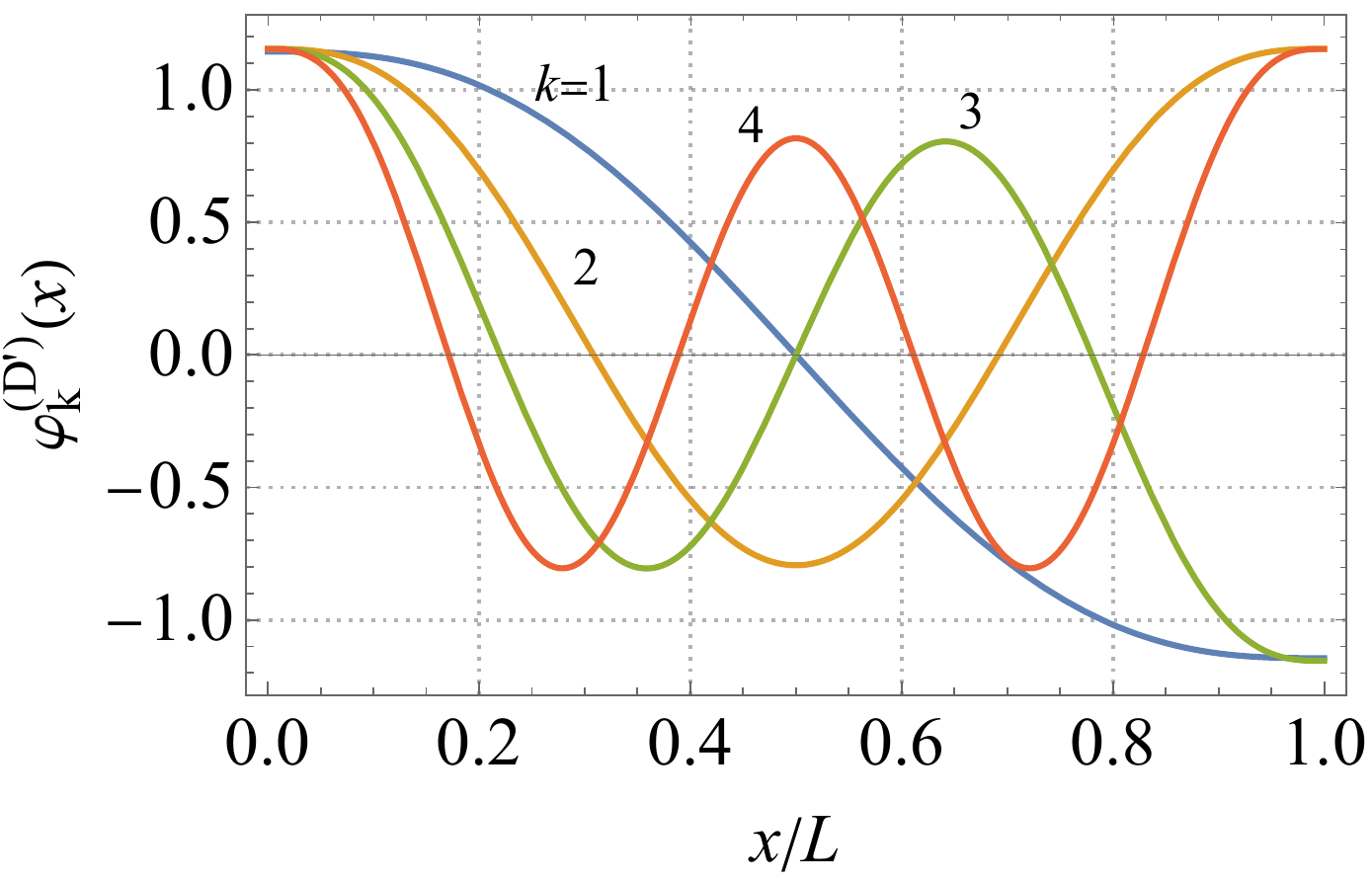}}
    \caption{(a) Eigenfunctions $\sigma\DirNoFl_k$ [\cref{eq4_DirFl0_eigenf}] for Dirichlet no flux \bcs [\cref{eq4_bcs_Dir,eq4_noflux_bcs}] for the four lowest modes $k=1,\ldots,4$. (b) Associated adjoint eigenfunctions $\varphi\DirNoFl_k$ given by \cref{eq4_adj_ansatz,eq4_adj_DirFl0_coeffsol}. For $k=0$ one has $\sigma\DirNoFl_{k=0}(x)=0$ and $\varphi\DirNoFl_{k=0}(x) = 2 \sqrt{\sfrac{2}{3}}$.}
    \label{fig_eigenfunc_mh_DirNoFlux}
\end{figure}

\subsubsection{Vanishing flux: adjoint eigenproblem}
\label{app_DirNoFl_eigenf_adj}
We now turn to the adjoint eigenvalue problem associated with Dirichlet no-flux \bcs, which is defined by \cref{eq4_adj_eigenvaleq,eq4_adj_bcs}.
The ansatz for the solution of the adjoint eigenvalue equation \eqref{eq4_adj_eigenvaleq} is of the same form as in \cref{eq4_ansatz}, i.e.,
\beq \varphi(x) = \tilde c_1 e^{x\tilde\gamma^{1/4}} + \tilde c_2 e^{-x\tilde\gamma^{1/4}} + \tilde c_3 \sin(x \tilde\gamma^{1/4}) + \tilde c_4 \cos(x \tilde\gamma^{1/4}).
\label{eq4_adj_ansatz}\eeq
The four conditions in \cref{eq4_adj_bcs} imply 
\beq \begin{pmatrix}
      1 		& -1 			& 1 			& 0 \\
      e^{\tilde\omega} & -e^{-\tilde\omega} & \cos(\tilde\omega) & -\sin(\tilde\omega) \\
      1			& 1			& 0		& -1 \\
      e^{\tilde\omega} & e^{-\tilde\omega} & -\sin(\tilde\omega) & -\cos(\tilde\omega)
     \end{pmatrix} 
     \begin{pmatrix}
      \tilde c_1 \\ \tilde c_2 \\ \tilde c_3 \\ \tilde c_4
     \end{pmatrix} = 
     \begin{pmatrix}
      0 \\ 0 \\ 0 \\ 0
     \end{pmatrix}
\label{eq4_adj_DirFl0_matrix}\eeq
for the coefficients $\tilde c_i$, where
\beq \tilde\omega\equiv L\tilde\gamma^{1/4}.
\eeq 
Existence of a nontrivial solution of \cref{eq4_DirFl0_matrix} implies the following determinant condition:
\beq \cos(\tilde\omega) \cosh(\tilde\omega) = 1.
\label{eq_adj_DirFl0_det}\eeq 
As anticipated, this relation coincides with \cref{eq_DirFl0_det} and, consequently, also the adjoint and the proper eigenvalues [see \cref{eq4_DirFl0_eigenval}] coincide:
\beq \tilde \omega_k = \omega_k.
\label{eq4_adj_DirFl0_eigenval}\eeq 
Proceeding as in \cref{app_DirNoFl_eigenf}, one obtains the nontrivial solutions of \cref{eq4_adj_DirFl0_matrix} as
\beq \begin{pmatrix}
       \tilde c_1,  \tilde c_2, \tilde c_3, \tilde c_4
     \end{pmatrix}_k
     =
     \begin{pmatrix}\displaystyle
       \frac{(-1)^k}{\sqrt{3+3 e^{2\omega_k}}}, &
       \displaystyle \frac{1}{\sqrt{6}}\sqrt{1+\tanh(\omega_k)}, &
       \displaystyle \frac{-(-1)^k+e^{\omega_k}}{\sqrt{3+3e^{2\omega_k}}}, &
       \displaystyle \frac{(-1)^k+e^{\omega_k}}{\sqrt{3+3e^{2\omega_k}}}
     \end{pmatrix}.
\label{eq4_adj_DirFl0_coeffsol}\eeq 
Since the eigenfunctions $\varphi_k(x)$ resulting from \cref{eq4_adj_ansatz,eq4_adj_DirFl0_coeffsol} are identical for $\pm \omega_k$, we consider henceforth only $\omega_k\geq 0$, i.e., $k\geq 0$.
As a consequence of \cref{eq4_adj_DirFl0_eigenval}, the orthogonality property in \cref{eq4_adj_ortho_eigen} follows. Specifically, one has (note that $\sigma$ and $\varphi$ are real-valued)
\beq \int_0^L \d x\, \sigma_m(x) \varphi_n(x) = \frac{L}{3}\left(1-\frac{(-1)^n}{\cosh(\omega_n)} \right) \delta_{mn}.
\label{eq4_adj_crossnorm}\eeq 
Furthermore, one readily proves the useful property 
\beq \int_0^L \d x\, \varphi_m(x) \varphi_n''(x) = -\frac{L}{3}\omega_n^2 \left(1-\frac{(-1)^n}{\cosh(\omega_n)}\right) \delta_{mn}.
\label{eq4_adj_eigenDx2_norm}\eeq
In \cref{fig_eigenfunc_mh_DirNoFlux}(b), the first few adjoint eigenfunctions $\varphi_k$ are illustrated.

\subsubsection{Vanishing chemical potential}
\label{app_DirCP_eigenf}

For completeness, we summarize here the solution of the eigenproblem for Dirichlet \bcs with a vanishing chemical potential at the boundaries (also called \emph{Dirichlet zero-$\mu$} \bcs).
Following the same steps as in \cref{app_DirNoFl_eigenf} renders the well-known normalized eigenfunctions 
\beq \sigma_k(x) = \sqrt{\frac{2}{L}} \sin(x \gamma_k^{1/4}),\qquad \gamma_k = \left(\frac{\pi k}{L}\right)^4,\qquad k=0,1,2,\ldots .
\label{eq4_DirCP0_eigenfunc}\eeq
Note that, since $\sigma_{k=0}(x)=0$, $k=0$ is not considered to be part of the actual eigenspectrum.
In summary, the solution of \cref{eq4_diffus_c} for Dirichlet zero-$\mu$ \bcs takes the well-known form
\beq h\DirCP(x,t) = \sum_{k=1}^\infty a_k e^{-\left(\frac{\pi k}{L}\right)^4 t} \sqrt{\frac{2}{L}}\sin\left(\frac{\pi k}{L} x\right),
\label{eq4_DirCP0_hsol}\eeq 
where the constants $a_k$ are determined by the initial conditions on $h\DirCP$.

Requiring a constant chemical potential at the boundaries generally leads to a mass loss during the time evolution:
\beq \int_0^L \d x\, h\DirCP(x,t) = \sum_{k=1}^\infty a_k e^{-\left(\frac{\pi k}{L}\right)^4 t} \times
\begin{cases} 
\frac{2L}{\pi k}, & \text{odd $k$},\\
0, & \text{even $k$}. 
\end{cases}
\eeq
One may wonder whether the coefficients $a_k$ can be chosen such that $h\DirCP$ [\cref{eq4_DirCP0_hsol}] satisfies no-flux \bcs [\cref{eq4_noflux_bcs}]: requiring a vanishing third derivative of $h\DirCP$ at the boundaries results in a relation involving the sum over all modes, e.g., for $x=0$ one has $0=\sum_{k=1}^\infty a_k \exp(-(\pi k/L)^4 t) (\pi k/L)^3$. As is readily seen, it is not possible to choose the coefficients $a_k$ such that no-flux \bcs are ensured during the \emph{whole} time evolution of $h\DirCP$. This requires, instead, a specific set of basis functions.

\subsection{Periodic \bcs}
In the case of periodic \bcs [\cref{eq4_bcs_per}], one has $c_1=c_2=0$ in \cref{eq4_ansatz} and $L\gamma_n^{1/4}=2\pi n$ with $n=0,1,2,\ldots$.
This yields the well-known series expansion
\beq 
h\pbc(x,t) = \sum_{k=-\infty}^\infty a_k e^{-\left(\frac{2\pi k}{L}\right)^4 t} \sqrt{\frac{1}{L}} e^{\frac{2\pi \im k}{L} x}.
\label{eq4_per_hsol}\eeq 
The parameters $a_k$ must fulfill $a_{-k}=a^*_k$ in order to ensure that $h\pbc$ is real-valued.
Since $\int_0^L dx\, h\pbc(x,t)=a_0$, the mass [\cref{eq_mass}] is conserved in time.

\subsection{Neumann \bcs}
Imposing Neumann \bcs [\cref{eq4_bcs_Neu}] in conjunction with a no-flux condition [\cref{eq4_noflux_bcs}] renders a solution of \cref{eq4_diffus_c} in terms of standard Neumann eigenfunctions:
\beq
h\ut{(N)}(x,t) = \sum_{k=0}^\infty a_k e^{-\left(\frac{\pi k}{L}\right)^4 t} \sqrt{\frac{2-\delta_{k,0}}{L}} \cos\left(\frac{\pi  k}{L} x\right) .
\eeq
We shall, however, not discuss Neumann \bcs further.

\section{Solution of weak-noise theory for the optimal profile}
\label{app_WNT_sol}
Here, the general solution of \cref{eq_EW_MFTr,eq_MH_MFTr} is determined, following the approach outlined in Ref.\ \cite{meerson_macroscopic_2016} for periodic \bcs.
Recall that a flat profile is assumed at the initial time [\cref{eq_init_cond}],
\beq h(x,t=0) = 0,
\label{eq6_h_initial}\eeq 
while the first-passage event at time $T$ is defined by the condition that $h$ attains its maximum height $M>0$ at the location $x_M$ [\cref{eq_firstpsg_cond}],
\beq h(x_M,T) = M.
\label{eq6_h_final}\eeq 
However, for actually determining the solution of WNT, we neither explicitly enforce that $h$ does not reach the height $M$ before $T$, nor that the profile stays below $M$ for all $x\neq x_M$. Consequently, one has to check at the end of the calculation that the obtained solution fulfills these conditions.
For sufficiently large $M$, this turns out to be the case.

We begin by casting \cref{eq_EW_MFTr,eq_MH_MFTr} into the common form
\begin{subequations}\begin{align}
\pd_t h &= (-\pd_x^2)^b\, [\pd_x^2 h+2p],\label{eq6_h_eq}\\
\pd_t p &= -(-\pd_x^2)^b \pd_x^2 p, \label{eq6_p_eq}
\end{align}\label{eq6_common}\end{subequations}
where $b=0$ for EW dynamics and $b=1$ for MH dynamics.
The profile $h(x,t)$ is assumed to fulfill either periodic or Dirichlet \bcs [see \cref{eq_H_pbc,eq_H_Dbc}]. For MH dynamics with Dirichlet \bcs, we additionally assume either a vanishing chemical potential [\cref{eq4_zerochempot_bcs}] or a vanishing flux [\cref{eq4_noflux_bcs}] at the boundaries. (In the main text, we focus only on the latter.) 
The profile is expanded into a set of eigenfunctions $\sigma_k$,
\beq 
h(x,t) = \sum_k h_k(t) \sigma_k(x),
\label{eq6_h_expansion}\eeq 
which are determined by the associated eigenvalue problem [see \cref{sec_eigenv_mh}],
\beq \pd_x^{z} \sigma_k(x) = \gamma_k \sigma_k(x),
\label{eq6_eigen}\eeq 
where the dynamic index $z=2b+2$.
The conjugate field $p$ satisfies the \bcs of the associated adjoint eigenproblem [see \cref{sec_eigenv_mh}] and is accordingly expanded in terms of the adjoint eigenfunctions $\varphi_k$ as
\beq 
p(x,t) = \sum_k p_k(t) \varphi_k(x).
\label{eq6_p_expansion}\eeq
The adjoint eigenfunctions $\varphi_k$ fulfill 
\beq \pd_x^{z} \varphi_k(x) = \gamma_k \varphi_k(x).
\label{eq6_adj_eigen}\eeq 
If the operator $\pd_x^{z}$ is self-adjoint on $[0,L]$, one has $\varphi_k=\sigma_k$.  
This is in particular the case for periodic or Dirichlet zero-$\mu$ \bcs, such that
\beq \varphi_k\ut{(p,D)}=\sigma_k\ut{(p,D)}.
\eeq 
In contrast, for Dirichlet no-flux \bcs on $h$, the operator $\pd_x^{4}$ is not self-adjoint.
In this case, the required adjoint eigenfunctions $\varphi_k\DirNoFl$, which fulfill Neumann zero-$\mu$ \bcs [see \cref{eq4_adj_bcs}], are provided in \cref{app_DirNoFl_eigenf_adj} \footnote{It turns out that the adjoint eigenmode $\varphi_{k=0}\DirNoFl$ does not contribute to the dynamics for Dirichlet no-flux \bcs and will therefore be neglected henceforth in the corresponding expansion in \cref{eq6_p_expansion}.}.

By construction, $\sigma_m$ and $\varphi_n$ are mutually orthogonal [see \cref{eq4_adj_ortho_eigen}]
\beq \int_0^L \d x\, \sigma_m^*(x)\varphi_n(x)  = \kappa_n\delta_{mn},
\label{eq6_ortho}\eeq 
where the star denotes complex conjugation and $\kappa_n$ is a real number.
Complex conjugation is necessary here in order to also take into account complex-valued eigenfunctions, which occur in the case of periodic \bcs [see \cref{eq4_per_hsol}].
We furthermore have
\beq \int_0^L \d x\, \varphi_m^*(x) \varphi_n''(x) = \epsilon_n\delta_{mn},
\label{eq6_orthoDx2}\eeq 
with a real number $\epsilon_n$.
The relevant properties of $\sigma_k$, $\varphi_k$ are summarized in \cref{tab_eigenfunc}.

\begin{table}[t]\begin{center}
\begin{tabular}{l|c|c|c}
    & periodic [\cref{eq4_bcs_per}] & Dirichlet zero-$\mu$ [Eqs.\ \eqref{eq4_bcs_Dir}, \eqref{eq4_zerochempot_bcs}] & Dirichlet no-flux [Eqs.\ \eqref{eq4_bcs_Dir}, \eqref{eq4_noflux_bcs}] ($b=1$)${}^\dagger$ \\
\hline
$\pd_x^{z}$ self-adjoint & yes & yes & no \\[2ex]
$\sigma_k$ & $\dps \frac{1}{\sqrt{L}} \exp\left(\frac{2\pi \im k}{L} x\right)$  & $\dps \sqrt{\frac{2}{L}} \sin\left(\frac{k\pi}{L}x\right)$ & $\sigma_k\DirNoFl$ [\cref{eq4_DirFl0_eigenf}] \\[4ex]
$\varphi_k$ & $\sigma_k$ & $\sigma_k$ & $\varphi_k\DirNoFl$ [\cref{eq4_adj_ansatz}] \\[3ex]
$k$ & $0,\pm 1,\pm 2,\ldots^\ddagger$ & $1,2,3, \ldots$ & $1,2,3, \ldots$ \\[3ex]
$\gamma_k$ [Eqs.\ \eqref{eq6_eigen}, \eqref{eq6_adj_eigen}] & $\dps (-1)^{b+1}\left(\frac{2\pi k}{L}\right)^{z}$  & $\dps (-1)^{b+1}\left(\frac{k\pi}{L}\right)^{z}$ & $(\omega_k/L)^4$ [\cref{eq4_DirFl0_eigenval}]  \\[3ex]
$\kappa_k$ [\cref{eq6_ortho}] & 1 & 1 & $\dps \frac{L}{3}\left(1-\frac{(-1)^k}{\cosh(L\gamma_k^{1/4})} \right)$ [\cref{eq4_adj_crossnorm}] \\[3ex]
$\epsilon_k$ [\cref{eq6_orthoDx2}] & \vtop{\hbox{\strut $\dps \left[-|\gamma_k|^{1/2}\right]^b\kappa_k$,}\hbox{\strut $\epsilon_0=0$}} & $\dps \left[-|\gamma_k|^{1/2}\right]^b\kappa_k$ & $-\gamma_k^{1/2} \kappa_k$ [\cref{eq4_adj_eigenDx2_norm}] \\[3ex]
\end{tabular}
\end{center}\caption{Eigenfunctions and related properties of the operator $\pd_x^{z}$ on the interval $[0,L]$ for various \bcs. The proper and adjoint eigenfunctions are denoted by $\sigma_k$ and $\varphi_k$, respectively, and they coincide if $\pd_x^{z}$ is self-adjoint. The dynamic index $z$ is related to the parameter $b$ via $z=2b+2$, with $b=0$ for EW dynamics and $b=1$ for MH dynamics [see \cref{eq6_common}]. ${}^\dagger$Dirichlet no-flux \bcs are considered only for $b=1$. Note that $\sigma_k\DirNoFl$ and $\varphi_k\DirNoFl$ are not normalized here, such that the system size $L$ appears in the corresponding expression for $\kappa_k$. $^\ddagger$Due to the mass constraint [\cref{eq_zero_vol}], the zero mode ($k=0$) is absent from the actual solution for periodic \bcs [see \cref{eq6_zeromode_pbc} below].}
\label{tab_eigenfunc}\end{table}

To proceed, we insert the expansions given in \cref{eq6_h_expansion,eq6_p_expansion} into \cref{eq6_common}, multiply \cref{eq6_h_eq} by $\varphi_k^*$, \cref{eq6_p_eq} by $\sigma_k^*$, and make use of the orthogonality properties in \cref{eq6_ortho,eq6_orthoDx2}. This yields ordinary differential equations for the coefficients $h_k$ and $p_k$:
\begin{subequations}\begin{align}
\dot h_k &= (-1)^b \left( \gamma_k h_k + 2p_k \hat\epsilon_k \right),\label{eq6_deq_hk}\\
\dot p_k &= (-1)^{b+1} \gamma_k p_k ,\label{eq6_deq_pk}
\end{align}\label{eq6_deq_coeffs}\end{subequations}
with 
\beq \hat\epsilon_k \equiv \begin{cases}
                       1, & b=0\\
                       \epsilon_k/\kappa_k, & b=1.
                      \end{cases}
\eeq 
Equation \eqref{eq6_deq_pk} is solved by 
\beq p_k(t) = B_k \exp\left[(-1)^{b+1} \gamma_k t\right],
\label{eq6_pk_sol0}\eeq 
with integration constants $B_k$ determined below.
The solution of \cref{eq6_deq_hk} follows as
\beq h_k(t) = \begin{cases}
\displaystyle  A_k \exp\left[(-1)^b\gamma_k t\right] - p_k(t)\frac{\hat\epsilon_k}{\gamma_k},\qquad & \gamma_k\neq 0\\
  A_k + (-1)^b 2 \hat\epsilon_k B_k t, &\gamma_k=0.
  \end{cases}
\label{eq6_hk_sol0}\eeq
As can be inferred from \cref{tab_eigenfunc}, the case $\gamma_k=0$ is only relevant for $k=0$ and periodic \bcs, where one obtains a linear dependence of $h_0$ on time for $b=0$ (EW dynamics), whereas $\hat\epsilon_0=0$ for $b=1$.
Imposing the initial condition in \cref{eq6_h_initial} and using \cref{eq6_hk_sol0,eq6_pk_sol0} yields
\beq B_k = \frac{\gamma_k}{\hat\epsilon_k} A_k,\qquad (\gamma_k\neq 0),
\eeq 
while for $\gamma_k=0$ ($k=0$), one obtains $A_0=0$ and $B_0$ is left undetermined.
Accordingly,
\beq h_k(t) = \begin{cases}
		2 A_k \sinh\left((-1)^b \gamma_k t\right),\qquad & \gamma_k\neq 0,\\
		(-1)^b 2 \hat\epsilon_0 B_0 t, & \gamma_k=0,
              \end{cases}
\label{eq6_hk_sol1}\eeq 
from which readily follows that $h_0(t)=0$ for periodic \bcs and MH dynamics. 
Expanding the profile at the final time $T$ as
\beq h(x,T) =  \sum_k H_k \sigma_k(x),
\label{eq6_h_final_expansion}\eeq
provides the relations
\beq A_k = \frac{H_k}{2\sinh\left((-1)^b \gamma_k T\right)},\qquad (\gamma_k\neq 0)
\eeq 
as well as $B_0= (-1)^b H_0/(2\hat\epsilon_0 T)$ (for  $\gamma_0=0$ and if $\hat\epsilon_0\neq 0$). 
Summarizing, in terms of the (yet undetermined) coefficients $H_k$, the solution of \cref{eq6_deq_coeffs} is given, for $\gamma_k\neq 0$, by
\begin{subequations}\begin{align}
h_k(t) &= H_k \frac{\sinh\left((-1)^b \gamma_k t\right)}{\sinh\left((-1)^b \gamma_k T\right)}, \\
p_k(t) &= H_k \frac{ \gamma_k \exp\left(-(-1)^{b} \gamma_k t\right)}{2\hat\epsilon_k \sinh\left((-1)^b \gamma_k T\right)}. \label{eq6_p_sol}
\end{align}\label{eq6_hp_sol}\end{subequations}
In the special case $\gamma_0=0$, $\hat\epsilon_0\neq 0$ ($k=0$), corresponding to EW dynamics with periodic \bcs, one has
\begin{subequations}\begin{align}
h_0(t) &= H_0 \frac{t}{T}, \\
p_0(t) &= (-1)^b \frac{H_0}{2\hat\epsilon_0 T},
\end{align}\label{eq6_hp_sol_k0_EW}\end{subequations}
whereas for $\gamma_0=0$, $\hat\epsilon_0=0$, corresponding to MH dynamics with periodic \bcs, one has
\begin{subequations}\begin{align}
h_0(t) &= 0, \\
p_0(t) &= \const.
\end{align}\label{eq6_hp_sol_k0_MH}\end{subequations}
In fact, performing the limit $\gamma_k\to 0$ in \cref{eq6_hp_sol} leads to the expressions in \cref{eq6_hp_sol_k0_EW}.
Furthermore, the fact that $h_0(t)=0$ for periodic \bcs and MH dynamics [see \cref{eq6_hk_sol1}] implies $H_0=0$ in this case.
This allows us to generally proceed by using \cref{eq6_hp_sol}, keeping in mind that $p_0(t)=0$ for periodic \bcs and MH dynamics [as this result does not readily follow from a limit of \cref{eq6_p_sol}].

The coefficients $H_k$ are determined by minimizing the (rescaled) action in \cref{eq_Sopt2_resc,eq_Sopt4}, 
\beq \Scal\opt[p] = (-1)^b \int_0^T \d t \int_0^L \d x\, p (\pd_x^{2b} p),
\label{eq6_Sopt}\eeq 
subject to the constraint in \cref{eq6_h_final}.
Inserting the expansion defined in \cref{eq6_p_expansion,eq6_p_sol} into $\Scal\opt$ and making use of the orthogonality property in \cref{eq6_orthoDx2} leads to 
\beq \Scal\opt = \sum_k \frac{\gamma_k \tilde\epsilon_k}{2\hat \epsilon_k^2\left[\exp\left(2(-1)^b \gamma_k T\right)-1\right]} |H_k|^2 \equiv \sum_k N_k(T) |H_k|^2,
\label{eq6_Sopt_expansion}\eeq 
where 
\beq \tilde \epsilon_k \equiv \begin{cases}
				\kappa_k,\qquad &b=0,\\
				\epsilon_k, &b=1,
                              \end{cases}
\eeq                               
and the quantity $N_k(T)$ is introduced as a shorthand notation.
Taking into account \cref{eq6_h_final_expansion}, the augmented action reads
\beq \tilde \Scal\opt = \Scal\opt - \lambda \left[h(x_M,T)-M\right] = \sum_k N_k(T) |H_k|^2 - \lambda\left[\sum_k H_k \sigma_k(x_M)-M\right],
\eeq 
where $\lambda$ is a Lagrange multiplier. Minimization of $\tilde \Scal\opt$ with respect to $H_k$, i.e., requiring  $0 = \sfrac{\delta \tilde \Scal\opt}{\delta H_k}$, results in 
\beq H_k^* = \frac{\lambda \sigma_k(x_M)}{2 N_k(T)}.
\label{eq6_hk_sol2}\eeq 
The complex conjugation in \cref{eq6_hk_sol2} is relevant only for periodic \bcs, where one has $H_k^* = H_{-k}$, $N_{-k}=N_k$, and $\varphi_{-k}=\varphi_k^*$ [which has also been used in \cref{eq6_Sopt_expansion}]; for the other \bcs, $H_k^*=H_k$.
Upon using \cref{eq6_h_final,eq6_h_final_expansion}, one obtains the constraint-induced value of the Lagrange multiplier,
\beq \lambda(T) = \frac{M}{\displaystyle \sum_k \frac{|\sigma_k(x_M)|^2}{2N_k(T)}}.
\label{eq6_Lagr_val}\eeq 
The solution of \cref{eq6_common} under the conditions in \cref{eq6_h_initial,eq6_h_final} is thus given by 
\beq h(x,t) = \frac{M}{Q(x_M,T,L)} \sum_k \frac{\hat\epsilon_k^2 \left[\exp\left(2(-1)^b \gamma_k T\right)-1\right]}{\gamma_k \tilde \epsilon_k} \frac{\sinh\left((-1)^b \gamma_k t\right)}{ \sinh\left((-1)^b \gamma_k T\right)} \sigma_k^*(x_M)\sigma_k(x)
\label{eq6_h_sol}\eeq 
with 
\beq Q(x_M,T,L) \equiv \sum_k \frac{|\sigma_k(x_M)|^2}{2 N_k(T)}  = \sum_k |\sigma_k(x_M)|^2 \frac{\hat\epsilon_k^2 \left[\exp\left(2(-1)^b\gamma_k T \right) -1\right] }{\gamma_k\tilde\epsilon_k}.
\label{eq6_Q}\eeq
It is useful to note that $\displaystyle H_k = \frac{M \sigma_k^*(x_M)}{2 Q(x_M,T,L) N_k(T)}$.
For the \bcs considered here and $k\neq 0$, one has $\hat \epsilon_k^2/\tilde \epsilon_k = \epsilon_k^b/\kappa_k^2$, $\epsilon_k^b=\epsilon_k$ as well as $\epsilon_k/\gamma_k< 0$ (see \cref{tab_eigenfunc}).
We emphasize that in general $Q(T/\tau)$ is only proportional to the function $\mathpzc{Q}(T/\tau)$ defined in \cref{eq_opt2_solh_Q_pbc,eq_opt2_solh_Q_Dir,eq_opt4_solh_Q_pbc,eq_opt4_solh_Q_DirNoFl} in the main text, because the latter results from \cref{eq6_h_sol} after performing some simplifications.
According to \cref{eq6_p_expansion,eq6_p_sol}, the conjugate field $p$ is given by
\beq p(x,t) = \frac{M}{Q(x_M,T,L)}\sum_k \frac{\exp\left( (-1)^b \gamma_k T\right)}{\kappa_k \exp\left((-1)^b \gamma_k t\right)} \sigma_k^*(x_M) \varphi_k(x).
\label{eq6_p_expr}\eeq 
Notably, this result implies that the initial and final configurations of $p(x,t)$ are fully determined by the corresponding ones for $h$ specified in \cref{eq6_h_initial,eq6_h_final}.
The optimal action in \cref{eq6_Sopt_expansion} reduces to
\beq \Scal\opt(x_M, M, T,L) = \frac{M^2}{2Q(x_M,T,L)} ,
\label{eq6_Sopt_expr}\eeq 
which is most easily proven by using \cref{eq6_p_sol} and the expression for $H_k$ stated after \cref{eq6_Q}.
Recall that the above results pertain to rescaled fields and time [see \cref{eq_h_rescaled}]. In particular, $\Scal\opt$ in \cref{eq6_Sopt_expr} gets multiplied by $\eta/D$ upon returning to dimensional variables [see \cref{eq_Sopt2_resc_relation}].

\subsection{Specialization to different \bcs}
\subsubsection{Periodic \bcs}
In the case of EW dynamics with periodic \bcs, the mass constraint in \cref{eq_zero_vol} is explicitly imposed. Since $\int_0^L \d x\, \exp(\im k x)=L\delta_{k,0}$ for $k=2\pi n /L$ with $n\in\mathbb{Z}$, this constraint implies
\beq h_{k=0}(t) = 0 = H_{k=0}
\label{eq6_zeromode_pbc}\eeq 
for the expansion coefficients defined in \cref{eq6_h_expansion,eq6_h_final_expansion}.
Since the profile $h(x,t)$ is real-valued, \cref{eq6_h_expansion} yields $h^* = \sum_{k=-\infty}^\infty h^*_k \exp(-2\pi \im k x/L) =  \sum_{k=-\infty}^\infty h_{-k} \exp(2\pi\im (-k) x/L) = h$ and thus
\beq h_k^* = h_{-k}.
\eeq
Furthermore, we have the symmetry property $N_k(T) = N_{-k}(T)$, as well as $\sigma_k(L/2) = (-1)^k/\sqrt{L}$ and $\sigma_{-k}(L/2)\sigma_{-k}(x) + \sigma_k(L/2)\sigma_k(x) = 2(-1)^k\cos(2\pi k x/L)/L = 2\cos(2\pi k (x/L-1/2))/L$.
Accordingly, \cref{eq6_Q,eq6_h_sol} can be written as
\beq h\pbc(x,t) = \frac{2M}{L Q\pbc(T,L)}\sum_{k=1}^\infty \frac{1-\exp\left(-2|\gamma_k| T\right)}{|\gamma_k|^{1-b/2}} \frac{\sinh\left(|\gamma_k| t\right)}{\sinh\left(|\gamma_k| T\right)} \cos\left(2\pi k(x/L-1/2)\right)
\label{eq6_h_pbc}\eeq 
with
\beq Q\pbc(T,L) = \frac{2}{L}\sum_{k=1}^\infty \frac{1-\exp\left(-2|\gamma_k| T\right)}{|\gamma_k|^{1-b/2}}.
\label{eq6_Q_pbc}\eeq
The factor 2 arises since the sum originally includes also negative $k$.
We have furthermore taken into account that, in the case of MH dynamics ($b=1$), the summand in \cref{eq6_h_pbc,eq6_Q_pbc} vanishes for $k= 0$ (which can be proven by carefully considering the limit $\gamma_k\to 0$), such that the zero mode is absent from the solution.
In fact, \cref{eq6_h_pbc} agrees with the expression obtained for MH dynamics in Ref.\ \cite{meerson_macroscopic_2016}.
In the case of EW dynamics without the mass constraint, the profile defined in \cref{eq6_h_pbc} would superimpose onto a linear center-of-mass motion according to \cref{eq6_hp_sol_k0_EW}.
 
\subsubsection{Dirichlet \bcs}
\label{app_std_Dirichlet_MH}
Both for standard and no-flux Dirichlet \bcs, \cref{eq6_h_sol} assumes the generic expression
\beq h\Dbc(x,t) = \frac{M}{Q\Dbc(x_M, T,L)} \sum_{k=1}^\infty \frac{1-\exp\left(-2|\gamma_k| T\right)}{|\gamma_k|^{1-b/2}\kappa_k} \frac{\sinh\left(|\gamma_k| t\right)}{\sinh\left(|\gamma_k| T\right)} \sigma_k\left(x_M\right) \sigma_k(x)
\label{eq6_h_Dir}\eeq 
with 
\beq Q\Dbc(x_M, T,L) = \sum_{k=1}^\infty \sigma_k^2\left(x_M\right) \frac{1-\exp\left(-2|\gamma_k| T\right)}{|\gamma_k|^{1-b/2}\kappa_k}.
\label{eq6_Q_Dir}\eeq
If a vanishing chemical potential is imposed at the boundaries, the eigenfunctions are given by the standard Dirichlet ones, $\sigma_k\DirCP(x) = \sqrt{2/L}\sin\left(\pi k x/L\right)$ with $\gamma_k=(\pi k/L)^4$. 
Taking $x_M=L/2$ [which is a convenient choice in the transient regime and minimizes the action in the equilibrium regime, see \cref{eq5_xM_DirCP}], one has
\beq  \sqrt{L/2}\, \sigma_k\DirCP\left(L/2\right) = 1,0,-1,0,1,\ldots
\label{eq6_eigen_Dir_halfpt}\eeq 
for $k=1,2,3,\ldots$, implying that only the odd modes contribute to the evolution of the profile.
Furthermore, we note the useful relation 
\beq \sigma_k\DirCP(L/2)\sigma_k\DirCP(x) = \frac{2}{L}\cos\left(\frac{\pi k}{L} \left(x-\frac{L}{2}\right)\right),\qquad k=1,3,5,\ldots .
\label{eq6_eigen_Dir_prodrep}\eeq 
In the case of Dirichlet no-flux \bcs, the corresponding eigenfunctions $\sigma_k\DirNoFl$ are reported in \cref{eq4_DirFl0_eigenf}. 
Here, one has $\sigma_k\DirNoFl(L/2)=0$ for odd $k$.  
In the equilibrium regime, $x_M$ as given in \cref{eq5_constr_minLoc} has to be used instead of $L/2$.

\begin{figure}[t]\centering
    \subfigure[]{\includegraphics[width=0.445\linewidth]{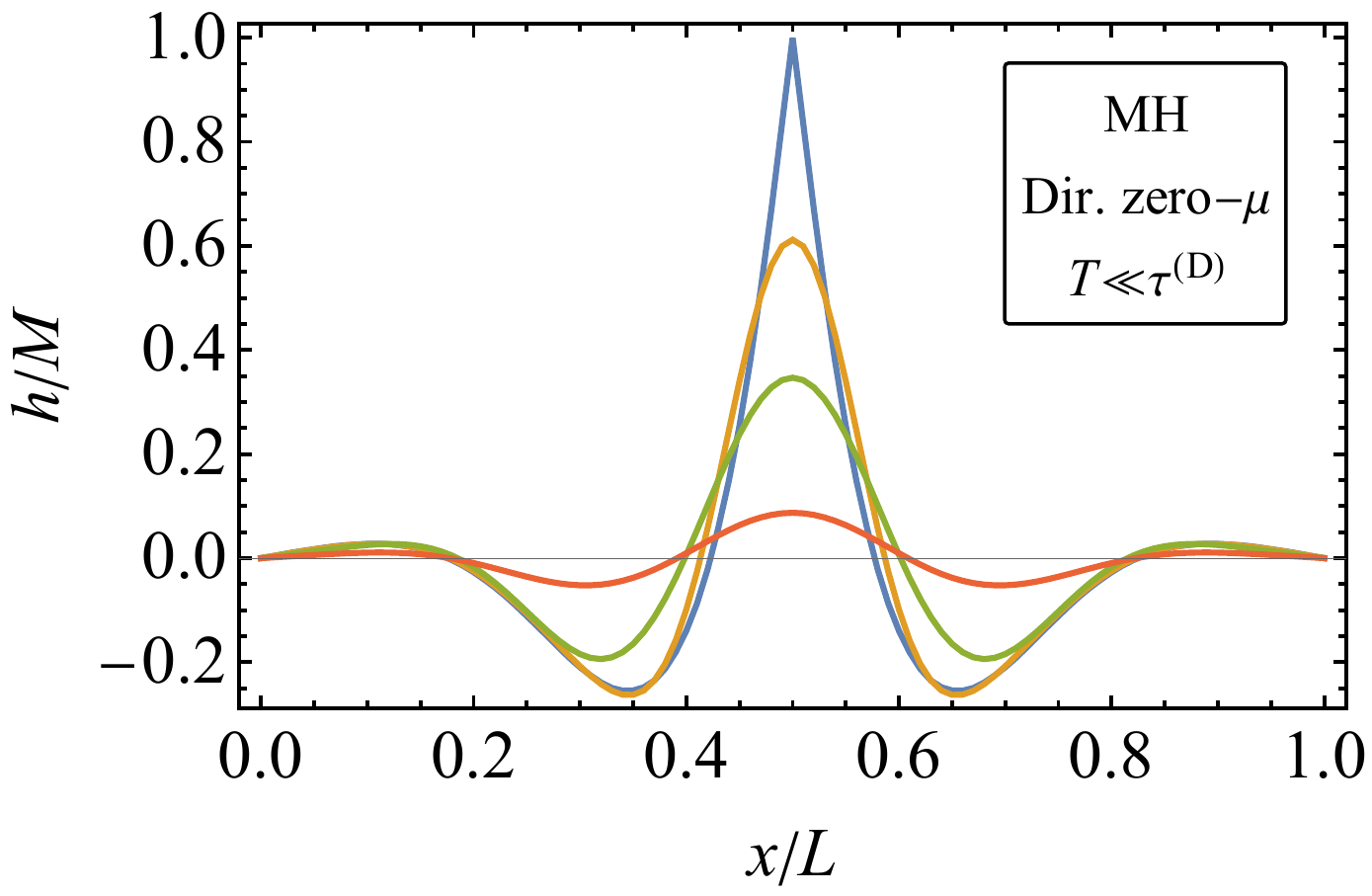}}\hfill 
    \subfigure[]{\includegraphics[width=0.43\linewidth]{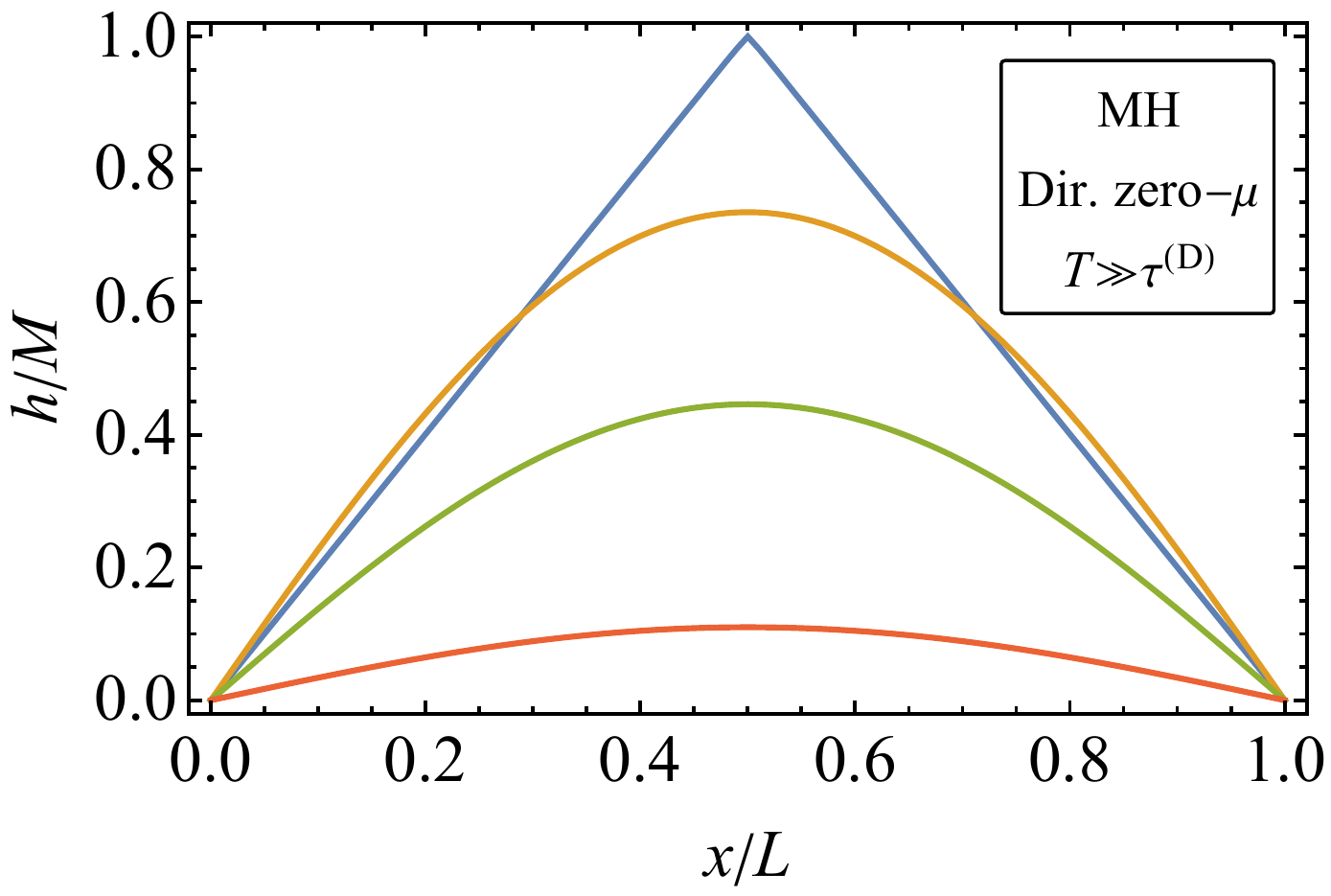}}
    \caption{(a,b) Time evolution of the optimal profile for the MH equation with Dirichlet \bcs and a vanishing chemical potential at the boundaries [\cref{eq6_h_Dir}] in (a) the transient regime ($T=10^{-3}\tau\DirCP$) and (b) the equilibrium regime ($T=100\tau\DirCP$). The curves correspond from center top to bottom to (a) $1-t/T = 0, 0.05, 0.3, 0.8$ and (b) $1-t/T=0, 10^{-5}, 10^{-4}, 5\times 10^{-4}$. The fundamental time scale is given by $\tau\DirCP=(L/\pi)^4$.}
    \label{fig_mft_prof_c_DirCP}
\end{figure}

The optimal profile $h\DirNoFl(x,t)$ for MH dynamics with Dirichlet no-flux \bcs is discussed in the main text [see \cref{eq_opt4_solh_DirNoFl}]. 
As a byproduct of the present analysis, we readily obtain the optimal profile $h\DirCP(x,t)$ for MH dynamics with Dirichlet zero-$\mu$ \bcs, which is illustrated in  \cref{fig_mft_prof_c_DirCP}. Mass is in general not conserved in this case.
Introducing the time scale 
\beq \tau\DirCP = \left(\frac{L}{\pi}\right)^4, \label{eq_c_timescale_DirCP}
\eeq 
the scaling form in \cref{eq_opt4_hscalform} applies with 
\beq \mathpzc{h}\DirCP(\upx,\upt,\upT) = \frac{1}{\mathpzc{Q}\DirCP(\upT)}\sum_{k=1,3,5,\ldots}^\infty \frac{1-\exp\left(-2 k^4 \upT\right)}{k^2} \frac{\sinh\left(k^4 \upt\right)}{\sinh\left(k^4 \upT\right)} \cos\left(\pi k \left(\upx-1/2\right)\right)
\label{eq_opt4_solh_DirCP}\eeq 
and  
\beq \mathpzc{Q}\DirCP(\upT) = \sum_{k=1,3,5,\ldots}^\infty \frac{1-\exp\left(-2 k^4 \upT\right)}{k^2}.
\label{eq_opt4_solh_Q_DirCP}\eeq
The above expressions for $\mathpzc{h}$ and $\mathpzc{Q}$ in fact coincide with the corresponding ones in the EW case [\cref{eq_opt2_solh_pbc,eq_opt2_solh_Dir}], except for the presence of $k^2$ instead of $k^4$.

\subsection{Limiting cases}
Introducing $\delta t\equiv T-t$, $\Gamma_k \equiv  \sfrac{\epsilon_k}{\gamma_k \kappa_k^2}$ and using \cref{tab_eigenfunc}, \cref{eq6_h_sol} can be simplified to
\beq h(x,\delta t) = \frac{M}{Q(T)}\sum_k \Gamma_k \left[ \exp\left(-|\gamma_k|(2T-\delta t)\right) - \exp\left(-|\gamma_k| \delta t\right) \right] \sigma_k^*(x_M)\sigma_k(x),
\label{eq6_h_simp}\eeq
where we suppressed further arguments of $Q$ and note that $\Gamma_k<0$ as well as $(-1)^b\gamma_k = -|\gamma_k|$. Here and in the following, $h$ is considered to be a function of $\delta t$ instead of $t$.
Specifically for $\delta t= 0$, \cref{eq6_h_simp} reduces to
\beq h(x, \delta t= 0) = \frac{M}{Q(T)} \sum_k \Gamma_k \left[\exp(-2|\gamma_k| T) - 1\right]  \sigma_k^*(x_M)\sigma_k(x).
\label{eq6_h_Dt0}\eeq 
Convenient analytical expressions for $h$ can be derived by replacing the sum in \cref{eq6_h_simp} by an integral using the Euler-Maclaurin formula. 
The error caused by this approximation is small if the summands in \cref{eq6_h_simp} vary significantly only over a few values of $k$. This is the case if $\delta t\ll \tau\simeq 1/|\gamma_1|$ (or, equivalently, $T\ll \tau$), since then the variation occurs for large $k$, where $|\gamma_k|\sim k^z$.
[For $T\to\infty$, on the other hand, the first term in \cref{eq6_h_simp} can be neglected, see \cref{sec_optprof_lateT}.]

\subsubsection{Transient regime ($T\ll \tau$)}
\label{sec_optprof_shortT}
\underline{Case $\delta t=0$.} We first consider the case $\delta t=0$.
For periodic \bcs, \cref{eq6_h_Dt0} becomes
\beq\begin{split}
h\pbc(x,\delta t=0)\big|_{T\ll \tau} &= \frac{2M}{L Q\pbc(T)} \left(\frac{L}{2\pi}\right)^2 \sum_{k=1}^\infty \frac{1-\exp\left[-(2\pi k\, (2T)^{1/z}/L)^z\right]}{k^2} \cos\left(\frac{2\pi k}{L}(x-L/2)\right) \\
&\simeq \frac{(2T)^{1/z} M }{\pi\, Q\pbc(T)} \int_0^\infty \d y \frac{1-e^{-y^z}}{y^2} \cos(y \xi) ,
\end{split}\eeq 
with the fundamental integral 
\beq \int_0^\infty \d y \frac{1-e^{-y^z}}{y^2} \cos(y \xi) = 
\begin{cases}\dps \sqrt{\pi} \exp\left(-\frac{\xi^2}{4}\right) + \onehalf \pi |\xi| \left[\mathrm{erf}\left(\frac{|\xi|}{2}\right)-1 \right] ,\qquad z=2,\\
\dps \Gamma\left(\frac{3}{4}\right) {}_1 F_3\left(-\frac{1}{4}; \frac{1}{4},\frac{1}{2}, \frac{3}{4}; \frac{\xi^4}{256}\right) + \frac{1}{8} \Gamma\left(\frac{1}{4}\right) \xi^2  {}_1 F_3\left(\frac{1}{4}; \frac{3}{4}, \frac{5}{4}, \frac{3}{2}; \frac{\xi^4}{256}\right) - \frac{\pi}{2}|\xi|,\qquad z=4,
\end{cases}
\label{eq6_integrals}\eeq 
and $\xi\equiv (x-L/2)/(2T)^{1/z}$.
Analogously, \cref{eq6_Q_pbc} evaluates to
\beq\begin{split} Q\pbc(T\ll\tau) &=  \frac{2}{L} \left(\frac{L}{2\pi}\right)^2 \sum_{k=1}^\infty \frac{1-\exp\left(-[2\pi k\, (2T)^{1/z}/L]^z\right)}{k^2} \\
&\simeq  \frac{(2T)^{1/z} }{\pi} \int_0^\infty \d y \frac{1-\exp(-y^z)}{y^2} = \frac{(2T)^{1/z}}{\pi} \Gamma\left(1-1/z\right),
\end{split}
\label{}\eeq
where, in the intermediate steps, the integration variable $k$ has been substituted by $y=2\pi k(2T)^{1/z}/L$. The lower integration boundary has been sent to zero since we consider $T\to 0$, noting that the associated error is negligible because the integrand vanishes for $y\to 0$.
Analogously, for Dirichlet zero-$\mu$ \bcs, using \cref{eq6_eigen_Dir_prodrep}, we obtain from \cref{eq6_h_Dir,eq6_Q_Dir,eq6_h_Dt0}:
\beq\begin{split} 
h\DirCP(x, \delta t=0)\big|_{T\ll \tau} &= \frac{2M}{L Q\Dbc(T)} \left(\frac{L}{\pi}\right)^2 \sum_{j=0}^\infty \frac{1-\exp\left(-\left[\pi (2j+1) (2T)^{1/z}/L\right]^z\right)}{(2j+1)^2} \cos\left(\frac{(2j+1)\pi}{L}(x-L/2)\right) \\
&= \frac{(2T)^{1/z} M }{\pi  Q\Dbc(T)} \int_0^\infty \d y \frac{1-\exp(-{y^z})}{y^2} \cos(y \xi ),
\end{split}\label{eq6_hDir_shortT}\eeq 
with  
\beq\begin{split} 
Q\DirCP(T\ll \tau) &=  \frac{2}{L}\left(\frac{L}{\pi}\right)^2 \sum_{j=0}^\infty \frac{1-\exp\left(-\left[\pi (2j+1) (2T)^{1/z}/L\right]^z\right)}{(2j+1)^2} \\
&= \frac{(2T)^{1/z} }{\pi} \int_0^\infty \d y \frac{1-\exp(-{y^z})}{y^2} = \frac{(2T)^{1/z} }{\pi} \Gamma\left(1-1/z\right).
\end{split}\eeq 
In order to evaluate the sum in \cref{eq6_h_Dt0} for Dirichlet no-flux \bcs, we assume a $k'$ such that, for $k\geq k'$, the eigenvalue $\gamma_k$ and the parameter $\kappa_k$ can be approximated by their respective asymptotic forms [see \cref{eq4_DirFl0_eigenval_approx,tab_eigenfunc}]
\beq \gamma\DirNoFl_k \simeq \left(\frac{(k+1/2)\pi}{L}\right)^4,\qquad \kappa_k\simeq \frac{L}{3}.
\eeq 
In the transient regime, we set $x_M=L/2$ [see \cref{eq6_Sopt_shortT} for justification] and thus have $\sigma_k\DirNoFl(L/2)=0$ for odd $k$.
For $T\ll (L/\omega_k)^4\ll \tau\DirNoFl$, terms with $k<k'$ in the sum in \cref{eq6_h_Dt0} are exponentially small and can be neglected.
For even $k$ with $k\geq k'$, we approximate $\sigma_k\DirNoFl$ by 
\beq \sigma\DirNoFl_k(x) \simeq (-1)^{k/2}\frac{2}{3} \cos\left[\pi\left(k+\onehalf\right)\left(x-\frac{L}{2}\right)\right].
\eeq 
While this approximation does not respect Dirichlet no-flux \bcs, it captures the oscillatory behavior of the actual $\sigma\DirNoFl_k$ well. A numerical comparison of the resulting scaling profile with the exact one justifies the above approximations \textit{a posteriori}.
Within the large $k$ approximation, we have $[\sigma_k\DirNoFl(L/2)]^2\simeq 2/3$ for even $k$. 
Accordingly, one obtains
\beq\begin{split} h\DirNoFl(x,\delta t=0)\big|_{T\ll \tau} &\simeq  \frac{2 M}{L Q\DirNoFl(T)} \left(\frac{L}{\pi}\right)^2 \sum_{k\geq k'}^\infty \frac{1-\exp\left(-\left[\pi(2j+1)(2T)^{1/4}/L\right]^4\right)}{(2j+1)^2} \cos\left(\frac{(2j+1)\pi}{L}(x-L/2)\right) \\
&\simeq \frac{(2T)^{1/4} M}{\pi Q\DirNoFl(T)} \int_0^\infty \d y \frac{1-\exp(-y^4)}{y^2} \cos(y\xi)
\end{split}\eeq
and analogously
\beq
Q\DirNoFl(T\ll \tau) = \frac{(2T)^{1/4} }{\pi} \Gamma\left(3/4\right).
\eeq 
As before, sending the lower integration boundary to zero is justified in the limit $T\to 0$.
In summary, in the transient regime, the asymptotic expressions of the static profiles $h(x,\delta t=0)$ for periodic and Dirichlet \bcs are identical and reduce to
\beq h(x,\delta t= 0)\big|_{T\ll \tau} = M \Hcal\left(\frac{x-L/2}{(2T)^{1/z}}\right),
\label{eq6_h_shortT}\eeq 
with the scaling function
\beq \Hcal(\xi)=
\begin{cases}\dps   \exp\left(-\frac{\xi^2}{4}\right) + \onehalf \sqrt{\pi} |\xi| \left[\mathrm{erf}\left(\frac{|\xi|}{2}\right)-1 \right] ,\qquad z=2,\\
\dps {}_1 F_3\left(-\frac{1}{4}; \frac{1}{4},\frac{1}{2}, \frac{3}{4}; \frac{\xi^4}{256}\right) + \xi^2 \frac{\Gamma\left(\frac{1}{4}\right)}{8\Gamma\left(\frac{3}{4}\right)} {}_1 F_3\left(\frac{1}{4}; \frac{3}{4}, \frac{5}{4}, \frac{3}{2}; \frac{\xi^4}{256}\right) - \frac{\pi}{2\Gamma\left(\frac{3}{4}\right)}|\xi|,\qquad z=4,
\end{cases}\label{eq6_h_shortT_scalF}\eeq 
which has the limits $\Hcal(0)=1$ and $\Hcal(\xi\to\infty)=0$.
The expression of $\Hcal$ for $z=4$ coincides with the result for periodic \bcs reported in Ref.\ \cite{meerson_macroscopic_2016}.
The profile given by \cref{eq6_h_shortT} does not respect mass conservation [\cref{eq_zero_vol}] for finite $T$. This can be readily shown by computing the mass using the last expression in \cref{eq6_hDir_shortT} before performing the integral over $y$.  
However, as $T\to 0$, the resulting error becomes negligible since the width of the profile rapidly shrinks.

The quantity $Q$ has been evaluated above for the particular choice $x_M=L/2$. Analogous calculations can in fact be performed for arbitrary $x_M$ with $0<x_M<L$, yielding
\beq Q(x_M, T\ll \tau) = (2T)^{1/z} \mathpzc{q}(x_M/(2T)^{1/z}),
\eeq 
with a scaling function $\mathpzc{q}$ that has the property $\mathpzc{q}(\zeta\to\infty)=\const$. 
Accordingly, the action in \cref{eq6_Sopt_expr} behaves as (see also Ref.\ \cite{meerson_macroscopic_2016})
\beq
\Scal\opt(x_M)\big|_{T\to 0} \propto T^{-1/z},
\label{eq6_Sopt_shortT}\eeq 
and becomes independent of $x_M$ for $0<x_M<L$ in the limit $T\to 0$. 
For $x_M\in \{0,L\}$, instead, Dirichlet \bcs imply $\sigma_k\ut{(D,D')}(x_M)=0$ for all $k$, such that $Q\ut{(D,D')}(x_M)$ [\cref{eq6_Q_Dir}] vanishes identically at the boundaries, resulting in a divergence of $\Scal\opt\ut{(D,D')}(x_M)$ for $x_M\in\{0,L\}$.
The fact that $\Scal\opt$ is independent of $x_M$ asymptotically in the transient regime justifies the choice $x_M=L/2$ made above.

\begin{figure}[t]\centering
    \subfigure[]{\includegraphics[width=0.425\linewidth]{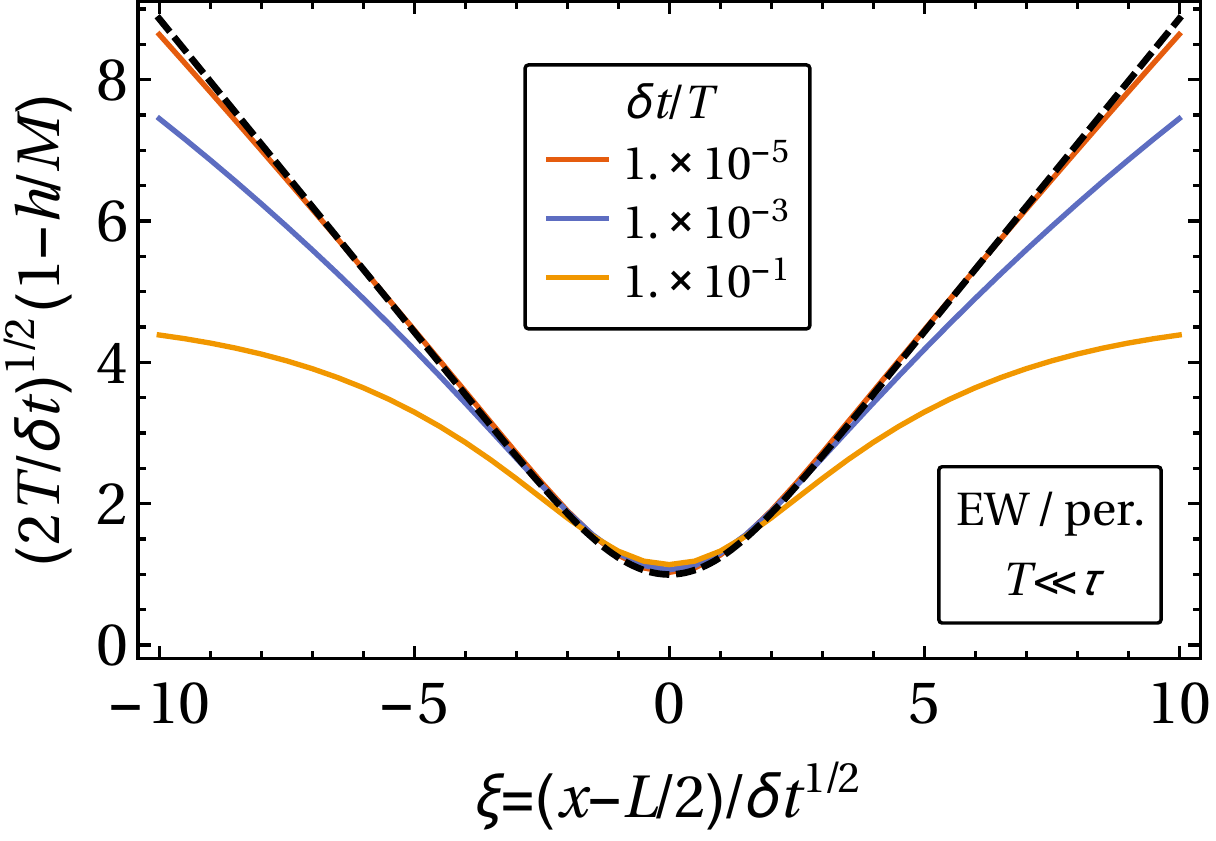}}\qquad 
    \subfigure[]{\includegraphics[width=0.43\linewidth]{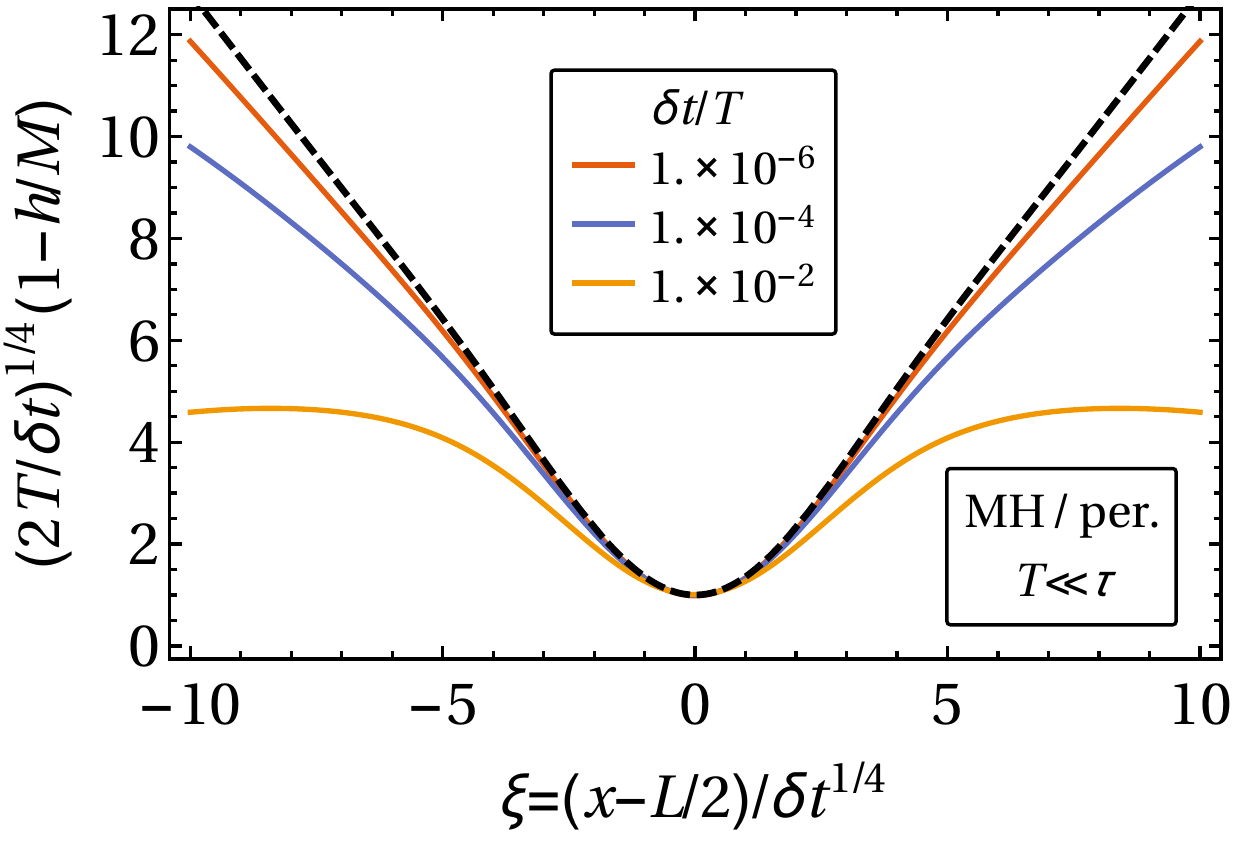}}
    \subfigure[]{\includegraphics[width=0.43\linewidth]{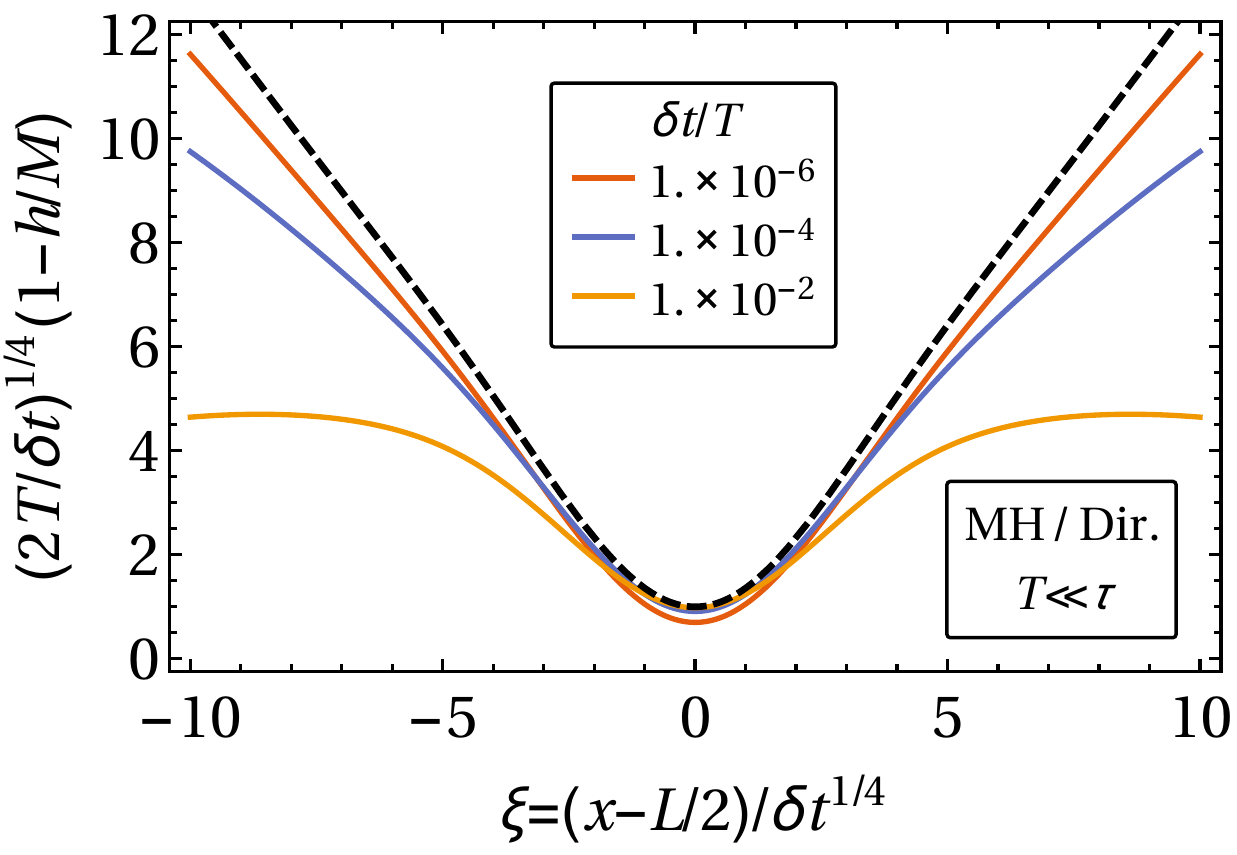}}
    \caption{Scaling behavior in the transient regime for a profile subject to (a) EW and (b) MH dynamics with periodic \bcs, and (c) MH dynamics with Dirichlet no-flux \bcs. The dashed black curve represents the scaling function \cref{eq6_h_asympt_scalF}, while the solid curves represent the full expression of the profile in \cref{eq6_h_pbc,eq6_h_Dir}, rescaled according to \cref{eq6_h_shortTdyn_asympt}. }
    \label{fig_mft_shortT_scaling_test}
\end{figure}

\underline{Case $\dt > 0$.} In order to obtain dynamic scaling profiles for nonzero $\delta t$ with $\delta t\ll \tau$ and $T\ll\tau$, we rewrite \cref{eq6_h_simp} as
\beq h(x,\delta t) = \frac{M}{Q(T)}\sum_k \Gamma_k \left\{ \left[\exp\left(-|\gamma_k|(2T-\delta t)\right) -1\right] + \left[1-\exp\left(-|\gamma_k| \delta t\right)\right]\right\} \sigma_k^*(x_M)\sigma_k(x).
\label{eq6_h_simp2}\eeq
Performing calculations analogous to those leading from \cref{eq6_h_Dt0} to \cref{eq6_h_shortT}, the corresponding dynamic scaling profile in the transient regime follows as
\beq h(x,\delta t\ll\tau)\big|_{T\ll \tau} = M \left(1-\frac{\dt}{2T}\right)^{1/z}  \Hcal\left(\frac{x-L/2}{(2T-\delta t)^{1/z}}\right) - M \left(\frac{\delta t}{2T}\right)^{1/z} \Hcal\left(\frac{x-L/2}{(\delta t)^{1/z}}\right).
\label{eq6_h_shortTdyn}\eeq 
For $x=L/2$ and $\dt\ll T$, \cref{eq6_h_shortTdyn} simplifies to $h(L/2,\dt)\simeq M-[\dt/(2T)]^{1/z}$. 
In order to obtain an analogous scaling form for $x\neq L/2$, we consider the expression
\beq \left(\frac{2T}{\dt}\right)^{1/z} \left[M- h(x,\dt)\right] \simeq \left(\frac{2T}{\dt}\right)^{1/z} M\left[1- \Hcal\left(\xi \left(\frac{\dt}{2T}\right)^{1/2}\right) + \left(\frac{\dt}{2T}\right)^{1/z} \Hcal(\xi)\right],\qquad \dt\ll T,
\label{eq6_h_shortTdyn_intermed}\eeq
where we introduced $\xi\equiv (x-L/2)/\dt^{1/z}$.
Expanding the r.h.s.\ in \cref{eq6_h_shortTdyn_intermed} to leading (i.e., zeroth) order in $\dt/T$, keeping $\xi$ fixed, yields the desired scaling form:
\beq h(x,\delta t\ll T)\big|_{T\ll \tau} \simeq M - M \left(\frac{\delta t}{2T}\right)^{1/z} \tilde \Hcal\left(\frac{x-L/2}{(\delta t)^{1/z}}\right),
\label{eq6_h_shortTdyn_asympt}\eeq 
with 
\beq \tilde\Hcal(\xi) = 
\begin{cases}\dps   \exp\left(-\frac{\xi^2}{4}\right) + \onehalf \sqrt{\pi}\, \xi\, \mathrm{erf}\left(\frac{\xi}{2}\right) ,\qquad z=2,\\
\dps {}_1 F_3\left(-\frac{1}{4}; \frac{1}{4},\frac{1}{2}, \frac{3}{4}; \frac{\xi^4}{256}\right) + \xi^2 \frac{\Gamma\left(\frac{1}{4}\right)}{8\Gamma\left(\frac{3}{4}\right)} {}_1 F_3\left(\frac{1}{4}; \frac{3}{4}, \frac{5}{4}, \frac{3}{2}; \frac{\xi^4}{256}\right) ,\qquad z=4.
\end{cases}\label{eq6_h_asympt_scalF}\eeq 
As shown in \cref{fig_mft_shortT_scaling_test}, the scaling form in \cref{eq6_h_shortTdyn_asympt} provides an accurate approximation to the full profiles [\cref{eq6_h_pbc,eq6_h_Dir}] in a region around $x_M$. The size of this region increases as $\delta t/T\to 0$.

\subsubsection{Equilibrium regime ($T\gg \tau$)}
\label{sec_optprof_lateT}
In the long-time limit, $T\to\infty$, the first term in the square brackets in \cref{eq6_h_simp} can be neglected, as can the exponential function in \cref{eq6_h_Dt0}.
Accordingly, $h$ becomes independent of $T$ and \cref{eq6_h_simp} reduces to
\beq h\eq(x,\dt)\equiv h(x,\delta t)\big|_{T\to\infty} = -\frac{M}{Q\eq} \sum_k \frac{\epsilon_k}{\gamma_k \kappa_k^2} \exp\left(-|\gamma_k| \delta t\right) \sigma_k(x_M)\sigma_k(x),
\label{eq6_eq_h}\eeq 
with [see \cref{eq6_Q}]
\beq Q\eq = -\sum_k |\sigma_k(x_M)|^2 \frac{\epsilon_k}{\gamma_k\kappa_k^2}.
\label{eq6_eq_Q}\eeq 
\underline{Case $\delta t=0$.} For $\delta t=0$, the expressions in \cref{eq6_eq_h,eq6_eq_Q} can be evaluated exactly in the case of periodic and Dirichlet zero-$\mu$ \bcs:
according to \cref{tab_eigenfunc}, we have
\beq Q\eq = \sum_k |\sigma_k(x_M)|^2 |\gamma_k|^{b/2-1}
\eeq
as well as
\beq h\eq(x,\delta t=0) = \frac{M}{Q\eq} \sum_k |\gamma_k|^{b/2-1} \sigma_k(x_M) \sigma_k(x),
\eeq 
with $|\gamma_k\pbc|^{b/2-1} = (2\pi k/L)^2$ for periodic and $|\gamma_k\DirCP|^{b/2-1}=(\pi k/L)^2$ for Dirichlet zero-$\mu$ \bcs, independently of the value of $b\in \{0,1\}$.
Specifically, one obtains, invoking known Fourier series representations (see, e.g., Ref.\ \cite{gradshteyn_table_2014})
\begin{subequations}\begin{align}
Q\eq\pbc &= 2L \sum_{k=1}^\infty \frac{1}{(2\pi k)^2} = \frac{L}{12}, \\
Q\eq\DirCP &= 2L \sum_{k=1,3,5,\ldots}^\infty \frac{1}{(\pi k)^2} = \frac{L}{4},
\end{align}\label{eq6_Q_eq}\end{subequations}
and analogously,
\begin{subequations}\begin{align}
h\eq\pbc(x,\delta t= 0) = \frac{2L M}{(2\pi)^2 Q\eq\pbc}\sum_{k=1}^\infty \frac{\cos(2\pi k(x/L-1/2))}{k^2} &=  M\left[ 1-6\Bigg|\frac{x}{L}-\onehalf\Bigg| + 6\left(\frac{x}{L}-\onehalf\right)^2 \right], \label{eq6_eq_prof_pbc}\\
h\eq\DirCP(x,\delta t= 0) = \frac{2 L M}{\pi^2 Q\eq\DirCP} \sum_{n=0}^\infty (-1)^n \frac{\sin((2n+1)\pi x/L)}{(2n+1)^2} &= M-M\left|1-\frac{2x}{L}\right|,
\end{align}\label{eq6_h_eq_limprof}\end{subequations}
where we used \cref{eq6_eigen_Dir_halfpt}.
These expressions coincide with the ones in \cref{eq5_prof_pbc,eq5_prof_DirCP} for the respective \bcs.
A direct proof of the equivalence between $h\eq\DirNoFl(x,\dt =0)$ and the expression in \cref{eq5_prof_DirNoFl} is not available owing to the non-algebraic dependence of $\omega$ on $k$ [see \cref{eq_DirFl0_det}]. 

\underline{Case $\delta t>0$.} For $T\to\infty$ and nonzero $\delta t\ll \tau$, asymptotic scaling profiles can be derived from \cref{eq6_eq_h} analogously to the calculation leading from \cref{eq6_h_Dt0} to \cref{eq6_h_shortT}.
In the conversion of the sum to an integral, however, possible divergences have to be taken care of.
In the case of periodic \bcs one obtains, taking $x_M=L/2$,
\beq\begin{split}
h\pbc(x,\delta t)\big|_{T\to \infty} &= \frac{2M}{L Q\pbc\eq} \left(\frac{L}{2\pi}\right)^2 \sum_{k=1}^\infty \frac{\exp\left[-(2\pi k\, \dt^{1/z}/L)^z\right]}{k^2} \cos\left(\frac{2\pi k}{L}(x-L/2)\right) \\
&\simeq \frac{(\delta t)^{1/z} M }{\pi\, Q\pbc\eq} \int_{Y_1}^\infty \d y \frac{e^{-y^z}}{y^2} \cos(y \xi) ,
\end{split}\label{eq6_h_pbc_lateT0}\eeq 
where $Y_1\equiv 2\pi(\delta t)^{1/z}/L$ and $\xi\equiv (x-L/2)/(\delta t)^{1/z}$ is a scaling variable. In order to take into account the singularity of the integral for $Y_1\to 0$, we write
\beq \int_{Y_1}^\infty \d y \frac{e^{-y^z}}{y^2} \cos(y \xi) = \int_{Y_1}^\infty \d z \frac{e^{-y^z}-1}{y^2} \cos(y \xi) + \int_{Y_1}^\infty \d y \frac{ \cos(y \xi)}{y^2} .
\eeq
In the first term on the r.h.s.\ the limit $Y_1\to 0$ can be performed, yielding \cref{eq6_integrals} up to a sign. For the second term, we obtain
\beq \int_{Y_1}^\infty \d y \frac{ \cos(y \xi)}{y^2} = \frac{\cos(\xi Y_1)}{Y_1} - \onehalf \pi |\xi| + \xi\, \mathrm{Si}(\xi Y_1),
\eeq
where $\mathrm{Si}$ is the sine integral \cite{olver_nist_2010}. Since $\xi Y_1 = 2\pi (x/L-1/2)$, expanding to first order in $(x/L-1/2)$, using $\mathrm{Si}(\zeta)\simeq \zeta + \Ocal(\zeta^2)$, we obtain
\beq \int_0^\infty \d y \frac{e^{-y^z}}{y^2} \cos(y \xi) \simeq  
\begin{cases}\dps \frac{1}{Y_1} -\sqrt{\pi} \exp\left(-\frac{\xi^2}{4}\right) - \onehalf \pi \, \xi\, \mathrm{erf}\left(\frac{\xi}{2}\right) ,\qquad z=2,\\
\dps \frac{1}{Y_1} -\Gamma\left(\frac{3}{4}\right) {}_1 F_3\left(-\frac{1}{4}; \frac{1}{4},\frac{1}{2}, \frac{3}{4}; \frac{\xi^4}{256}\right) - \frac{1}{8} \Gamma\left(\frac{1}{4}\right) \xi^2  {}_1 F_3\left(\frac{1}{4}; \frac{3}{4}, \frac{5}{4}, \frac{3}{2}; \frac{\xi^4}{256}\right) ,\qquad z=4.
\end{cases}
\label{eq6_integrals_lateT}\eeq 
For consistency in the approximation, we calculate $Q\eq$ in \cref{eq6_Q_eq} in an analogous fashion, obtaining
\beq Q\eq\pbc \simeq  \frac{L}{2\pi^2} \int_1^\infty \d k\, k^{-2} = \frac{L}{2\pi^2}.
\label{eq6_Qeq_approx}\eeq 
Inserting \cref{eq6_integrals_lateT,eq6_Qeq_approx} in \cref{eq6_h_pbc_lateT0} yields
\beq h(x,\delta t)\big|_{T\to\infty} \simeq M -  M (\delta t)^{1/z} \Gamma(1-1/z) \tilde\Hcal\left(\frac{x-L/2}{\delta t^{1/z}}\right),
\label{eq6_h_lateT_full}\eeq 
with the scaling function $\tilde\Hcal$ given in \cref{eq6_h_asympt_scalF}.
Hence, asymptotically, the scaling functions in the transient and the equilibrium regime are identical.
The calculation proceeds analogously for Dirichlet \bcs, yielding for $h\eq\DirCP$ the same result as in \cref{eq6_h_lateT_full}.
Moreover, \cref{eq6_h_lateT_full} applies also to Dirichlet no-flux \bcs, since in the asymptotic regime, i.e., for $\xi\lesssim \Ocal(1)$ with $\delta t\ll \tau$, the precise value of $x_M$ is irrelevant, despite \cref{eq_xM_DirNoFl}.

\subsubsection{Effect of an upper mode cutoff}

Above results pertain to a continuum system, which can sustain an infinite number of eigenmodes.
Conversely, the presence of a minimal length scale in the system (e.g., a lattice constant) gives rise to an upper bound on the mode spectrum.
Accordingly, the sums in \cref{eq6_h_sol,eq6_Q} are bounded by a maximum mode index $k\cro$. Associated with this mode is a relaxation rate $\gamma_{k\cro}$, which defines a \emph{cross-over time}
\beq \tau\cro \equiv \frac{1}{\gamma_{k\cro}}.
\eeq 
In a system with a mode cutoff, for times $\delta t\ll \tau\cro$ and $\delta t\ll T$, \cref{eq6_h_simp} can be approximated as
\beq\begin{split} h(x,\delta t \lesssim \tau\cro) 
&\simeq \frac{M}{Q(T)}\sum_k^{k\cro} \Gamma_k \big[ \exp\left(-2|\gamma_k| T\right) - 1 + |\gamma_k| \delta t \big] \sigma_k^*(x_M)\sigma_k(x) \\
&= h(x,0) + \delta t \frac{M}{Q(T)}\sum_k^{k\cro} \Gamma_k  |\gamma_k| \sigma_k^*(x_M)\sigma_k(x)\,  , 
\end{split}\label{eq6_h_cutoff_small_dt}\eeq
where $h(x,0)$ is the static profile defined in \cref{eq6_h_Dt0}.
Note that the second term in the last line of \cref{eq6_h_cutoff_small_dt} is negative owing to the sign of $\Gamma_k$.
Hence, for a bounded mode spectrum, the algebraic time evolution (with exponent $1/z$) of the profile described by \cref{eq6_h_shortTdyn_asympt,eq6_h_lateT_full} crosses over to a linear one in $\delta t$ for small times, $\delta t\lesssim \tau\cro$.
This behavior applies both in the transient and the equilibrium regime, independently from the \bcs.

%

\end{document}